\documentclass[12pt]{article}
\usepackage{graphicx} % Required for inserting images
\usepackage{float}
\usepackage[authoryear,round]{natbib} % Required for generating bibliography in APA 7th style
\usepackage{amsmath}
\usepackage{graphicx,psfrag,epsf}
\usepackage{enumerate}
\usepackage{authblk}
\usepackage[colorlinks=true,linkcolor=blue,citecolor=blue,urlcolor=blue]{hyperref} 
\usepackage{setspace}
\usepackage{titlesec}
\usepackage{chngcntr}
\usepackage{setspace}
\usepackage{etoolbox}
\usepackage{booktabs}
\usepackage{multirow}
\usepackage{pdflscape}
\usepackage{threeparttable} % For adding notes to the table

\title{Improving the Estimation of Site-Specific Effects and their Distribution in Multisite Trials}

\author[1]{JoonHo Lee\thanks{jlee296@ua.edu}}
\author[2]{Jonathan Che}
\author[3]{Sophia Rabe-Hesketh}
\author[3]{Avi Feller}
\author[2]{Luke Miratrix}

\affil[1]{University of Alabama}
\affil[2]{Harvard University}
\affil[3]{University of California, Berkeley}

\begin{document}
\maketitle

%%% Set double spacing & Indentation
\begin{doublespace}
\setlength{\parindent}{2em}

%%%%%%%%%%%%%%%%%%%%%%%%%%%%%%%%%%%%%%
%%% Abstract & Keywords
%%%%%%%%%%%%%%%%%%%%%%%%%%%%%%%%%%%%%%

%%% Abstract
\begin{abstract}
In multisite trials, researchers are often interested in several inferential goals: estimating treatment effects for each site, ranking these effects, and studying their distribution. This study seeks to identify optimal methods for estimating these targets. Through a comprehensive simulation study, we assess two strategies and their combined effects: semiparametric modeling of the prior distribution, and alternative posterior summary methods tailored to minimize specific loss functions. Our findings highlight that the success of different estimation strategies depends largely on the amount of within-site and between-site information available from the data. We discuss how our results can guide balancing the trade-offs associated with shrinkage in limited data environments.
\end{abstract}

%%% Keywords
\noindent%
{\it Keywords:} multisite trials, site-specific effects, finite-population estimand, Dirichlet process mixture; constrained Bayes; triple-goal estimator; heterogeneous treatment effects

\clearpage

%%%%%%%%%%%%%%%%%%%%%%%%%%%%%%%%%%%%%%
%%% Introduction
%%%%%%%%%%%%%%%%%%%%%%%%%%%%%%%%%%%%%%

\section{Introduction}

%%% Background + super poulation inferential goals

Multisite trials, which conduct (nearly) identical randomized experiments simultaneously at multiple sites, have become increasingly prevalent as an experimental design in education \citep{spybrook2009examination, spybrook2014detecting, raudenbush2015learning}. Multisite trials can be viewed as a form of planned meta-analysis \citep{bloom2017using}. By synthesizing evidence from multiple contexts to estimate the average treatment effect, multisite trials can establish a more robust foundation for general policy recommendations \citep{meager2019understanding}. Understanding the extent of treatment effect variation allows researchers to determine how an intervention may perform in diverse contexts \citep[e.g.,][]{sabol2022exploring, schochet2022children}.

%%% Finite-population estimands
Expanding on such meta-analytic objectives of multisite trials, researchers occasionally direct their attention to the sites in the sample itself. This approach pursues several inferential goals, including (a) estimating individual site-specific treatment effects, (b) ranking the sites, and (c) analyzing the distribution of site-specific effects. Institutional performance evaluation studies have pioneered the examination of such fine-grained, site-specific parameters. For instance, school or teacher effectiveness studies directly aim to estimate individual school- or teacher-specific “value-added” effect parameters \citep{angrist2017leveraging, chetty2014measuringa, mccaffrey2004models, mountjoy2021returns}. There has also been a sustained interest in generating rankings or league tables based on estimated effectiveness for schools or other service providers \citep{goldstein1996league, lockwood2002uncertainty, mogstad2020inference}. Performance evaluation objectives also include estimating empirical distribution functions (EDFs) or histograms of site-specific parameters \citep{kline2022systemic}, identifying outliers or hot spots \citep{ohlssen2007hierarchical}, and determining the proportion of sites with effects surpassing a threshold \citep{gu2023invidious}. These practices can be effectively adapted to contexts that investigate site-specific treatment effects in multisite trial settings.

%%% Research aim - early introduction
%%% Standard approaches and their limitations

In this paper, we conduct a comprehensive simulation study to examine how different analytic strategies can enhance the estimation of site-specific effects under various conditions. As we cannot directly observe the true individual site-specific effects, we must model them using observed data. Standard modeling approaches depend on parametric distributional assumptions in random-effect multilevel models. Generally, random effects are assumed to follow a normal distribution, and empirical Bayes (EB) estimates are used for the random effects themselves \citep[e.g.,][]{raudenbush2015learning, weiss2017much}. When adopting a fully Bayesian approach, empirical Bayes is replaced by the marginal posterior means (PM) of the random effects.

%%% Flexible modeling approches on prior G

Assuming a normal distribution for random effects can be problematic when the true distribution is far from normal. For instance, \cite{mcculloch2011misspecifying} discovered that when the true distribution is multi-modal or long-tailed, the distribution of the EB predictions may reflect the assumed Gaussian distribution rather than the true distribution of effects. To address this issue and safeguard against model misspecification, researchers have proposed more flexible distributional assumptions for random effects. These alternatives encompass continuous parametric non-Gaussian distributions \citep{liu2008skew}; arbitrary discrete distributions obtained using nonparametric maximum likelihood estimation \citep{rabe2003correcting}; and mixture distributions \citep{ghidey2004smooth, paddock2006flexible, verbeke1996linear}.

%%% Posterior summary method approaches

Instead of relaxing the normality assumption, some approaches replace EB or PM estimators with alternative posterior summaries, such as constrained Bayes \citep[CB,][]{ghosh1992constrained} and triple-goal \citep[GR\footnote{The abbreviation "GR" denotes the dual inferential objectives: the EDF ($G$) and the rank ($R$) of site-specific parameters, commonly referenced as the triple-goal estimator in studies such as \cite{paddock2006flexible}.},][]{shen1998triple} estimators. These alternatives, designed to correct shrinkage-induced underdispersion in PM estimates, directly adjust the loss function minimized by the estimator in order to target specific inferential goals. Such strategies have received less attention compared to flexible modeling of the random-effects distribution.

%%% Joint application of two strategies

In practice, the joint application of these two strategies has been relatively rare, with only a few notable exceptions \citep[e.g.,][]{paddock2006flexible, lockwood2018flexible}. The costs and benefits of these strategies have not been systematically compared in previous simulation studies exploring similar topics \citep[e.g.,][]{kontopantelis2012performance, paddock2006flexible, rubio2018methodological}. For instance, if the true distribution is not Gaussian and the inferential goal is to estimate the empirical distribution of site-specific effects, a question arises about which approach performs better: (a) combining the misspecified Gaussian model for random effects with a targeted posterior summary method (CB or GR), or (b) utilizing the flexible semiparametric model for the prior in conjunction with the misselected posterior mean estimator that solely aims at the optimal estimation of individual site-specific effects. To our knowledge, no prior studies have compared these two strategies within the context of multisite trials, thereby revealing an unexplored area in the literature that warrants further investigation.

%%% Research aims and contributions

This paper offers four important contributions. First, our simulations focus on educational applications, which are often characterized by small sample sizes. Beyond the standard relative comparisons between methods, we offer practical guidelines for practitioners in these low-data settings, acknowledging that even the best method may deliver suboptimal results. Second, we bring attention to the frequently overlooked distinction between super population and finite-population estimands (e.g., the distribution of impacts across sites in the sample vs. the distribution of sites in the population where the sample came from), which serves to frame and motivate different estimators. Third, rather than solely focusing on one strategy, we examine the effects of combining flexible random-effects distributions with alternative posterior summaries. Finally, we apply insights gleaned from our simulation to real-world multisite trial data, illustrating how the selection of analytic strategies can considerably influence site-specific estimates.

%%% Paper organization

This paper is structured as follows. First, we introduce standard approaches for modeling super population and finite-population effects. Next, we discuss inferential goals, identify potential threats to inferences for site-specific effects, and outline two strategies to improve these inferences. We then provide a comprehensive description of our simulation study’s design and results, and follow with a real-data example that demonstrates the application of these strategies. Finally, we summarize our findings and consider their practical implications.

%%%%%%%%%%%%%%%%%%%%%%%%%%%%%%%%%%%%%%
%%% 2. Standard approaches to 
%%%     modeling site-specific effects
%%%%%%%%%%%%%%%%%%%%%%%%%%%%%%%%%%%%%%

\section{Standard approaches to modeling site-specific effects}

%%%%%%%%%%%%%%%%%%%%%%%%%%%%%%%%%%%%%%
%%% 2.1. Basic Setup: the Rubin (1981) model
%%%

\subsection{Basic setup: the Rubin (1981) model}

%%% Rubin model setup

Suppose a multisite trial consists of $J$ sites, indexed by $j=1, \ldots, J$, where identical treatments are administered within each site. The analytic setting considered in this paper is based on the Rubin (1981) model for parallel randomized experiments, also known as a random-effects meta-analysis \citep{raudenbush1985empirical}. For data inputs, this model requires only the observed or estimated effects $\widehat{\tau}_j$ from each of the $J$ sites, along with their corresponding squared standard errors $\widehat{se}_j^2$. The $\widehat{\tau}_j$ are estimating $\tau_j$, the true average impact in site $j$; under the potential outcomes framework, this would be the difference in the average outcomes of units if all were treated minus if all were control; we would make the usual assumptions about non-interference of treatment and a well-defined treatment (SUTVA) both within and across sites to make the estimand well defined (see, e.g., \cite{imbens2015causal} for further discussion). The $\widehat{\tau}_j$ and $\widehat{s e}_j^2$ values are obtained through maximum likelihood (ML) estimation using data exclusively from site $j$. Our meta-analytic approach closely aligns with the fixed intercept, random coefficient (FIRC) estimator \citep{raudenbush2015learning}, as it focuses on modeling only the true, randomly varying treatment effect $\tau_j$. This approach effectively circumvents some of the strong joint distributional assumptions typically made between $\tau_j$ and a randomly varying intercept or control group mean \citep{bloom2017using}.

The first stage of the hierarchical model describes the relationship between the observed statistics $\widehat{\tau}_j$ and the parameter $\tau_j$:

\begin{equation*}
\widehat{\tau}_j \mid \tau_j, \widehat{s e}_j^2 \sim N\left(\tau_j, \widehat{s e}_j^2\right) \quad j=1, \ldots, J
\end{equation*}

\noindent where
\begin{equation}\label{eq:1_Rubin}
    \widehat{s e}_j^2=\left(\frac{1}{p_j}+\frac{1}{1-p_j}\right) \cdot \frac{s_j^2\left(1-R^2\right)}{n_j}.
\end{equation}

\noindent When estimating average impacts, the central limit theorem applied to the individuals within the site provides the normality, or something close to it. For the standard error, $n_j$ denotes the sample size and $p_j$ represents the proportion of units treated in site $j$. $s_j^2$ signifies the (assumed constant across treatment arms) variance in outcomes within treatment arms for a given site. The coefficient $R^2$ reflects the proportion of variance in the outcome explained by unit-level covariates. For example, if $R^2=0.6$, e.g., due to having a reasonable pre-test score of an academic outcome, a $60\%$ improvement in the precision of $\widehat{\tau}_j$ due to covariate adjustment can be expected for a fixed $n_j$.

The second stage assumes that the site average impacts $\tau_j$ are independent and identically distributed ($i.i.d.$) with a specific prior distribution\footnote{We refer to $G$ as a `prior distribution' to be consistent with standard practices in Bayesian hierarchical models \citep{gelman2013bayesian, rabe-hesketh_multilevel_2022}. In this framework, $G$ represents our prior knowledge or beliefs about the parameter $\tau_j$ before observing the data from site $j$. In contrast, within the frequentist framework, $G$ is interpreted as a distributional assumption regarding $\tau_j$.} $G$,

\begin{equation}\label{eq:2_Rubin}
\tau_j \mid \tau, \sigma^2 \sim N\left(\tau, \sigma^2\right) \quad j=1, \ldots, J.
\end{equation}

\noindent The prior distribution is generally unknown but is classically specified as a Gaussian distribution $G$ with two hyperparameters: $\tau$, the average treatment effect across the sites, and $\sigma^2$, the variance in true site-specific effects $\tau_j$ across sites, both defined at the super population level. Here $\tau_j$ represents the unadjusted intent-to-treat (ITT) effect within each site. Consequently, the variance $\sigma^2$ encapsulates the aggregate variation in treatment effects across sites, stemming from two main sources: ``compositional" differences in unit-level covariates across sites and ``contextual" differences in site-level covariates or unobserved compositional differences \citep{lu2023you, rudolph2018composition}. One could redefine $\tau_j$ to be the conditional or adjusted site-level average treatment effect, potentially estimated by including covariate by treatment interaction terms; this would typically results in a reduced $\sigma^2$. This reduction occurs because such an analysis would account for and removes the variation attributed to compositional differences across sites, though $\sigma^2$ could also increase in certain scenarios as highlighted by \cite{lu2023you}. In the work of this paper, this distinction is immaterial: we seek to estimate, with $\widehat{\tau}_j$, the distribution defined by the $\tau_j$.

%%% The key insight of the Rubin model

The Bayesian hierarchical framework plays a vital role in separating the genuine heterogeneity in true effects $\tau_j$ across different sites ($\sigma^2$) from the sampling variation of each estimated effect $\widehat{\tau}_j$ around its true effect within sites ($\widehat{s e}_j^2$). A key focus in our analytic setting is to assess the relative magnitudes of $\sigma^2$ and $\widehat{s e}_j^2$, aiming to understand how this signal-to-noise ratio influences the efficacy of various estimation strategies. Particularly when $\widehat{s e}_j^2$ is relatively large, the $\widehat{\tau}_j$ values will generally be overdispersed compared to true $\tau_j$ values. This can lead to a misrepresentation of the rank order of effects across different sites; e.g., when the $\widehat{s e}_j^2$ vary due to factors such as site size, the smaller sample sizes might produce more extreme estimates. When estimating conditional effects, incorporating unit-level covariates into our analysis could concurrently reduce both $\sigma^2$ and $\widehat{s e}_j^2$, although these reductions might occur at different rates, or in some cases, $\sigma^2$ might even increase. This introduces an additional layer of complexity to the comparative analysis of these two variances. Therefore, to ensure focused and clear exposition and formulas regarding the comparisons of $\sigma^2$ and $\widehat{s e} j^2$, our study deliberately defers the inclusion of conditional effects and covariates interacted with treatment to future research; covariates for precision gain will simply decrease the standard errors and increase the information of the estimates regarding the estimands and all our results carry.

%%%%%%%%%%%%%%%%%%%%%%%%%%%%%%%%%%%%%%
%%% 2.2. Site-specific parameter estimation
%%%

\subsection{Site-specific parameter estimation}

%%% Rubin model estimation

In the special case of the normality assumption for $\tau_j$ the conditional posterior distribution of $\tau_j$ given the hyperparameters $\tau$ and $\sigma^2$ is normal and the corresponding conditional posterior mean and variance of $\tau_j$ have simple closed-form expressions:

\begin{equation*}
\tau_j \mid \tau, \sigma^2, \hat{\tau}_j \sim N\left(\tau_j^*, V_j\right) \quad j=1, \ldots, J,
\end{equation*}

\noindent where

\begin{equation}\label{eq:3_shrinkage}
\tau_j^*=\tau + \left(\hat{\tau}_j-\tau\right) \cdot \frac{\sigma^2}{\sigma^2+\widehat{s e}_j^2}, \quad V_j=\frac{1}{\sigma^2}+\frac{1}{\widehat{s e}_j^2}
\end{equation}

%%% Shrinkage factor

\noindent In other words, the posterior mean effect, $\tau_j^*$, is the observed mean effect, $\hat{\tau}_j$, shrunk toward the prior mean effect, $\tau$ \citep{gelman2013bayesian}. The weight $S_j=\sigma^2 /\left(\sigma^2+\widehat{s e}_j^2\right)$, commonly referred to as the shrinkage factor, indicates the reliability of the ML estimator $\hat{\tau}_j$, and is defined as the proportion of the variance of the ML estimator attributable to genuine underlying heterogeneity across sites \citep{rabe-hesketh_multilevel_2022}. If $\widehat{s e}_j^2=0$, the ML estimator $\hat{\tau}_j$ is perfectly precise or reliable, and the posterior mean and the ML estimator are identical $\left(\tau_j^*=\hat{\tau}_j\right)$. A large $\widehat{s e}_j^2$ indicates the data is less informative, relative to the variation in $\tau$, resulting in the posterior mean effect, $\tau_j^*$, being more significantly shrunk toward the prior mean effect, $\tau$.

%%% Average reliability

To assess the magnitude of $\sigma^2$ relative to the sampling error $\widehat{s e}_j^2$ for each site, we use the shrinkage factor $S_j$ as a direct measure for site $j$. For a comprehensive evaluation across all $J$ sites, we calculate the geometric mean of $\widehat{s e}_j^2$, yielding an average measure of reliability for the ML estimator $\hat{\tau}_j$, which we call the informativeness ($I$):

\begin{equation}\label{eq:4_I-level}
I = \frac{\sigma^2}{\sigma^2 + \exp \left(\frac{1}{J} \sum_{j=1}^{J} \ln (\widehat {se}_j^2) \right)}.
\end{equation}

\noindent The $I$ value ranges between 0 and 1 . Higher values of $I$ indicate the $\hat{\tau}_j$ 's provide greater information about the $\tau_j$'s, on average, as compared to lower values.

%%%%%%%%%%%%%%%%%%%%%%%%%%%%%%%%%%%%%%
%%% 2.3. Super population vs. finite-population estimands
%%%

\subsection{Super population vs. finite-population estimands}

%%% Recovering G for super population inference

The Rubin model highlights two estimands of interest: the super population distribution of site-specific effects, $G$, and the finite-population distribution, $\left\{\tau_j\right\}_{j=1}^J$, for the $J$ sites in the experiment. These two estimands correspond to distinct research objectives. Super population inference aims to recover $G$, the effect distribution for the broader population of sites from which the $J$ experimental sites were sampled. This task is referred to as a \textit{Bayes deconvolution} problem, wherein the observed, noisy sample estimates $\left\{\hat{\tau}_j, \widehat{se}_j^2\right\}_{j=1}^J$ are used to compute an estimate $\widehat{G}$ of $G$ \citep{efron2016empirical}. Performance evaluation studies typically focus on this research objective, as they often have sufficiently large $J$ \citep[e.g.,][]{gilraine2020new}. In contrast, finite-population inference examines the distribution of the true site-effect parameters for the specified $J$ sites, taking these sites as the entire finite population of interest.

%%% Issues with super population estimands

In multisite trial settings, recovering the super population estimand $G$ can be challenging without assumptions about the form of $G$. For a normal model of $G$, only two hyperparameters, $\tau$ and $\sigma^2$, require estimation in the deconvolution process. However, when the prior $G$ is unknown, the performance of deconvolution estimators heavily relies on the shape of the underlying $G$ and the number of sites $J$ \citep{mcculloch2011misspecifying}. Specifically, the mathematical guarantees of flexible estimators are typically asymptotic in nature, necessitating a large $J$ \citep{jiang2009general}. In the context of multisite trials in education, however, $J$ is often small to moderate \citep{weiss2017much}, rendering such asymptotics potentially less applicable. The core concern is that a random sample of 30 to 40 sites' $\tau_j$, even if obtained without measurement error, may not give a sharp picture of the target distribution. Furthermore, it may be unreasonable to assume that the experimental sites in many multisite trials are adequately representative of the super population \citep{stuart2011use}. For example, \cite{tipton2021toward} found that schools involved in grants from the Institute of Education Science were predominantly located in large urban districts, geographically close to one another and to the study's principal investigators.

%%% Focusing on the full finite-population distribution in this study

In this paper, we focus on finite-population estimands, represented by $\left\{\tau_j\right\}_{j=1}^J$. This approach helps us circumvent the risks associated with generalizing treatment effect estimates from a sample of sites to a super population when systematic discrepancies exist between them. Moreover, concentrating on a finite-population of sites typically results in more precise effect estimates compared to situations where researchers aim to generalize to a super population \citep{miratrix2021applied}. This approach is statistically less demanding, allowing for more relevant recommendations in low-data settings for multisite trials.

%%%%%%%%%%%%%%%%%%%%%%%%%%%%%%%%%%%%%%
%%% 3. Improving inferences for site-specific effects
%%%%%%%%%%%%%%%%%%%%%%%%%%%%%%%%%%%%%%

\section{Improving inferences for site-specific effects}

%%%%%%%%%%%%%%%%%%%%%%%%%%%%%%%%%%%%%%
%%% 3.1. Inferential goals and threats to 
%%%      inferences for $\tau_j$
%%%

\subsection{Inferential goals and threats to inferences for \texorpdfstring{$\tau_j$}{Lg}}\label{section:3-1_goals}

%%% Three inferential goals

\cite{shen1998triple} identify three common inferential goals regarding a set of site-specific parameters: (1) estimating the individual site-specific effect parameters, $\tau_j$, (2) ranking the sites based on $\tau_j$, and (3) estimating the empirical distribution function (EDF) of the $\tau_j$'s. Each of these three goals can be associated with a different loss function, which may be minimized by a distinct estimator. The loss function typically associated with the first goal is the mean-squared error loss (MSEL):

%%% MSEL and RMSE

\begin{equation}
\mathrm{MSEL}=\frac{1}{J} \cdot \sum_{j=1}^J\left(a_j-\tau_j\right)^2
\end{equation}

\noindent where $a_j$ is the estimate of $\tau_j$ generated by a candidate estimator. In practice, we measure the root-mean-squared error, $\mathrm{RMSE}=\sqrt{\mathrm{MSEL}}$, which is more interpretable because it preserves the scale of the $\tau_j$ values. The theoretical posterior means of $\tau_j, a_j=\mathrm{E}\left[\tau_j \mid \hat{\tau}_j\right]$ for $j=1, \ldots, J$, minimize the expected MSEL and RMSE. For example, when the $\tau_j$ values are indeed normally distributed, setting $a_j=\tau_j^*$ from equation (\ref{eq:3_shrinkage}) minimizes the MSEL and RMSE.

%%% MSELR and MSELP

For the second inferential goal, we can choose an estimator for the vector of ranks of $\tau_j$ that minimizes the mean squared error loss of the ranks (MSELR),

\begin{equation}
\operatorname{MSELR}=\frac{1}{J} \cdot \sum_{j=1}^J\left(\mathrm{A}_j-\mathrm{R}_j\right)^2
\end{equation}

\noindent where, $\mathrm{R}_j=\sum_{i=1}^J I\left(\tau_i \leq \tau_j\right)$ is the true rank of $\tau_j$, $I(\cdot)$ is the indicator function, and $\mathrm{A}_j$ is an estimated rank. In practice, we measure the mean squared error loss of the percentiles (MSELP), 

\begin{equation}
\text { MSELP }=\frac{1}{J} \cdot \sum_{j=1}^J\left(\frac{\mathrm{A}_j}{J}-\frac{\mathrm{R}_j}{J}\right)^2=\frac{1}{J^2} \cdot \text { MSELR, }
\end{equation}

\noindent where $\mathrm{A}_j / J$ and $\mathrm{R}_j / J$ correspond to the estimated and true percentiles, respectively. The MSELP is less sensitive to changes in $J$, so it can be compared across settings with different numbers of sites. The theoretical expected posterior percentiles in decimal form $\mathrm{A}_j / J=\mathrm{E}\left[\mathrm{R}_j / J \mid \hat{\tau}_j\right]$ minimize expected MSELP. These expected posterior percentiles may not correspond to the percentiles based on the posterior means, and indeed, \cite{goldstein1996league} show that ranks based on posterior means can be suboptimal in general.

%%% ISEL

For the third inferential goal, \cite{shen1998triple} suggest minimizing the integrated squared-error loss (ISEL):

\begin{equation}\label{eq:8_ISEL}
\operatorname{ISEL}\left(\mathrm{A}, \mathrm{G}_J\right)=\int\left\{A(t)-G_J(t)\right\}^2 d t
\end{equation}

\noindent where $G_J(t)=J^{-1} \cdot \sum I_{\left\{\tau_j \leq t\right\}}$ is the true EDF and $A(t)$ is an estimated EDF, $-\infty<t<\infty$. They show that $\bar{G}_J(t)=J^{-1} \cdot \sum_{j=1}^J \operatorname{Pr}\left(\tau_j \leq t \mid \hat{\tau}_j\right)$ minimizes the expected ISEL. The third inferential goal is related to the deconvolution problem, where the goal is to use observed data to recover an unknown prior density $G$ \citep{efron2016empirical}. We focus on finite-population inference in this paper, so our goal is to use the observed $\hat{\tau}_j$ values to infer the EDF of the true $\tau_j$ values.

%%% Threats to valid finite-sample estimation

The three inferential goals, their associated loss functions, and their optimal estimators reveal two primary challenges in achieving valid finite-population estimation. The first challenge is model misspecification, which arises when we assume an incorrect parametric form for the super population distribution $G$. If the true distribution $G$ is not Gaussian, assuming normality may result in insensitivity to skewness, long tails, multimodality, and other complexities in the EDF of the $\tau_j$'s \citep{mcculloch2011misspecifying}. Consequently, estimators based on the standard Rubin model become unreliable, especially for the third inferential goal. The second challenge emerges when, for a given goal, an unsuitable estimator is chosen, even when the prior distribution $G$ is accurately specified in a model. The optimal estimator is contingent upon the chosen loss function or inferential goal. However, practitioners often use the same set of posterior mean effect estimates for all three goals, leading to suboptimal outcomes for at least some of them. For instance, the EDF of posterior mean effect estimates will tend to be underdispersed compared to the EDF of $\tau_j$ due to shrinkage towards the prior mean effect $\tau$. Conversely, the EDF of raw observed ML effect estimates $\hat{\tau}_j$ is overdispersed because of sampling error \citep{mislevy1992estimating}.

%%%%%%%%%%%%%%%%%%%%%%%%%%%%%%%%%%%%%%
%%% 3.2. Strategies to improve inferences 
%%%      for a distribution of $\tau_j$
%%%

\subsection{Strategies to improve inferences for the distribution of \texorpdfstring{$\tau_j$}{Lg}}

%%% Introduction to two strategies

To address the threats outlined in the previous section, two strategies can be employed. The first strategy entails adopting flexible semiparametric or nonparametric specifications for the prior distribution $G$ to protect against model misspecification \citep{paddock2006flexible}. One prominent Bayesian nonparametric specification is the Dirichlet process (DP) prior, which has gained widespread use in the literature \citep{paganin2022computational}. The second strategy focuses on employing posterior summary methods, such as the constrained Bayes or the triple-goal estimators. These estimators are designed to directly target the loss functions associated with specific inferential goals. The following sections discuss these two strategies and their implications for addressing the challenges in finite-population estimation.

%%%%%%%%%%%%%%%%%%%%%%%%%%%%%%%%%%%%%%
%%% 3.3. Posterior summary methods: 
%%%      constrained Bayes and triple-goal estimators
%%%

\subsection{Posterior summary methods: Constrained Bayes and triple-goal estimators}

%%% The CB estimator

Our primary target of interest is $\left\{\tau_j\right\}_{j=1}^J$, representing the EDF of the $\tau_j$'s. Both CB and GR generate estimates of the $\tau_j$ that are not as overly shrunk as the simple posterior means. The CB estimator directly modifies the posterior mean estimates $\tau_j^*$ to avoid under-dispersion, rescaling them to ensure that the estimated variance of site-specific effect estimates aligns with the estimated marginal variance of the latent parameters $\tau_j$. In equation (\ref{eq:3_shrinkage}), the posterior mean of $\tau_j$ is denoted as $\tau_j^* \equiv \mathrm{E}\left(\tau_j \mid \hat{\tau}_j\right)$, and the posterior variance of $\tau_j$ as $V_j \equiv \operatorname{Var}\left(\tau_j \mid \hat{\tau}_j\right)$. We introduce $\bar{\tau} \equiv J^{-1} \sum \tau_j^*$ to represent the finite-population mean of the posterior mean estimates and $v$ as the corresponding finite-population variance. The CB estimates are then:

\begin{equation}\label{eq:CB_eq1}
\tau_j^{\mathrm{CB}}=\bar{\tau}+\left(\hat{\tau}_j-\bar{\tau}\right) \cdot S_j \cdot \frac{\hat{\sigma}}{\sqrt{v}}
\end{equation}

\noindent Here, $S_j$ is an estimate of the shrinkage factor, defined in equation (\ref{eq:3_shrinkage}) as $\sigma^2 /\left(\sigma^2+\widehat{s e}_j^2\right)$. The estimate of the cross-site impact standard deviation $(\sigma)$ and the variance of the posterior means $(v)$ are:

\begin{equation}\label{eq:CB_eq2}
\hat{\sigma}=\sqrt{\frac{\sum V_j}{J}+v}, \quad v=\frac{\sum\left(\tau_j^*-\bar{\tau}\right)^2}{J-1} .
\end{equation}

\noindent Posterior mean estimates tend to be under-dispersed because their variance $v$ is always less than the estimated marginal variance $\hat{\sigma}^2$, as demonstrated in equation (\ref{eq:CB_eq2}). By rescaling the site-specific estimates by $\hat{\sigma} / \sqrt{v}$, the CB estimator in equation (\ref{eq:CB_eq1}) attempts to "reverse" overshrinkage towards the finite-population prior mean effect $\bar{\tau}$ driven by $S_j$. This adjustment expands the gap between $\tau_j^*$ and $\bar{\tau}$ because $\left(\hat{\tau}_j-\bar{\tau}\right) \cdot S_j$ is equal to $\tau_j^*-\bar{\tau}$. An empirical Bayes version of the CB estimator was also introduced by \cite{bloom2017using}.

%%% The triple-goal (GR) estimator

In contrast, the GR estimator employs evenly spaced percentiles of the estimated $G$ to generate its estimates. The GR estimator is constructed via a three-stage process. In the first stage, an estimate of the distribution function that minimizes ISEL is constructed as

\begin{equation}
\bar{G}_J(t)=J^{-1} \cdot \sum_{j=1}^J \operatorname{Pr}\left(\tau_j \leq t \mid \hat{\tau}_j\right)
\end{equation}

\noindent The smooth function $\bar{G}_J(t)$ is then discretized to an EDF that places a mass of $1 / J$ at $J$ evenly spaced discrete points $\widehat{U}_k=\bar{G}_J^{-1}\left(\frac{2 k-1}{2 J}\right)$ for $k=1, \ldots, J$. In the second stage, a rank for each site $j$ is computed as

\begin{equation}
\bar{R}_j=\sum_{k=1}^J \operatorname{Pr}\left(\tau_j \leq \tau_k \mid \hat{\boldsymbol{\tau}}\right)
\end{equation}

\noindent The generally non-integer-valued ranks $\bar{R}_j$ are then discretized to integer-valued ranks as $\hat{R}_j= \operatorname{rank}(\bar{R}_j)$. In the final stage, the GR estimate $\tau_j^{\mathrm{GR}}$ for each site $j$ is computed as $\tau_j^{\mathrm{GR}}=\widehat{U}_{\hat{R}_j}$. Intuitively, the GR estimator first finds a good estimate $\bar{G}_J(t)$ of the full distribution of $\tau_j$ values and then produces estimates for the $J$ sites as $J$ evenly spaced quantiles of $\bar{G}_J(t)$, in an appropriate order.

%%%%%%%%%%%%%%%%%%%%%%%%%%%%%%%%%%%%%%
%%% 3.4. Relaxing distributional assumption 
%%%      for the prior $G$: Dirichlet process mixture
%%%

\subsection{Relaxing the distributional assumption for the prior \texorpdfstring{$G$}{Lg}: Dirichlet process mixture}

%%% Basic DPM model specification

Instead of presupposing a known parametric form for the prior distribution $G$ in equation (\ref{eq:2_Rubin}), a Dirichlet process (DP) can be employed as the prior for $G$ to accommodate uncertainty regarding its shape. Conceptually, the DP model can be viewed as an infinite mixture model in which both the priors and the data contribute to determining the number of observed mixture components.

The DP prior is characterized by two hyperparameters: a base distribution $G_0$ and a precision parameter $\alpha$ \citep{antoniak1974mixtures}. A two-stage hierarchical model, incorporating the DP prior, can be formulated as follows:

\begin{equation}\label{eq:12_DPM}
\begin{gathered}
\hat{\tau}_j \mid \tau_j, \widehat{s e}_j^2 \sim N\left(\tau_j, \widehat{s e}_j^2\right), \quad j=1, \ldots, J, \\
\tau_j \mid G \stackrel{\text { i.i.d. }}{\sim} G, \\
G \mid \alpha, G_0 \sim \operatorname{DP}\left(\alpha, G_0\right) .
\end{gathered}
\end{equation}

\noindent $G$ is a discrete distribution, with positive probability on a countably infinite number of points. It models the distribution of site impacts, therefore, as a set of \textit{clusters}, where groups of sites will all have the same impact. In a given finite sample we would observe $K$ such clusters, with $K$ potentially being less than $J$.

%%% Hyperpriors - $G_0$ and $\alpha$

To finalize the model specification, we must define the base distribution $G_0$ and establish a hyperprior for the parameter $\alpha$. $G_0$ serves as the initial best guess for the shape of the prior distribution $G$. Typically, we assume $G_0$ follows a Gaussian distribution with mean $\tau$ and variance $\sigma^2$. For computational convenience, the base distribution $G_0$ is typically constructed as the product of conjugate distributions, such as a normal distribution for the mean parameter $\tau$ and an inverse-gamma distribution for the variance parameter $\sigma^2$ \citep{paganin2022computational}.

%%% Precision parameter $\alpha$ & Meaning of K

The precision hyperparameter $\alpha$ controls the degree to which $G$ converges toward $G_0$ (see the Supplemental Material). The larger the $\alpha$, the larger the number of unique observed site impact estimates $K$. It is worth noting that $K$ does not always precisely correspond to the number of mixture components, $C$, representing latent subpopulations with substantive meaning, as would be defined in a finite mixture modeling approach. However, $K$ can be reasonably regarded as an upper limit for $C$ \citep{ishwaran2000markov}.

%%% The hyperprior for $\alpha$
%%% Debate around hyperprior selection

The hyperprior for $\alpha$ plays a pivotal role in determining $K$, which in turn shapes the posterior distribution over clusters. Using a Gamma distribution - with $a$ as the shape parameter and $b$ as the rate parameter - as a hyperprior for $\alpha$ is common practice, attributable to its computational advantages \citep{escobar1995bayesian}. While the choices of parameters $a$ and $b$ can significantly influence the posterior distribution of $\alpha$, thus impacting clustering behavior \citep{dorazio2009selecting, murugiah2012selecting}, other studies suggest that data are typically sufficient to produce a concentrated posterior, thereby reducing the influence of the hyperprior even when a high-variance prior for $\alpha$ is used \citep{leslie2007general, gelman2013bayesian}.

%%% The first option to specify the $\alpha$ prior: Diffuse prior

Our objective is to assess the sensitivity under two distinct hyperprior options for $\alpha$, particularly in the context of varying levels of data informativeness. The first hyperprior, explored by \cite{antonelli2016mitigating}, is relatively diffuse, allowing for numerous clusters. This strategy, which we refer to as the DP-diffuse model, is applicable when prior knowledge about $K$ is absent. The DP-diffuse model chooses values of $a$ and $b$ such that $\alpha$ is centered between 1 and $J$, with a large variance to assign a priori mass to a wide range of possible $\alpha$ values. For instance, with $J=50$, we might set the mean of the $\alpha$ distribution to 25 and its variance to 250 , ten times the mean. With these initial values, we can use the moments of a Gamma distribution to derive the corresponding hyperparameters $a=2.5$ and $b=0.1$, based on the formulas $\mathrm{E}(\alpha \mid a, b)=$ $a / b$ and $\operatorname{Var}(\alpha \mid a, b)=a / b^2$.

%%% The second option to specify the $\alpha$ prior: Informative prior

This research proposes a novel approach, employing a $\chi^2$ distribution to formulate an informative hyperprior for $\alpha$. Assume we have $\operatorname{Pr}(K)$, a prior that encapsulates our knowledge of the distribution of $K$. \citet{dorazio2009selecting} provides an expression for $\operatorname{Pr}(K \mid J, a, b)$, the probability mass function for the prior distribution $K$ induced by a Gamma $(a, b)$ prior for $\alpha$ and $J$. We can then obtain optimal values for $a$ and $b$ by minimizing the discrepancy between the encoded prior $\operatorname{Pr}(K)$ and the $\alpha$-induced prior $\operatorname{Pr}(K \mid J, a, b)$ using the Kullback-Leibler divergence measure.

\begin{figure}[t]
    \centering
    \includegraphics[width=\textwidth]{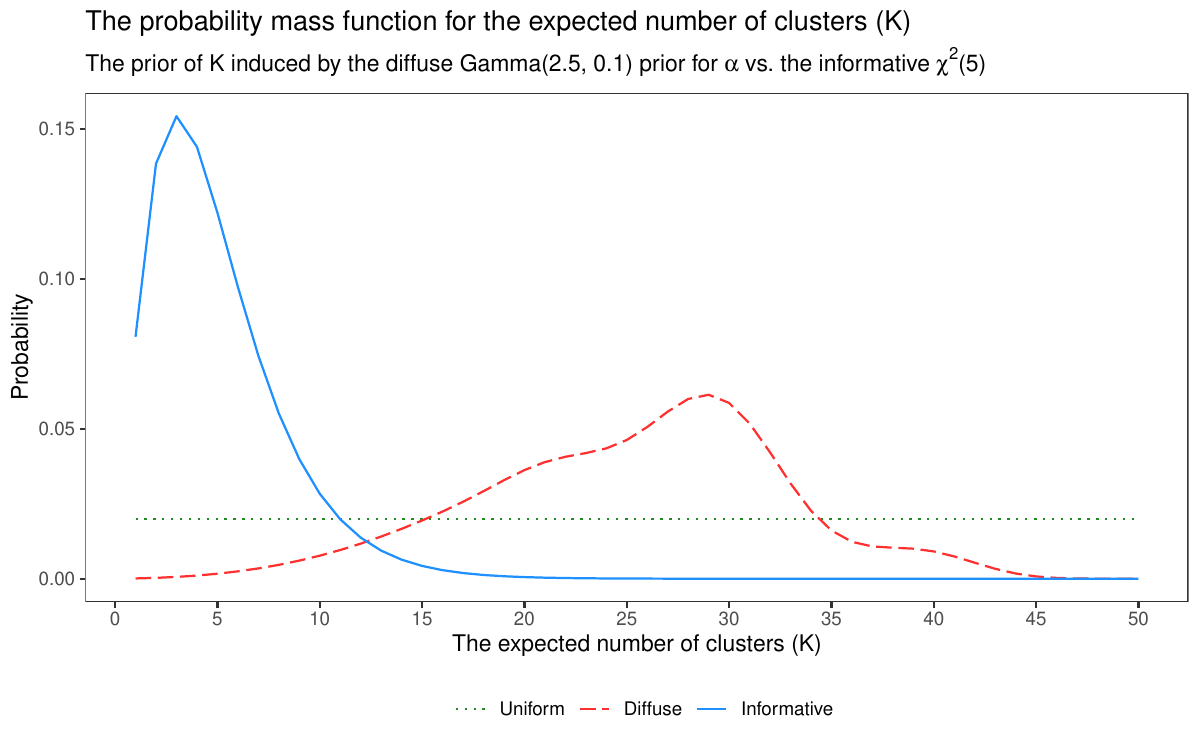}
    \caption{The distributions of the expected number of clusters $K$ for the DP-diffuse and DP-inform models ($J = 50$)}
    \label{fig:figure01}
\end{figure}

We propose setting $\operatorname{Pr}(K) \sim \chi^2(\mathrm{df}=u)$ to more intuitively encapsulate our prior
knowledge on $K$ and its inherent uncertainty. This strategy is primarily motivated by the fact that the $\chi_u^2$ distribution has a mean and variance of $u$ and $2 u$, respectively. For instance, if we regard five clusters ($K=5$) among 50 sites as a reasonable approximation of the underlying distribution, we could assume that $\operatorname{Pr}(K)$ follows a $\chi^2(5)$ distribution and specify a $\operatorname{Gamma}(a, b)$ that closely matches $\chi^2(5)$. In this case, the appropriate Gamma distribution would have parameters $(a, b)=(1.60,1.22)$ (see the Supplemental Material). This approach, which we term the DP-inform model, is particularly effective for crafting an informative prior for $K$, especially when aiming to impose near-zero probabilities beyond a certain threshold (e.g., $K=25$ as shown in Figure~\ref{fig:figure01}) and to clarify the prior mean and variance of $K$.

In Figure~\ref{fig:figure01}, we observe that the two DPM models represent contrasting beliefs regarding $K$. The DP-diffuse model favors a smoother $G$ with approximately 25-30 clusters for 50 sites, while the DP-inform model, based on $\chi^2(5)$, is more favorable to a more uneven $G$, assuming that about five clusters are needed for 50 sites. Further details about the choices of hyperpriors and their implementations can be found in the Supplemental Material.

%%%%%%%%%%%%%%%%%%%%%%%%%%%%%%%%%%%%%%
%%% 4. A simulation study
%%%%%%%%%%%%%%%%%%%%%%%%%%%%%%%%%%%%%%

\section{A simulation study}

%%%%%%%%%%%%%%%%%%%%%%%%%%%%%%%%%%%%%%
%%% 4.1. Simulation design
%%%

\subsection{Simulation design}

%%% Mirroring typical data for multisite trials in education

To compare the performance of different models and posterior summary methods, we conduct a comprehensive Monte Carlo simulation. We generate data by systematically varying five factors: (a) the number of sites ($J$), (b) the average number of observations per site ($\bar{n}_j$), (c) the coefficient of variation in site sizes ($\mathrm{CV}$), (d) the true cross-site population distribution $(G)$ of $\tau_j$, and (e) the true cross-site population standard deviation of impacts ($\sigma$). We calibrate our simulated datasets to mirror typical data settings for multisite trials in education, as described by \cite{weiss2017much}.

%%% The number of sites ($J$)

We consider five values for the number of sites in our study: $J = 25, 50, 75, 100$, and $300$. The number of random assignment blocks reported in the 16 studies from \cite{weiss2017much} ranged from 20 to 356, with a median of 78. Although multisite trials typically involve a small to moderate numbers of sites, we also wanted to investigate model performance for larger multisite trials, such as the Head Start Impact Study \citep[e.g.,][]{bloom2015quantifying}.

%%% Average site sizes and coefficient of variation → within-site sampling error

We vary the average number of observations per site in our study as $\bar{n}_j=10,20,40,80$, and $160$. \cite{weiss2017much} reported a range of mean site sizes, from 11 in the Head Start Impact Study to 1,176 in the Welfare-to-Work Program, with the 25th and 75th percentiles at 75 and 163, respectively. Larger site sizes also capture contexts where we are using baseline covariates to increase precision, as both size and $R^2$ directly impact site-level uncertainty, which is what drives our overall performance (see equation (\ref{eq:1_Rubin}) and the definition of informativeness in equation (\ref{eq:4_I-level})). We control the variation in site sizes by setting the coefficient of variation of the site sizes to be $\mathrm{CV}=0.00,0.25,0.50,0.75$ by sampling $n_j$ from a gamma distribution with a mean of $\bar{n}_j$ and a standard deviation of $\bar{n}_j \cdot \mathrm{CV}$. We then truncate $n_j$ at the lower limit of $n_j=5$. Finally, we compute the within-site sampling variances as: $\widehat{s e}_j^2=4 / n_j$ (this is assuming we are in an effect size metric with a control-side standard deviation of 1 \textit{within} site).

%%% Cross-site impact variation

We vary the cross-site impact standard deviation as $\sigma = 0.05, 0.10, 0.15, 0.20,$ and $0.25$ in effect size units. \cite{weiss2017much} report that the cross-site standard deviation of program impact, in effect size units, ranges between 0.00 and 0.35, with most values falling between 0.10 and 0.25. They define 0.05, 0.15, and 0.25 as having a modest, moderate, and substantial impact variation, respectively. In conjunction with the range of $\widehat{se}_j^2$ determined by our selection of $\bar{n}_j$, $\sigma$ determines the magnitude of the average reliability of the ML estimates, denoted by $I$. Across our simulation conditions, $I$ varies between 0.01 and 0.71, with a mean of 0.25.

%%% True population distribution of $\tau_j$ (G)

Finally, we consider three different population distributions $G$ for $\tau_j$: Gaussian, a mixture of two Gaussians, and asymmetric Laplace (AL). The Gaussian mixture represents a scenario where the actual $G$ is multimodal, while the AL distribution exemplifies a situation where the true $G$ is skewed and exhibits a long tail. We rescale the simulated $\tau_j$ so that the variance of $G$ equals $\sigma$. Additional details on the data-generation process can be found in the Supplemental Material. 

%%% Final simulated datasets

Using the five design factors ($J$, $\bar{n}_j$, $\text{CV}$, $\sigma$, and $G$), we generate 1,500 simulation conditions, each with 100 replications. We fit each replication with three data-analytic models: Gaussian, DP-diffuse, and DP-inform, and, for each model, use Markov Chain Monte Carlo to obtain 4,000 posterior samples from the joint posterior distributions of the J site-specific effects. Finally, we apply all three posterior summary methods (PM, CB, and GR) to each model fit. This gives 9 estimators for each generated dataset.

%%%%%%%%%%%%%%%%%%%%%%%%%%%%%%%%%%%%%%
%%% 4.2. Simulation analysis
%%%

\subsection{Simulation analysis}

To analyze the simulation results, we separate the three true population distributions $G$ into samples of 450,000 observations each, generated as a factorial combination of the four remaining design factors ($J$, $\bar{n}_j$, $\text{CV}$, and $\sigma$), two data-analytic factors (data-analytic model and posterior summary method), and 100 replications. We construct 50,000 cluster (i.e., dataset) identifiers based on the four design factors and 100 replications to account for the statistical dependence induced by analyzing the same simulated dataset.

We use meta-model regressions \citep{skrondal2000design} to examine the relationship between performance criteria and simulation factors. Unlike visual or descriptive methods, meta-models allow for a more precise detection of significant patterns while accounting for uncertainty from Monte Carlo error \citep{boomsma2013reporting}. We focus on the three performance criteria outlined in Section 3.1 as our target outcome variables: RMSE, ISEL, and MSELP. To ease the interpretation of the meta-model regression coefficients, we apply a logarithmic transformation to the outcome variables. As for the explanatory variables, we create a set of dummy variables representing the six design and data-analytic factors, along with their two-way interaction terms. The reference groups are $J = 25$, $\bar{n}_j = 10$, $\sigma = 0.05$, $\text{CV} = 0.00$, the Gaussian model, and the PM summary method. Lastly, since we include each dataset nine times in each meta-model regression (corresponding to every model/posterior summary method combination), we cluster our standard errors on datasets.

%%%%%%%%%%%%%%%%%%%%%%%%%%%%%%%%%%%%%%
%%% 5. Results: Effects of data-generating factors
%%%%%%%%%%%%%%%%%%%%%%%%%%%%%%%%%%%%%%

\section{Results: Effects of data-generating factors}

%%% Section introduction

Our simulation produces nine (three performance criteria $\times$ three true population distributions $G$) meta-model regressions, each summarizing a six-factor factorial experiment. Rather than exhaustively exploring these results, we divide our analyses into two simple, practical sections. We first explore how the four design factors ($J$, $\bar{n}_j$, $\text{CV}$, and $\sigma$) generally affect the performance of each model and posterior summary method. We then directly compare the model/method combinations, using hypothetical case studies to illustrate how they may perform relative to each other in real-world applications.

%%% Figure 2 main effects plot

\begin{figure}[H]
    \centering
    \includegraphics[width=\textwidth]{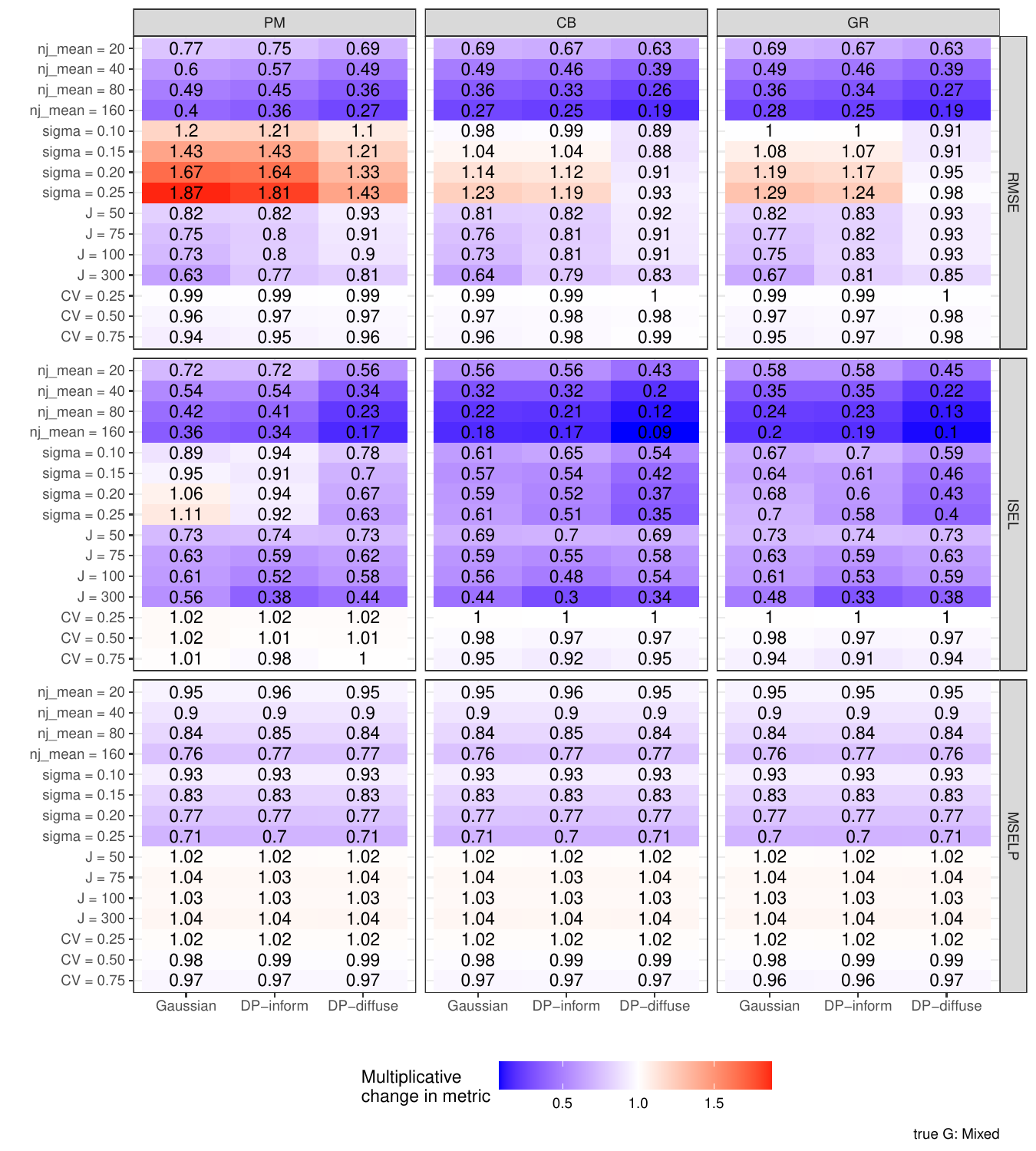}
    \caption{Predicted multiplicative change in average RMSE, ISEL, and MSELP (row blocks) by the meta-model regression for the 9 estimators (columns), when shifting a single data-generating factor from the base condition ($J = 25$, $\bar{n}_j=10$, $\sigma = 0.05$, and $\text{CV}=0.00$). The true $G$ is assumed to be a Gaussian mixture}
    \label{fig:figure03}
\end{figure}

Figure~\ref{fig:figure03} illustrates how the design factors affect the outcome metric for each model/method combination, under the Gaussian mixture data-generating distribution. Results for the Gaussian and AL data-generating distributions are qualitatively similar (see the Supplemental Material). The values within each cell indicate the predicted multiplicative change in average RMSE, ISEL, or MSELP by the meta-model regressions. This change corresponds to altering a single data-generating factor from the base condition ($J = 25$, $\bar{n}_j = 10$, $\sigma = 0.05$, $\text{CV} = 0.00$) to the value indicated by the row. These values are provided for each model, posterior summary method, and outcome metric, as indicated by the column, column panel, and row panel, respectively. Values less than 1 indicate improvement compared to the base condition. For example, the value 0.77 in the top-left cell indicates that for the Gaussian model using the PM estimator, increasing $\bar{n}_j$ from 10 to 20 reduces the RMSE by approximately 23\% on average, when all other design factors are held at their baseline levels. 

%%% Overall pattern: Importance of average site size

While the metrics respond differently to changes in the data-generating factors, we highlight some overall patterns in Figure~\ref{fig:figure03}. First, increasing $\bar{n}_j$ significantly improves all three metrics. The three metrics all measure how well some feature of the finite population distribution of site-level effects $\{\tau_j\}_{j=1}^J$ is estimated; increasing $\bar{n}_j$ increases the average precision of the site-level effect estimates $\{\hat{\tau}_j\}_{j=1}^J$, so it naturally improves these finite population metrics. Second, changing $\text{CV}$ has little effect on any of the metrics, indicating that the variation in site sizes matters much less than the average site size.

%%% Metric-specific pattern - RMSE

The RMSE metric is sensitive to changes in both $J$ and $\sigma$. When $J$ increases, the RMSE decreases. Adding sites increases the sample size used to estimate the cross-site variation. This improves the models’ estimates of the cross-site variation, which in turn improves their site-specific effect estimates. Conversely, increasing $\sigma$ generally results in a higher RMSE. The models all apply shrinkage to the raw site-specific estimates, which biases the estimates but typically reduces their RMSE by significantly decreasing their variance. As the variation $\sigma$ of the true site-specific effects increases, however, there is more room for bias as the sites are more spread out, and the benefits of variance reduction will no longer outweigh this bias cost. Given this, with larger $\sigma$, the models shrink less. For example, equation (\ref{eq:3_shrinkage}) illustrates how the estimates produced by the Gaussian/PM combination approach the initial raw site-specific estimates as $\sigma$ increases, given a fixed average site size. Overall, RMSE tends to be larger when $\sigma$ is larger, particularly for models and methods that exhibit the most shrinkage in the base case. As observed in Figure~\ref{fig:figure03}, this effect is strongest for the Gaussian/PM combination and weaker for the DP-diffuse model and the CB and GR summary methods.

%%% Metric-specific pattern - ISEL

In terms of the ISEL metric (for measuring how close our estimated distribution is to the truth), increasing $J$ improves results. Similar to the RMSE, a larger $J$ allows for better estimation of cross-site variation, thereby enhancing the estimates. Adding more sites is equivalent to drawing additional noisy samples from the true EDF, which provides information about the true EDF's shape. The DP-inform model is particularly effective at incorporating this information, although the other models also show improvement.

In contrast to the RMSE, higher $\sigma$ values result in better ISEL. Shrinkage causes systematic underestimation of large true effects and overestimation of small true effects. While this generally improves RMSE, which does not differentiate between overestimation and underestimation, excessive shrinkage can create significant discrepancies between the true and estimated EDFs, resulting in poor ISEL values. Consequently, ISEL rewards estimates that are not overly shrunken toward the overall mean. As $\sigma$ increases, Figure~\ref{fig:figure03} demonstrates how the CB/GR methods are better able to maintain correspondingly diffuse EDF estimates, while the PM method struggles in this regard.

%%% Metric-specific pattern - MSELP

Lastly, we observe that MSELP (measuring how well we rank the sites) improves as either $\bar{n}_j$ or $\sigma$ increases and remains largely unaffected by $J$, $\text{CV}$ , model, and summary method. Generally, the primary factor influencing order estimation is the average reliability of the ML estimates, $I$, which quantifies the relative magnitude of between-site and within-site variations. When most variation occurs between sites, it is easier to distinguish and accurately rank them. Increasing $\sigma$ raises the proportion of total variation in the data that is between sites (rather than within sites), thereby enhancing MSELP. Similarly, a higher $\bar{n}_j$ reduces the impact of within-site variation, which also improves MSELP. Employing different models and summary methods seldom alters the ordering of site-specific effect estimates, and thus, does not affect MSELP.

%%%%%%%%%%%%%%%%%%%%%%%%%%%%%%%%%%%%%%
%%% 6. Results: Case studies of best model-method choices
%%%%%%%%%%%%%%%%%%%%%%%%%%%%%%%%%%%%%%

\section{Results: Case studies of best model-method choices}

%%% Section introduction

\begin{figure}[H]
    \centering
    \includegraphics[width=\textwidth]{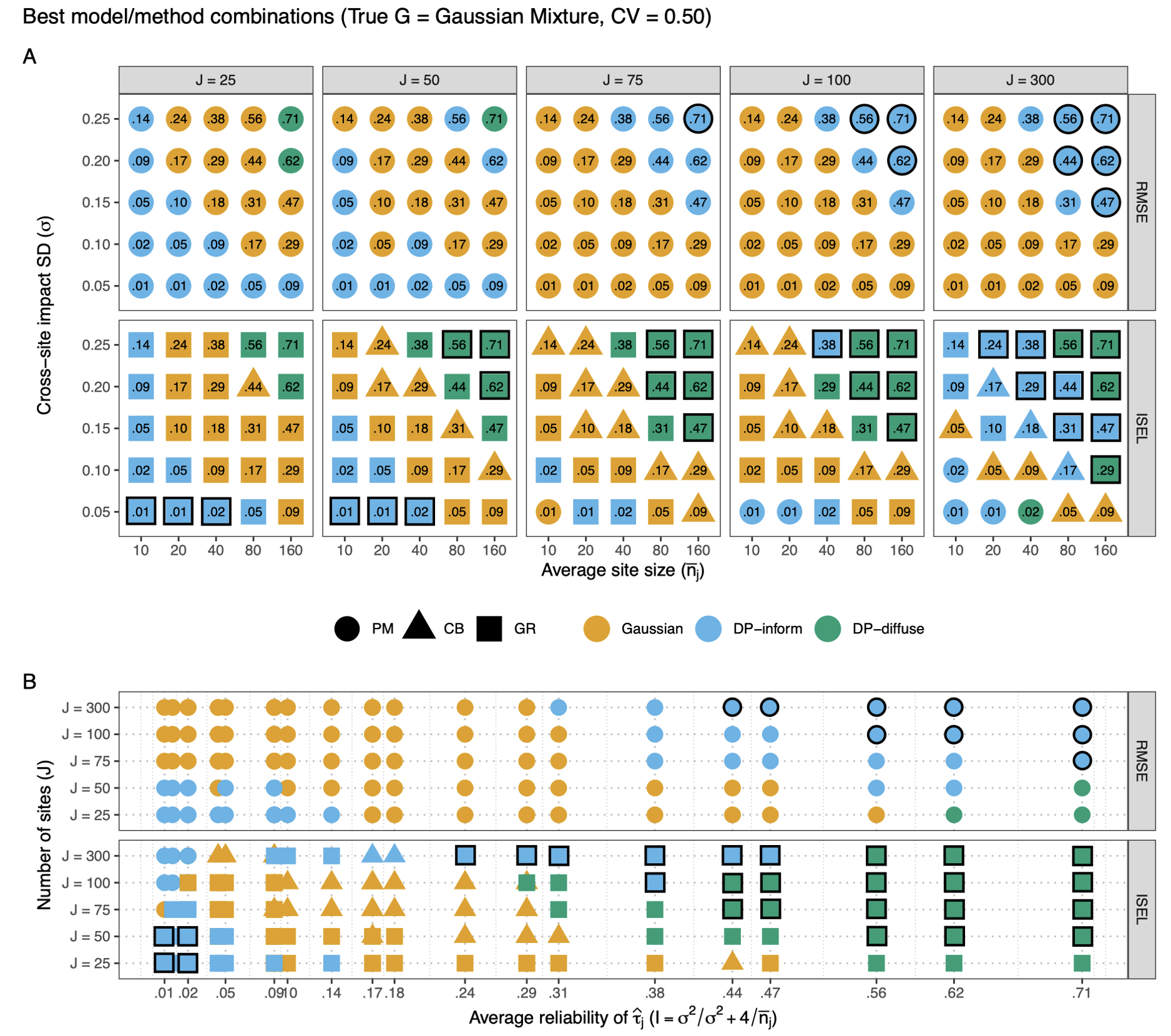}
    \caption{The best model-method combinations with the lowest average metric value (RMSE and ISEL) across the 100 simulation replications, for the Gaussian mixture data-generating distribution. Panel A presents the average reliability value of $\hat{\tau}_j$ ($I$) within each facet. Panel B offers a different organization of the same results where the x-axis corresponds to the level of $I$. If the DP models significantly outperform the Gaussian model, accounting for Monte Carlo errors, the cell is outlined with a thick black boundary. MSELP results have been omitted, as the choice of model/method combinations was found to have no effect on these metrics.}
    \label{fig:figure04}
\end{figure}

We now directly compare the models and posterior summary methods to assess the most effective combinations under various design settings. Figure~\ref{fig:figure04} illustrates the best model-method combinations, as indicated by the average outcome metric predicted by the meta-model regressions, for the Gaussian mixture data-generating distribution across different values of $\bar{n}_j$, $\sigma$, and $J$ (with $\text{CV}=0.50$ held constant). In the figure, both the shape and color at each design setting represent the model-method combination yielding the lowest average error metric value, while the numbers indicate the value of $I$ for each respective design setting.

%%% Overall pattern #1. Best choices for posterior summary method

Although Figure~\ref{fig:figure04} may be subject to some simulation noise, it uncovers several intriguing patterns for the best posterior summary method. As anticipated, the posterior mean (PM) summary method consistently outperforms other methods in terms of RMSE. Since the PM method is designed to minimize RMSE, practitioners aiming to optimize RMSE should always employ the posterior mean summary method. Conversely, PM is seldom the best summary method for ISEL. The CB and GR summary methods typically exhibit less shrinkage than the PM method, leading to more accurate representations of the EDF of site-specific effects.

%%% Overall pattern #2. Best choices for model for prior G
%%% Three factors that influence model performance

Figure~\ref{fig:figure04} additionally highlights patterns for the best prior model for $G$. We observe that these patterns are predominantly influenced by the average reliability of $\hat{\tau}_j$, represented as $I$. As $\bar{n}_j$ or $\sigma$ increase, the value of $I$ also increases, ranging from 0.01 to 0.71, indicating that the data becomes more informative. The number of sites $J$ introduces an extra dimension of data informativeness; the patterns alter as $J$ increases, since a higher number of sites supply more information, even if the value of $I$ remains constant. Consequently, the model's performance depends on three variables that determine the overall informativeness of the data: $\bar{n}_j$, $\sigma$, and $J$.

%%% When do the DP-models perform best? -> in highly informative settings

The DP models typically outperform the Gaussian model in highly informative settings. We observe that the DP-inform model tends to yield the lowest RMSE when $I$ is approximately greater than 0.44. Meanwhile, the DP-diffuse model performs best in terms of ISEL. Notably, this relationship is influenced by the number of sites, $J$. For example, when $J=300$, the DP models significantly outperform the Gaussian model in terms of ISEL if the $I$ level is above roughly 0.20. For $J=75$ or $100$, the DP models require an $I$ level exceeding approximately 0.40 to significantly outperform the Gaussian model. With a low $J$ of 25, the DP models’ performance on ISEL is never significantly better than the Gaussian model, even at the maximum $I$ level we explored ($I=0.71$). The Supplemental Material displays the 95\% prediction intervals around the meta-model predicted values, accounting for Monte Carlo error, to assess the statistical significance of the DP models' superior performance.

\begin{figure}[H]
    \centering
    \includegraphics[width=\textwidth]{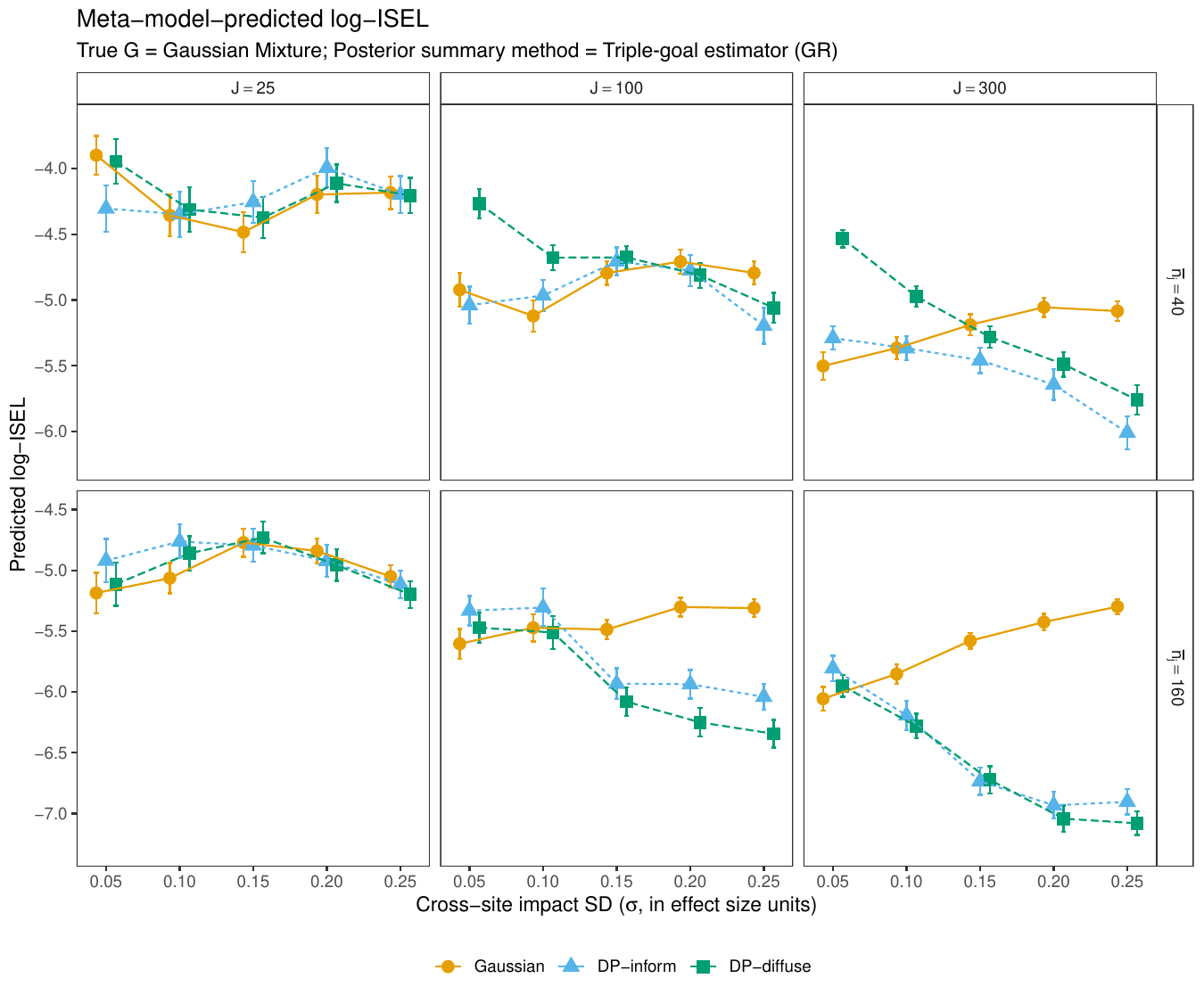}
    \caption{The meta-model predicted integrated squared error loss (ISEL) on a natural log scale, when the true $G$ is a Gaussian mixture and the triple-goal estimator (GR) is used as the posterior summary method (Figure~\ref{fig:figure_c02} in the Supplemental Material illustrates all possible combinations of $J$ and $\bar{n}_j$). Plots include 95\% prediction intervals around predicted values.}
    \label{fig:figure05}
\end{figure}

%%% Less informative settings -> Gaussian model works well

In less informative settings, however, Figure~\ref{fig:figure04} demonstrates that the Gaussian model often has better performance. We note that while the DP-inform model performs the best when $J \leq 50$ and $I$ is approximately less than 0.05, it only marginally outperforms the Gaussian model in these situations (refer to the Supplemental Material). Despite the true data-generating distribution not being Gaussian in these simulations, the Gaussian model combined with a suitable posterior summary method outperforms the DP models in settings with low-to-moderate informativeness, which is frequently the case in educational applications.

%%% Between-site information vs. within-site information

It is important to highlight that the performance of DP models is sensitive to the extent of between-site information present in the data. Figure~\ref{fig:figure05} demonstrates the substantial improvement in the ISEL of the DP models (all employing the GR estimator) as both $\sigma$ and $J$ increase. Conversely, the Gaussian model's imposed shape constraint on $G$ prevents it from effectively using the additional cross-site information, even when site sizes are substantially large ($\bar{n}_j = 160$). In scenarios characterized by a considerable amount of between-site information, the DP models' adaptive clustering facilitates the recovery of site-effect distributions with significantly higher accuracy compared to the Gaussian model. 

%%%%%%%%%%%%%%%%%%%%%%%%%%%%%%%%%%%%%%
%%% 6.1. Case 1: Large between-site information & 
%%%      Large within-site information 
%%%      (Highly informative scenario)
%%%

\subsection{Case 1: Large between-site information \& Large within-site information}

\begin{figure}[H]
    \centering
    \includegraphics[width=\textwidth]{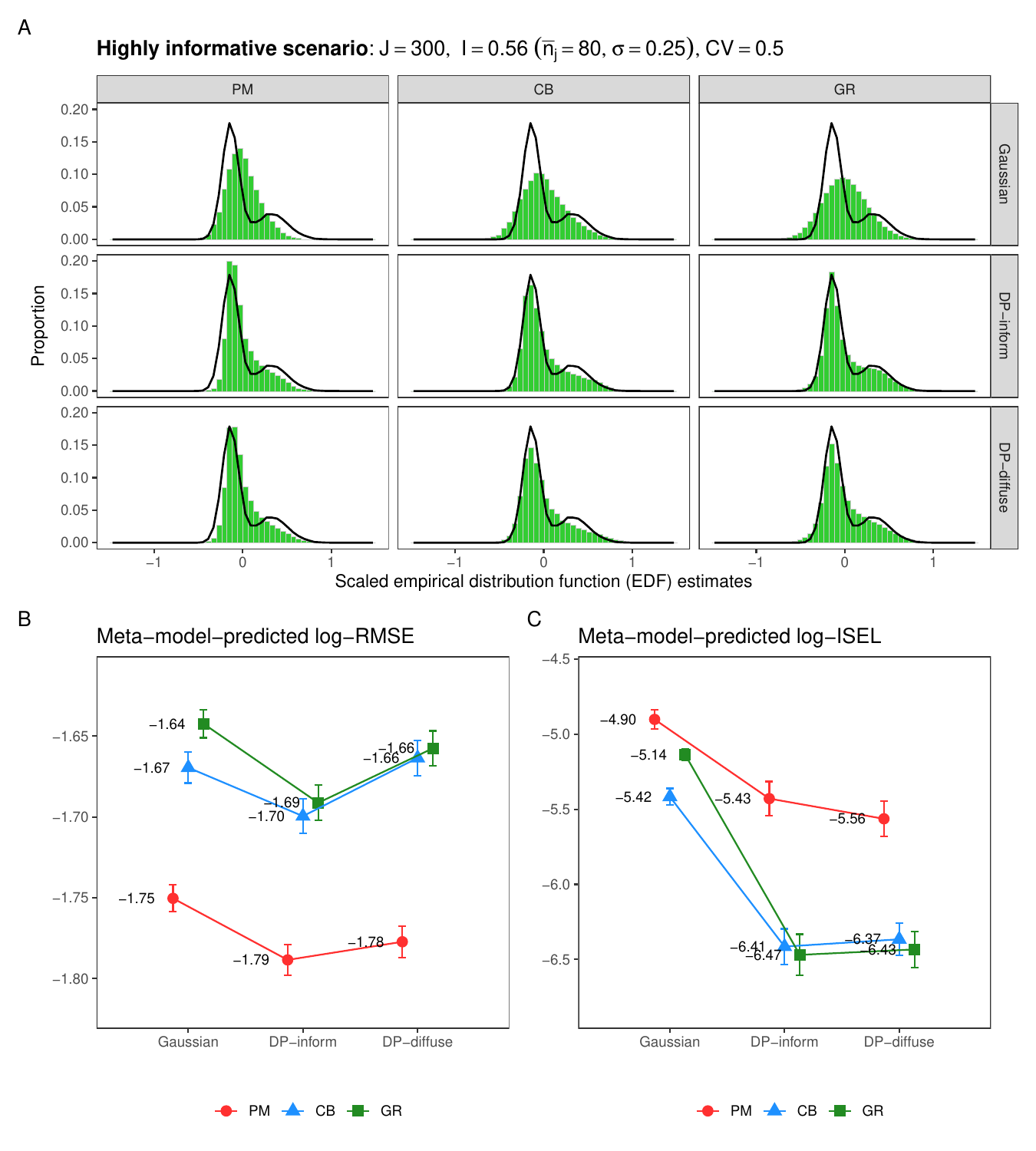}
    \caption{Scaled EDF estimates and meta-model-predicted log-RMSE and ISEL for each model-method combination with 95\% predictive intervals. Simulation based on highly informative scenario. True $G$ is Gaussian mixture, \text{CV} is fixed at 0.5.}
    \label{fig:figure06}
\end{figure}

We now examine a series of case studies to assess the relative performance of different model-method combinations in various real-world situations. Figure~\ref{fig:figure06} displays the performance of each combination in a simulation scenario characterized by high informativeness, encompassing considerable between-site information ($J = 300$, $\sigma = 0.25$) and significant within-site information ($\bar{n}_j = 80$). We note that \cite{weiss2017much} do not provide directly comparable cases, highlighting the challenge of acquiring highly informative data at both the within- and between-site levels in real-world multisite trials.

Panel A of Figure~\ref{fig:figure06} plots histograms of the estimates made by the different model-method combinations against the super population density of true site-specific effects. Panel A aggregates estimates across all simulations, so it does not show how well these approaches estimate the true finite-population distribution of site-specific effects for any given simulation sample; nonetheless, it illustrates the general behaviors of each of these approaches. We see that in this highly informative setting, the DP models can consistently adapt to the sampled sites, while the Gaussian estimator tends to shrink estimates toward the overall average. Similarly, even in this highly informative setting, the PM estimator tends to shrink estimates much more than the CB or GR estimators.

To summarize these results, we visualize the log-outcomes predicted by the meta-model regressions in Panels B and C of Figure~\ref{fig:figure06}. In this highly informative setting, the DP models outperform the Gaussian model in terms of both RMSE and ISEL. For example, using the DP-inform model with the GR estimator significantly improves ISEL compared to the Gaussian model with the same estimator. We see no significant differences in performance between the two DP models, suggesting that with sufficient information, the DP models are quite data-adaptive and insensitive to the choice of prior on $\alpha$.

%%%%%%%%%%%%%%%%%%%%%%%%%%%%%%%%%%%%%%
%%% 6.2. Case 2: Large between-site information & 
%%%      Small within-site information 
%%%

\subsection{Case 2: Large between-site information \& Small within-site information}

\begin{figure}[H]
    \centering
    \includegraphics[width=\textwidth]{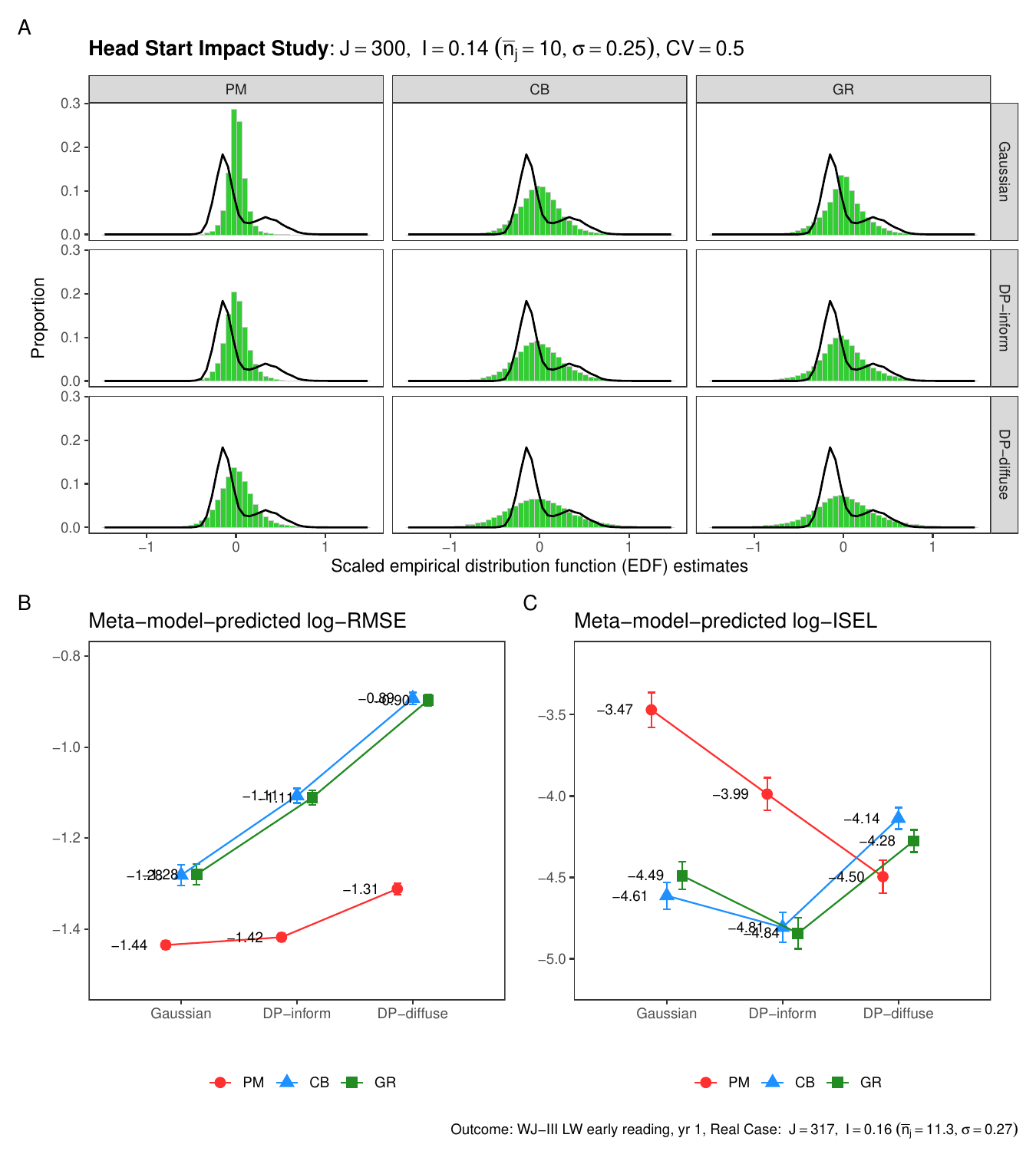}
    \caption{Scaled EDF estimates and meta-model-predicted log-RMSE and ISEL for each model-method combination with 95\% predictive intervals, when the data-generation is based on the scenario simulating the Head Start Impact Study ($J = 317$, $\bar{n}_j=11.3$, and $\sigma=0.30$ for the outcome, the Woodcock-Johnson III Letter-Word Identification subscale score for year 1). True $G$ is Gaussian mixture, while \text{CV} is fixed at 0.5. }
    \label{fig:figure07}
\end{figure}

Figure~\ref{fig:figure07} shows results in a scenario similar to the Head Start Impact Study \citep{puma2010headstart}, which has significant between-site information ($J = 317$, $\sigma = 0.30$) but limited within-site information ($\bar{n}_j = 11.3$). In Panel A, we see how the lack of within-site information leads even the DP models to lean more heavily on the Gaussian base distribution, leading to roughly Gaussian histograms when aggregated across all simulations.

When there is significantly more between-site information than within-site information, shrinkage toward the overall mean effect can considerably influence model estimates. For instance, the DP-diffuse model is found to be unsuitable for both RMSE and ISEL due to undershrinkage, which can be attributed to the model’s assumption of a large number of clusters and the substantial between-site information. Consequently, this leads to excessive weight being placed on a broad spectrum of potential site-specific effects. Conversely, the Gaussian model exhibits excessive shrinkage as a result of its shape constraints. In this scenario, the DP-inform model demonstrates strong performance in terms of ISEL, as its flexible shape allows it to adapt appropriately to the considerable between-site information, while its limited number of assumed clusters prevents undershrinkage.

In general, the choice of posterior summary method has a more pronounced impact than model selection. Using the PM summary, RMSE improves by 16-42\% compared to CB or GR methods. In contrast, only a 1\% RMSE difference is observed between the Gaussian and DP-inform models with PM. For ISEL, CB or GR methods outperform PM by 1.85 to 2.14 times, a more significant gain than the 20-35\% improvement when choosing the DP-inform over the Gaussian model with CB/GR methods.

%%%%%%%%%%%%%%%%%%%%%%%%%%%%%%%%%%%%%%
%%% 6.3. Case 3: Small between-site information & 
%%%      Large within-site information 
%%%

\subsection{Case 3: Small between-site information \& Large within-site information}\label{subsec:6-3}

\begin{figure}[H]
    \centering
    \includegraphics[width=\textwidth]{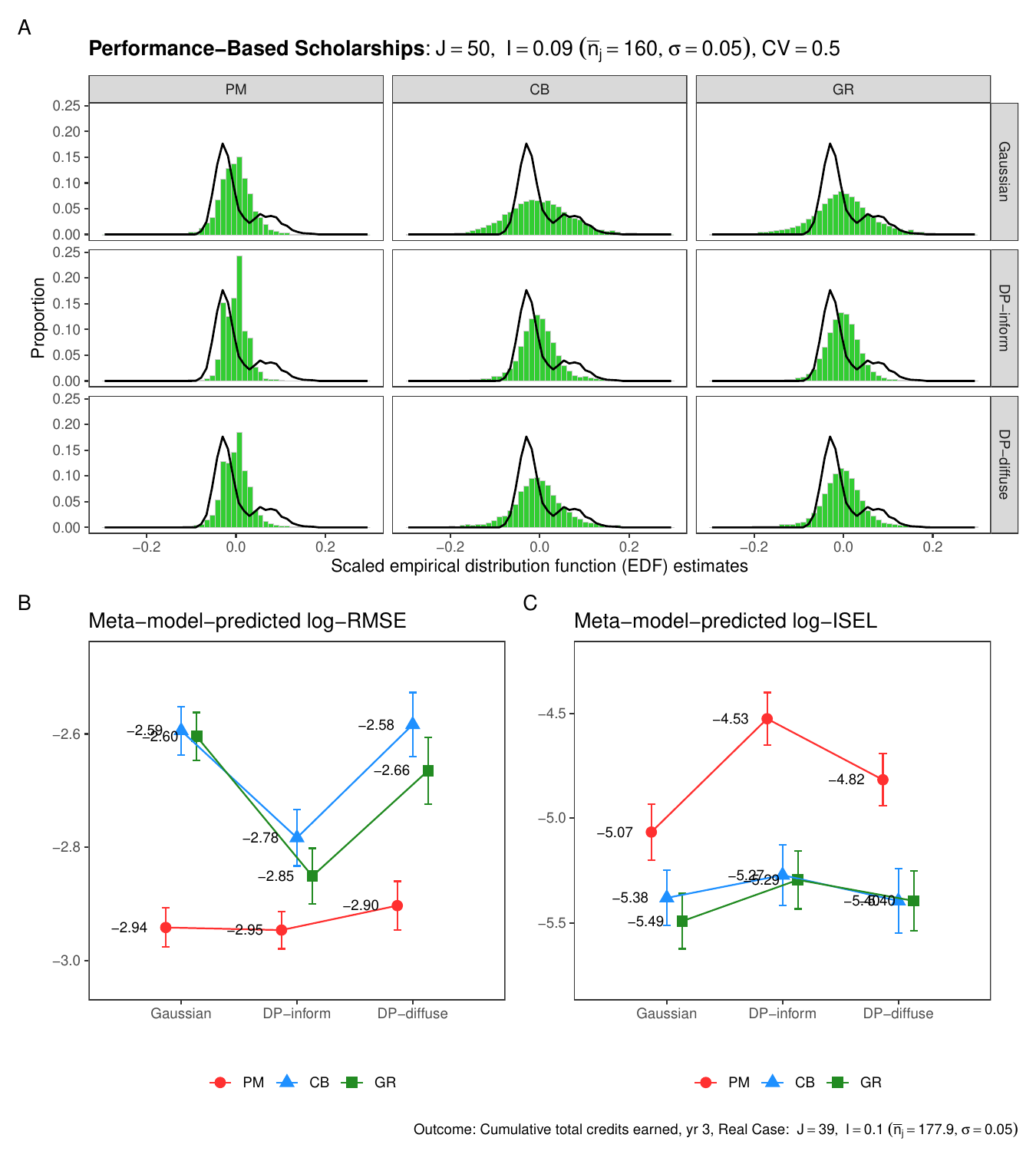}
    \caption{Scaled EDF estimates and meta-model-predicted log-RMSE and ISEL for each model-method combination with 95\% predictive intervals. Data-generating process based on the Performance-Based Scholarships study ($J = 39$, $\bar{n}_j=177.9$, and $\sigma=0.05$ for the outcome, cumulative total credits earned for year 3). True $G$ is Gaussian mixture, while \text{CV} is fixed at 0.5.}
    \label{fig:figure08}
\end{figure}

Figure~\ref{fig:figure08} presents simulation results for a setting akin to the Performance-Based Scholarships study, characterized by limited between-site information ($J = 39$, $\sigma = 0.05$) but abundant within-site information ($\bar{n}_j = 177.9$). In this scenario, the choice of model does not substantially influence performance once a suitable posterior summary method is employed (i.e., PM for RMSE, CB/GR for ISEL). The data-adaptive nature of the DP models is evident; they exhibit less shrinkage than the Gaussian model in Case 2, where $\sigma$ is large, 

\noindent but more shrinkage in the current scenario, where $\sigma$ is small, even with the relatively significant within-site information. As demonstrated in Figure~\ref{fig:figure05}, the true cross-site impact variation, $\sigma$, tends to have a considerable impact on the DP models, particularly when $J$ is moderate to large. In situations where $\sigma$ is estimated to be very small (i.e., $\sigma \leq 0.10$), Gaussian models may help protect results from excessive shrinkage. Similar patterns are evident in scenarios with $J=75$ or $100$ (see the Supplemental Material). 

%%%%%%%%%%%%%%%%%%%%%%%%%%%%%%%%%%%%%%
%%% 7. Results: Results: Real data example
%%%%%%%%%%%%%%%%%%%%%%%%%%%%%%%%%%%%%%

\section{Results: Real data example}

%%% Section and data introduction

In this section, we apply the strategies from this paper to study site-specific effects in the Conditional Subsidies for School Attendance program in Bogota, Colombia \citep{barrera2019medium}. The program conducted two conditional cash transfer (CCT) experiments in San Cristobal and Suba, covering 13,491 participants across 99 sites. In San Cristobal, eligible students in grades 6 to 11 were randomly assigned to a ``basic” treatment, ``savings” treatment, or control group. In contrast, in Suba, eligible participants in grades 9 to 11 were randomly assigned to a tertiary treatment or control group. We focus on the primary outcomes of secondary school enrollment and graduation in San Cristobal, with results for the Suba district available in the Supplemental Material.

%%% Multisite trial design characteristics

We first extract the design characteristics of the multisite trials, presented in Table 1. In San Cristobal, a total of 6,506 participants were nested within 38 sites, with an average site size of 171.2. Site sizes varied considerably, ranging from 23 to 484, with a coefficient of variation (\text{CV}) of 0.67. We combine the basic and savings treatments for analysis, resulting in an average treatment rate of 67\% within each site. 

\begin{table}[ht]\label{table1}
\centering
\begin{threeparttable} % Wrap the tabular environment with threeparttable
\caption{Multisite trial design characteristics of the Conditional Subsidies for School Attendance study and estimates of the mean and standard deviation of the distribution of site-specific treatment effects in effect size units (San Cristobal district).}
\begin{tabular}{ccc}
\toprule
 & \multicolumn{2}{c}{Basic/Savings experiment} \\
 & On-time secondary              & Secondary   \\
 & enrollment                     & graduation  \\ \midrule
Average treatment effect $\left(\hat{\tau}_d\right)$ & 0.07 & 0.05 \\
Cross-site effect SD $\left(\hat{\sigma}_d\right)$ & 0.04 & 0.06 \\
Geometric mean of $\widehat{s e}_j^{\prime} s$ & 0.21 & 0.24 \\
Average reliability of $\widehat{\tau}_j(I)$ & 0.04 & 0.06 \\
Total sample size $(N)$ & \multicolumn{2}{c}{6,506} \\
Number of $\operatorname{sites}(J)$ & \multicolumn{2}{c}{38} \\
Average site size $\left(\bar{n}_j\right)$ & \multicolumn{2}{c}{171.2} \\
Coefficient of variation of site sizes (CV) & \multicolumn{2}{c}{0.67} \\
Range of site sizes & \multicolumn{2}{c}{$(23,484)$} \\
Average proportion of units treated $\left(\bar{p}_j\right)$ & \multicolumn{2}{c}{0.67} \\
\bottomrule
\end{tabular}
\begin{tablenotes} % Add notes to the table
\item\textit{Note}: We excluded sites with extreme probabilities, that is, necessitating both $n_j p_j$ and $n_j (1 - p_j)$ to be at least 8, where $n_j$ represents the site size and $p_j$ denotes the proportion of participants treated in site $j$. Detailed distributions of $n_j$ and $p_j$ and comparisons of outcome means by experimental group are available in the Supplemental Material.
\end{tablenotes}
\end{threeparttable}
\end{table}

%%% Assessing data informativeness

We then apply the basic Rubin model (equations~(\ref{eq:1_Rubin}) and~(\ref{eq:2_Rubin})) to estimate the cross-site impact variation ($\sigma$) and the sampling variation within sites. The average reliability of $\hat{\tau}_j$ ($I$) for each outcome is crucial for assessing data informativeness within and between sites, guiding the choice of the optimal model/method combination. Since all four outcomes are binary, we summarize the observed site-specific effects using the log odds ratio (\text{logOR}), yielding $\tau$ and $\sigma$ estimates on the logOR scale. To facilitate interpretation, we transformed these estimates to the standard mean difference scale, known as Cohen's $d$. Specifically, we converted using $\hat{\tau}_d=$ $\widehat{\operatorname{logOR}} \cdot(\sqrt{3} / \pi)$ for the average treatment effect, and $\hat{\sigma}_d=\left[\operatorname{Var}(\widehat{\operatorname{logOR}}) \cdot\left(3 / \pi^2\right)\right]^{-1 / 2}$ for the cross-site effect standard deviation \citep{borenstein2009converting}.

%%% Estimation results - San Cristobal

San Cristobal's multisite CCT experiment exhibited small between-site information ($J = 38$, $\sigma = 0.04$ or $0.06$) and large within-site information ($\bar{n}_j = 171.2$). The estimated average reliability of $\hat{\tau}_j$ for this experiment were 0.04 and 0.06 for secondary on-time enrollment and graduation, respectively. As in Section~\ref{subsec:6-3}, this led to significant shrinkage toward the prior mean effect $\tau$, emphasizing the importance of choosing the appropriate model-method combination for controlling shrinkage and achieving the desired inferential goals.

\begin{figure}[H]
    \centering
    \includegraphics[width=\textwidth]{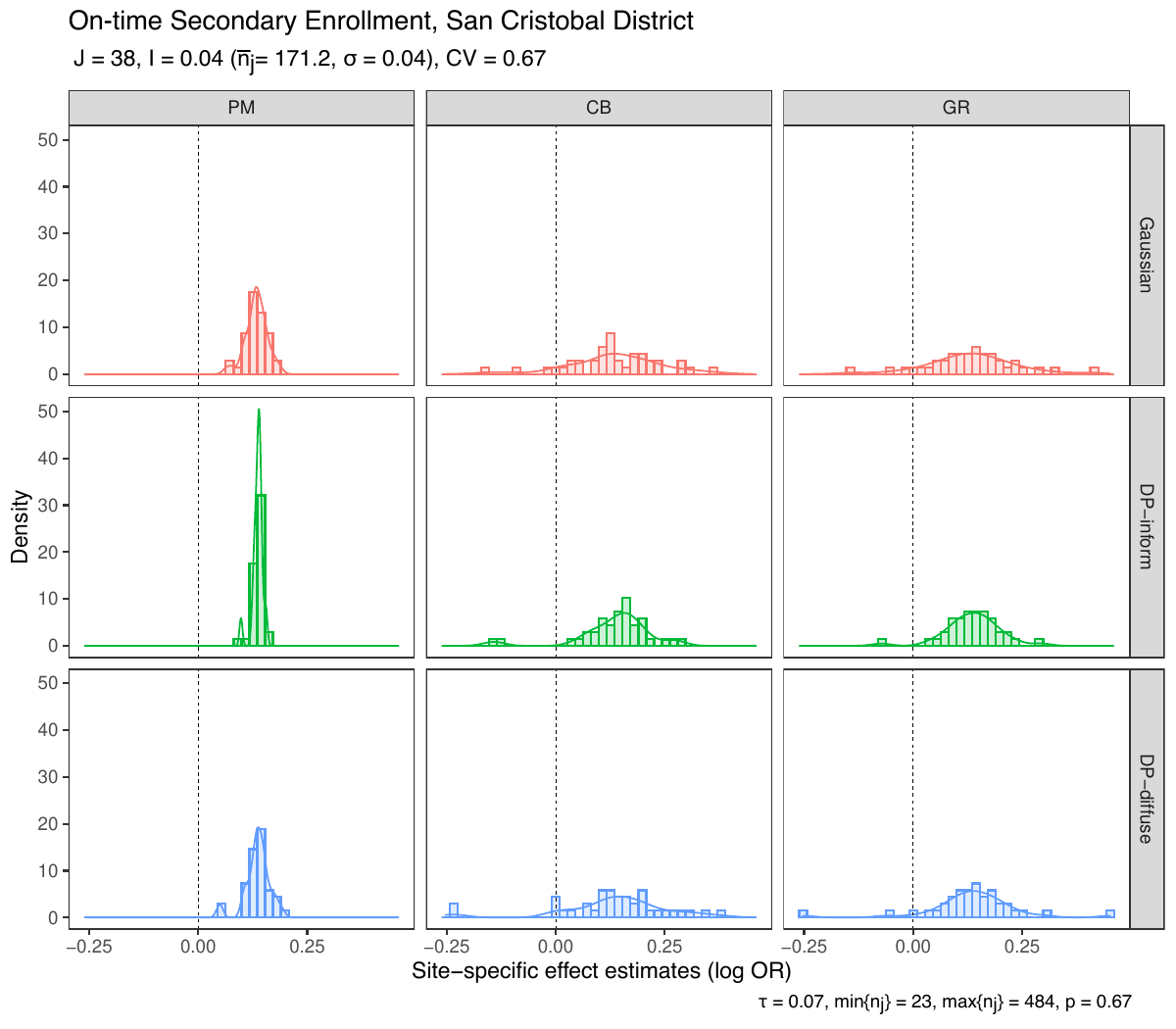}
    \caption{Distribution of site-specific treatment effect estimates on on-time secondary enrollment in San Cristobal district. The top-right distribution is the preferred estimator, given the characteristics of the data.}
    \label{fig:figure09}
\end{figure}

Figure~\ref{fig:figure09} highlights the considerable differences in the distributions of site-specific effects estimated by the model/method combinations. The posterior mean summary method and the DP-inform model notably shrink estimates more strongly than the other methods and models. Interestingly, the DP-diffuse model preserves the extreme estimates around -0.25 and 0.40 by assigning greater weight to a broad range of possible $\tau_j$ values, unlike the other two models. We also observe that the GR summary method generates smoother site-effect histograms due to its use of quantile estimates. 

From the third case study, discussed in Section~\ref{subsec:6-3}, we may be more inclined to trust the EDF estimates produced by the model-method combination least prone to excessive shrinkage. A simulated case study based on the San Cristobal experiment scenario, detailed in the Supplemental Material, indicates that the Gaussian model with the PM and GR summary methods demonstrates the best performance for RMSE and ISEL, respectively. If extreme effects were expected or of particular interest, we might choose to use the DP-diffuse model instead.

%%%%%%%%%%%%%%%%%%%%%%%%%%%%%%%%%%%%%%
%%% 8. Discussion
%%%%%%%%%%%%%%%%%%%%%%%%%%%%%%%%%%%%%%

\section{Discussion}

%%% Restate the research question or objective of the study

This study primarily aimed to identify optimal approaches for enhancing inferences about site-specific effect estimates to address various inferential goals. We conducted a systematic simulation study evaluating the performance of combinations of two strategies: flexible prior modeling for $G$ and posterior summary methods designed to minimize specific loss functions. We focused on the low-data environments commonly encountered in multisite trials. Below we discuss the key insights derived from this study.

%%% Main finding #1: Data informativeness for different inferential goals

The most influential factor for all inferential goals was the average reliability, or informativeness $I$, of raw site-specific effect estimates. In the data analysis stage, researchers should initially estimate the average reliability of $\hat{\tau}_j$ ($I$), as it dictates data informativeness and guides the best model-method combination. Increasing $\bar{n}_j$, which enhances the average precision of the $\hat{\tau}_j$ estimates, significantly improved the results for all three loss functions: RMSE, ISEL, and MSELP. Thus, when confronted with budgetary or logistical constraints during the design phase of multisite trials, it is important to prioritize increasing the average number of subjects per site (irrespective of the variation in site sizes) when the focus is on estimating site-specific effects. This holds especially true if the goal is to rank the $\tau_j$ values, given that MSELP was only affected by $\bar{n}_j$ and $\sigma$.

%%% Main finding #2: Best model-method combinations for different settings
%%% High informative settings
%%% DP-model's sensitivity to between-site information

In scenarios characterized by high data informativeness or a large average reliability of $\hat{\tau}_j$ ($I$), semiparametric Dirichlet Process models typically outperform Gaussian models. Nonetheless, we also found that certain requirements must be satisfied to use DP priors with a reasonable degree of confidence in practical applications. First, the performance of DP models is particularly sensitive to the amount of between-site information in the data. For a fixed average site size, the ability of DP models to recover the EDF of $\tau_j$'s is greatly enhanced by a large $\sigma$ and a large number of sites, $J$. Specifically, the level of cross-site impact variation, $\sigma$, dramatically alters the DP model's relative performance on ISEL compared to the Gaussian model. The case studies also demonstrated that when the value of $I$ is low (around $I=0.10$), the estimates of $\tau_j$ obtained using the DP models exhibited shrinkage behavior that was sensitive to the value of $\sigma$, shrinking less when $\sigma$ was large and more when $\sigma$ was small. These results indicate the necessity of assessing whether sufficient cross-site information ($\sigma$ and $J$) is available to effectively apply semiparametric estimation in practice. Flexible models such as DP models and recent NPMLE-based deconvolution methods \citep{armstrong2022robust, gu2023invidious} are designed to adapt to cluster-level data, and their mathematical guarantees are asymptotic in nature. Consequently, their suitability for analyzing real-world multisite trial data — where obtaining highly informative data both within and between sites is extremely challenging — must be carefully considered.

%%% Joint usage matters, even in high informative settings

Second, even in highly informative settings with abundant between-site information, DP priors must be used jointly with targeted posterior summary methods to achieve specific inferential goals. Our simulations demonstrate that, even in highly informative settings, the posterior mean method combined with DP priors consistently outperforms both CB and GR in terms of RMSE, while it is rarely the best estimator for ISEL. We emphasize the joint use of flexible distributions for $G$ and targeted posterior summary methods because estimators that produce optimal estimates of the collection of individual site-specific effects may produce poor estimates of the shape of their distribution, and vice versa.

%%% Low informative settings
%%% Posterior summary methods matter more
%%% Gaussisan models may be preferrable

In scenarios with low-to-moderate data informativeness, often observed in multisite trials with an average reliability below 0.20, it is much more important to target the posterior summary to the metric of interest than it is to choose a flexible prior model for $G$. In simulations with these conditions, the Gaussian model, paired with an appropriate posterior summary method, performed on par with or better than the DP models, even when the true $G$ was non-Gaussian. Practically, researchers might lean towards Gaussian models when data is limited. While the performance of DP models is data-adaptive and insensitive to the choice of prior on $\alpha$ when sufficient information is available, the choice can substantially affect the degree of shrinkage in uninformative settings. As a result, the Gaussian model offers a somewhat less assumption-heavy approach for estimation, which may be preferable in such cases.

%%% Limitation & future work
%%% Site size independent of site-level effects

A key limitation of this study is the assumption that site size is independent of site-specific treatment effects, i.e., zero correlation between $\tau_j$ and $\widehat{se}_{j}^{2}$ in our data-generation and analytic methods. In general, smaller school sites may be less (or more) efficient than larger school sites in implementing efficacious interventions and thus tend to have smaller (or larger) $\tau_j$ values \citep{angrist2022methods}. One advantage of the Bayesian model-based approach is the ability to estimate models with and without restrictions on the bivariate distribution of $\tau_j$ and $\widehat{se}_{j}^{2}$, and to assess the sensitivity of site-specific effect estimates to this assumption. Alternatively, we can treat $\widehat{se}_{j}^{2}$ as a covariate in the conditional distribution of $G$, or apply \cite{brown2008season}’s variance-stabilizing transform to $\tau_j$ to achieve approximately constant sampling variance.

%%% Concluding remark

In conclusion, no single estimation strategy universally excels across varied data scenarios and inferential goals. Flexible semiparametric models are not a panacea, given the complex interplay between study design characteristics and modeling choices. Researchers should carefully select methods tailored to their specific inferential goals, taking into account the amount of within-site and between-site information. Ideally, multisite trials should be prospectively designed to address inferential goals related to site-specific treatment effects.

%%% End of double spacing
\end{doublespace}

%%%%%%%%%%%%%%%%%%%%%%%%%%%%%%%%%%%%%%
%%% References
%%%%%%%%%%%%%%%%%%%%%%%%%%%%%%%%%%%%%%

\clearpage
\bibliographystyle{apalike}
\bibliography{bibliography}

%%%%%%%%%%%%%%%%%%%%%%%%%%%%%%%%%%%%%%
%%% Supplemental Material Cover Page
%%%%%%%%%%%%%%%%%%%%%%%%%%%%%%%%%%%%%%

\clearpage
\thispagestyle{empty} % Removes page number
\begin{center}
    \vspace*{2cm} % Adjust vertical spacing as needed
    {\Large\bfseries Supplemental Material for\\[0.5cm] ‘Improving the Estimation of Site-Specific Effects and their Distribution in Multisite Trials’\par}
    \vspace{3cm} % Adjust vertical spacing as needed
    % Include any additional details here, such as author names, institution, etc., if required
    \vfill % Pushes the following text to the bottom of the page
    % You can include the date, version, or any other information you deem necessary
\end{center}
\clearpage % Ends the current page

%%%%%%%%%%%%%%%%%%%%%%%%%%%%%%%%%%%%%%
%%% Appendix
%%%%%%%%%%%%%%%%%%%%%%%%%%%%%%%%%%%%%%

%%% Setup - New numbers for components
\appendix
\titleformat{\section}{\normalfont\Large\bfseries}{Appendix \Alph{section}}{1em}{}
\renewcommand{\thesubsection}{\Alph{section}.\arabic{subsection}}
\renewcommand{\theequation}{\Alph{section}.\arabic{equation}}
\numberwithin{equation}{section}
\renewcommand{\thefigure}{Figure \Alph{section}.\arabic{figure}}
\counterwithin{figure}{section}
\renewcommand{\thetable}{Table \Alph{section}.\arabic{table}}
\counterwithin{table}{section}

%%% Set double spacing & Indentation
\begin{doublespace}
\setlength{\parindent}{2em}

%%%%%%%%%%%%%%%%%%%%%%%%%%%%%%%%%%%%%%
%%% Appendix A. Simulation details
%%%%%%%%%%%%%%%%%%%%%%%%%%%%%%%%%%%%%%

\clearpage
\section{Simulation details}

%%%%%%%%%%%%%%%%%%%%%%%%%%%%%%%%%%%%%%
%%% A.1. Connecting sampling errors to site sizes
%%%

\subsection{Connecting sampling errors to site sizes}\label{subsec:B-1}

For a simple design-based estimator for site-specific effects $\hat{\tau}_j$, the sampling variation of each $\hat{\tau}_j$ around its true effect $\tau_j$ within sites will be approximately

\begin{equation}
    \widehat{s e}_j^2=\frac{1}{n_j} \cdot\left(\frac{1}{p_j}+\frac{1}{1-p_j}\right) \cdot \operatorname{Var}(Y) \cdot\left(1-R^2\right)
\end{equation}

\noindent where $n_j$ is sample size at site $j$ and $p_j$ is the proportion of units treated in site $j$, which \cite{rosenbaum1983central} refer to as the \textit{propensity score}. $\text{Var}(Y)$ represents the assumed constant variation in outcome for a given treatment arm in a given site. $R^2$ is the explanatory power of level-1 covariates. This expression is Neyman's classic formula under the assumption of homoskedasticity \citep{miratrix2021applied}. 

In effect size units, $\text{Var}(Y)=1$. If we further assume constant proportion treated, for convenience, we have

\begin{equation}
    \widehat{se}_{j}^{2}= \frac{1}{n_j} \cdot \left(\frac{1}{p} + \frac{1}{1-p}\right) \cdot(1-R^2) = \frac{1}{n_j}\kappa
\end{equation}

\noindent for some constant $\kappa$. If $p = 0.5$ and $R^2=0.6$ with a reasonable pre-test score, $\kappa=2.4$. Without level-1 covariates, $\kappa$ equals 4. Now we can obtain a vector of simulated $\widehat{se}_{j}^{2}$’s, denoted as $\hat{\text{E}}$, by directly controlling two constants, $p$ and $R^2$, and one variable, $n_j$. Our data-generating function for $\hat{\text{E}}$ follows this relationship assuming $p = 0.5$ and $R^2 = 0$. Consequently, the magnitude of the sampling errors were set to be $\widehat{se}_j^2 = 4/n_j$.

%%%%%%%%%%%%%%%%%%%%%%%%%%%%%%%%%%%%%%
%%% B.2. The average reliability $I$ 
%%%      under the simulation conditions of this study
%%%

\subsection{The average reliability \texorpdfstring{$I$}{Lg} under the simulation conditions of this study}\label{subsec:B-2}

The average reliability $I$ determines how informative the first-stage ML estimates $\hat{\tau}_j$’s are on average. $I$ is calculated as $\sigma^2/(\sigma^2+\text{GM}(\widehat{se}_j^2))$ where $\text{GM}(\widehat{se}_j^2)$ is the geometric mean of $\widehat{se}_j^2$, $\text{exp}(\sum^{J}_{j =1}{\text{ln}(\widehat{se}^2_j}))$. Consequently, a large average reliability value indicates less noisy and more informative observed ML estimates $\hat{\tau}_j$’s relative to the degree of cross-site impact heterogeneity.

\begin{figure}[H]
    \centering
    \includegraphics[width=\textwidth]{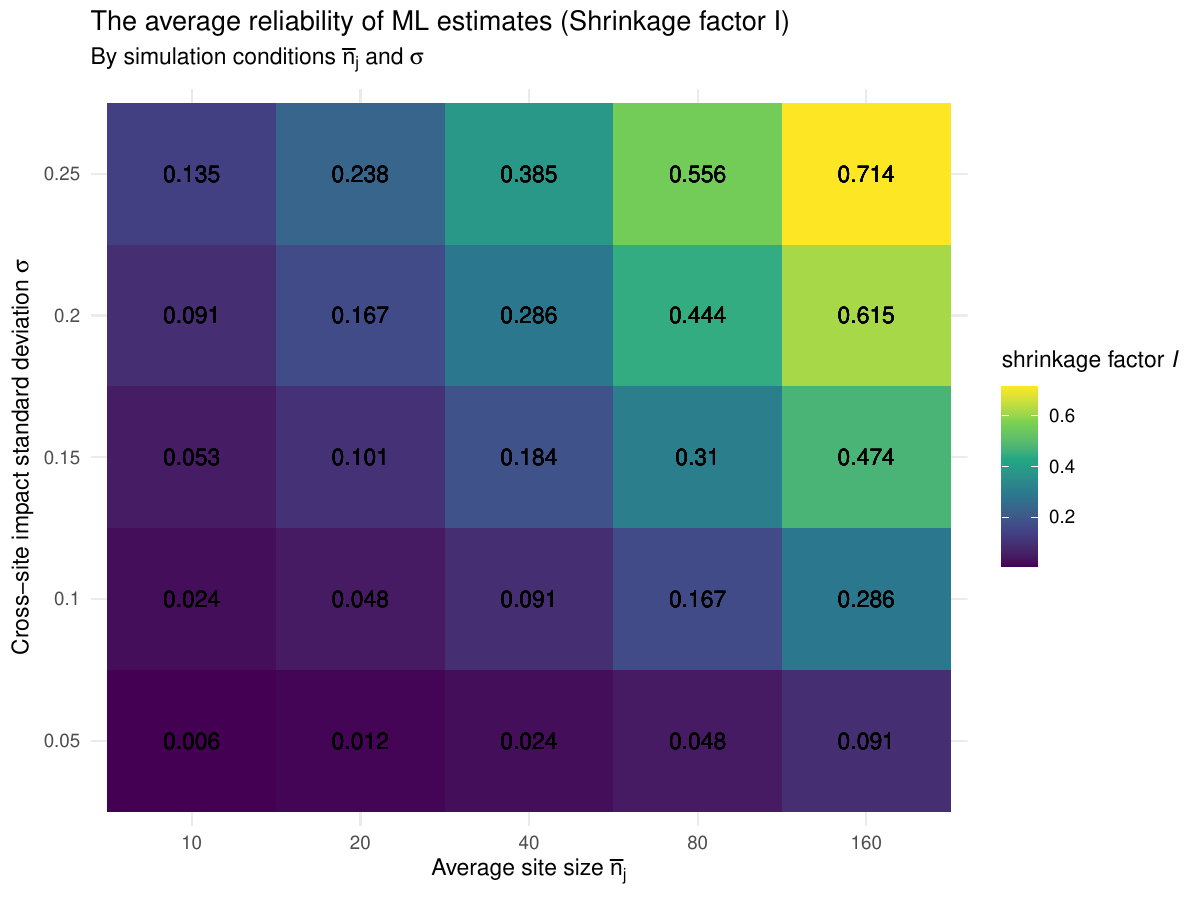}
    \caption{The magnitude of the average reliability of ML estimates ($I$) implied by the simulation conditions of the cross-site impact standard deviation ($\sigma$) and the average site size $\bar{n}_j$ Note: The proportion treated within each site ($p$) is assumed to be 0.5. No within-site level-1 covariates are included ($R^2=0$).
}
    \label{fig:figure_b01}
\end{figure}

Figure~\ref{fig:figure_b01} illustrates the magnitudes of $I$ assumed by the simulation conditions we chose for $\sigma$ and $\bar{n}_j$. The assumed $I$ level is equal to 0.714 if the average size of sites is 160 and the cross-site impact standard deviation is 0.25. In this situation, the cross-site impact variance is approximately 2.5 times the average within-site sampling variance ($\sigma^2/\text{GM}(\widehat{se}_j^2)=2.5$). In contrast, if $I = 0.184$, the average $\widehat{se}_j^2$ is roughly four times as large as $\sigma^2$ ($\sigma^2/\text{GM}(\widehat{se}_j^2)=0.225$), resulting in noisy $\hat{\tau}_j$’s. Since $I = 0.184$ is the median of the possible $I$ levels shown in Figure~\ref{fig:figure_b01}, half of our simulation conditions assume quite noisy and low informative data environments, which are often encountered in practical applications with small site sizes.

%%%%%%%%%%%%%%%%%%%%%%%%%%%%%%%%%%%%%%
%%% B.3. Data-generating models for $G$
%%%

\subsection{Data-generating models for \texorpdfstring{$G$}{Lg}}\label{subsec:B-3}

\begin{figure}[H]
    \centering
    \includegraphics[width=\textwidth]{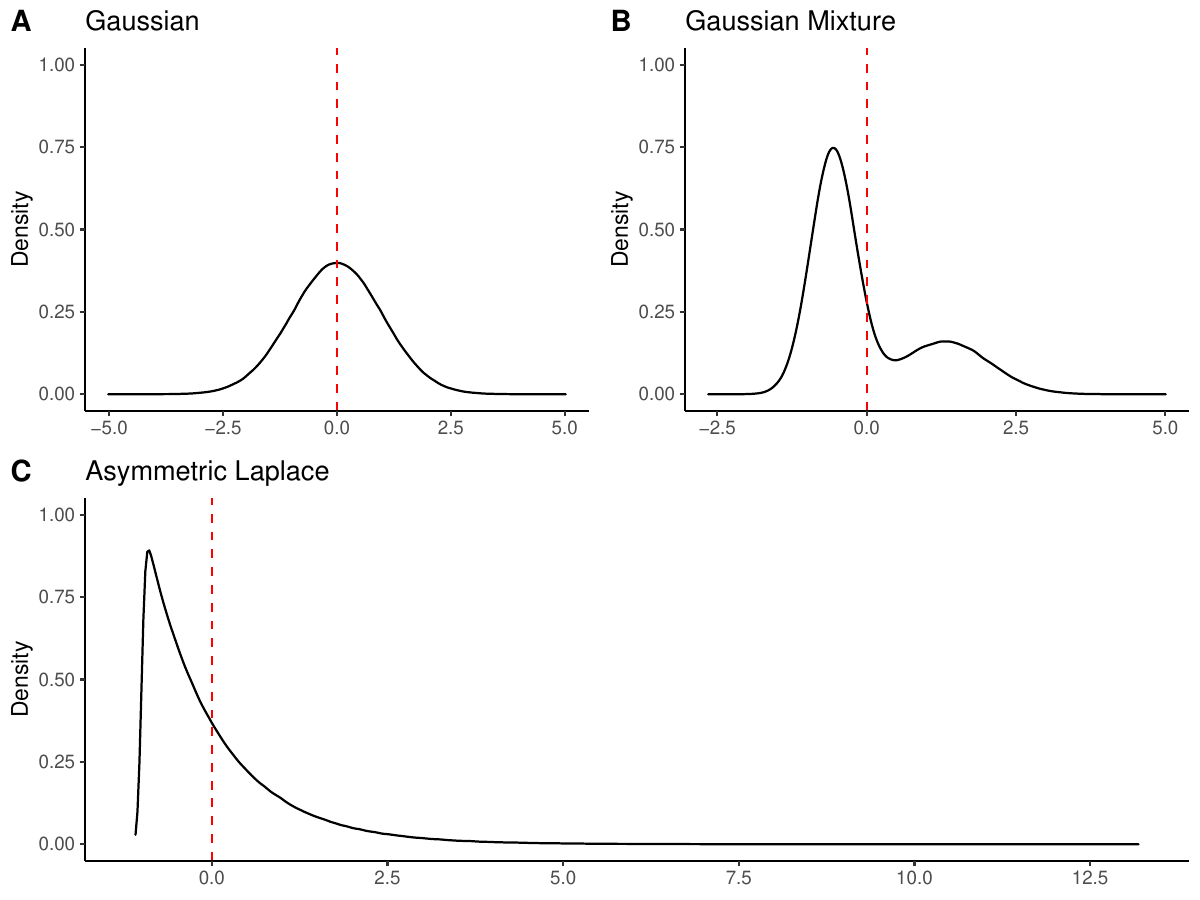}
    \caption{The true cross-site population distribution of $\tau_j$’s ($G$): Gaussian, Gaussian mixture, and asymmetric Laplace distributions}
    \label{fig:figure02}
\end{figure}

Figure~\ref{fig:figure02} illustrates three different population distribution $G$ for $\tau_j$: Gaussian, a mixture of two Gaussians, and asymmetric Laplace (AL).

Note that the Gaussian distribution in Figure~\ref{fig:figure02} has a mean of zero and a variance of one ($\tau_j \sim N(0, 1)$). For comparability with the Gaussian model, the Gaussian mixture and AL models were also normalized to have zero means and unit variances. Then, we rescaled the $G$ by multiplying the sampled distributions by $\sigma$.

Suppose that the Gaussian mixture data-generating model has two mixture components, $N(\tau_1, \sigma_1^2)$ and $N(\tau_2, \sigma_2^2)$, with a mixture weight for the first component, $w$. To force this mixture distribution to have mean 0 and variance 1, we define a normalizing factor denoted as $C$:

\begin{equation}
    C=\left[wu^2 + (1-w) + w(1-w)\delta^2 \right]^{1/2},
\end{equation}

\noindent where $\delta = \tau_2 - \tau_1$ and $u = \sigma_2^2 / \sigma_1^2$. Then, with probability $w$, $\tau_j$ is simulated from the first normalized component $N(- \frac{w\delta}{C}, \frac{1}{C})$, otherwise from the second normalized component $N(- \frac{(1-w)\delta}{C}, \frac{u}{C})$ with probability of $1-w$.

To simulate $G$ from the AL distribution with zero mean and unit variance, the location and scale parameters, $\mu$ and $\psi$, are adjusted as follows as function of the skewness parameter $\rho$:

\begin{equation*}
    \tau_j |\mu, \psi, \rho \sim \text{AL}(\mu, \psi, \rho) \quad j = 1, ..., J
\end{equation*}

\noindent where

\begin{equation}
    \mu = - \frac{\psi(1-\rho^2)}{\sqrt{2}\rho}, \quad \psi = \left[\frac{2\rho^2}{1+\rho^4} \right]^{1/2}.
\end{equation}

\noindent The skewness parameter $\rho$ is set to 0.1 so that the resulting $G$ can have a right-skewed distribution with a long tail. This AL data-generating model is useful to evaluate whether the two strategies for the improved $\tau_j$ inferences help to recover large but rare site-level effects.

%%%%%%%%%%%%%%%%%%%%%%%%%%%%%%%%%%%%%%
%%% B.4. Data-analytic models 
%%%

\subsection{Data-analytic models }\label{subsec:B-4}

We fit three models to each of the simulated datasets: (a) $G$ is standard Gaussian model, (b) $G$ is a Dirichlet process mixture (DPM) model with a diffuse $\alpha$ prior (DP-diffuse), and (c) $G$ is a DPM model with an informative $\alpha$ prior (DP-inform). These three models share the same first stage model specifications: $\hat{\tau}_j \mid \tau_j, \widehat{s e}_j^2 \sim N(\tau_j, \widehat{s e}_j^2)$. The models differ based on the second stage specifications for modeling $G$, $\tau_j \mid \boldsymbol{\theta} \sim G$ where $\boldsymbol{\theta}$ represents a vector of hyperparameters for $G$.

The Gaussian model assumes that $G \sim N(\tau, \sigma^2)$ where hyperpriors are set to be vague: $\tau \sim N(0,100)$ and $\sigma \sim \operatorname{Cauchy}(0,5)$. We avoid using the inverse-gamma prior for $\sigma^2$ because it does not have any proper limiting posterior distribution particularly for the simulation settings we have, where the number of sites $J$ is small or the site-level variation $\sigma^2$ is small by design \citep{gelman2006prior}. The two DPM models assume that $G \sim \operatorname{DP}\left(\alpha, G_0\right)$ where $G_0$ is a Gaussian base distribution and $\alpha$ is a precision parameter, and differ in their priors for the precision parameter. Both DPM models, DP-diffuse and DP-inform, share the same base distribution $G_0 \sim N(\tau_j \mid \tau, \sigma^2)$ with non-informative hyperpriors as in the Gaussian model: $\tau \sim(0,100)$ and $\sigma \sim \operatorname{Cauchy}(0,5)$.

The two DPM models differ in their specification of the $\operatorname{Gamma}(a, b)$ priors for $\alpha$. In the DP-diffuse models, $a$ and $b$ are chosen such that the mean and variance of $\alpha$ are $\mathrm{E}(\alpha \mid a, b) = J / 2$ and $\operatorname{Var}(\alpha \mid a, b) = J / 5$, respectively. For the five choices of $J = 25, 50, 75, 100,$ and $300$, the pairs of $(a, b)$ for the DP-diffuse model are $(1.25,0.1), (2.5,0.1), (3.75,0.1), (5.0, 0.1)$, and $(15,0.1)$, respectively. Conversely, for the DP-inform models, $a$ and $b$ are chosen as the solution that minimizes the KL distance of the distribution of the number of clusters to the $\chi^2$ distribution with $\mathrm{df}=J/10$. This means that the DP-inform models assume the expected number of clusters $K$ is $J/10$. The obtained solutions were $(1.24,0.64), (1.60,1.22), (2.72, 1.36), (3.88, 1.44)$, and $(9.32, 0.88)$ for the five choices of $J = 25, 50, 75, 100,$ and $300$, respectively. Section~\ref{subsec:A-3} outlines the detailed procedure for selecting $a$ and $b$ in the DP-inform models.

%%%%%%%%%%%%%%%%%%%%%%%%%%%%%%%%%%%%%%
%%% B.5. Meta-model regressions
%%%

\subsection{Meta-model regressions}\label{subsec:B-5}

Meta-model regressions offer a comprehensive summary of results by regressing performance metrics on the experimental factors. In contrast to visual or descriptive analyses, meta-models deliver a model-based inferential analysis, guided by experimental design, that enables the accurate identification of significant patterns and precise estimation of their magnitudes \citep{boomsma2013reporting, paxton2001monte}. 

Our meta-model regressions follow the form: $\log(\text{performance})=\beta_0+X \beta+\epsilon$. We focus on the three performance criteria as our target outcome variables: RMSE, ISEL, and MSELP. To ease the interpretation of the meta-model regression coefficients, we apply a logarithmic transformation to the outcome variables. As for the explanatory variables $X$, we create a set of dummy variables representing the six design and data-analytic factors, along with their two-way interaction terms. The reference groups are $J=25, \bar{n}_j=10, \sigma=0.05, \mathrm{CV}=0.00$, the Gaussian model, and the PM summary method.

Figure~\ref{fig:figure03} visualizes the difference in predicted log-outcome between the given condition and the baseline condition, i.e., $\log(\text{outcome})-\hat{\beta}_0=\left(\hat{\beta}_0+X \hat{\beta}\right)-\hat{\beta}_0=X \hat{\beta}$, where $X$ corresponds to the given design factor values and $\hat{\beta}_0$ corresponds to the average log-outcome in the baseline condition. Recall that for $x$ and $y$ close to each other, $\log (x)-\log (y) \approx \frac{x-y}{y}$, i.e., the difference in log-outcomes approximates the percent change. In Figure~\ref{fig:figure03}, the estimated differences in log outcomes were exponentiated to yield the multiplicative factor of change in average performance metrics.

Lastly, since we include each dataset nine times in each meta-model regression (corresponding to every model/posterior summary method combination), we cluster our standard errors on datasets. This approach results in a model akin to a repeated-measures ANOVA, with data-analytic model and summary method serving as within-subject factors, and the remaining design variables acting as between-subject factors.

%%%%%%%%%%%%%%%%%%%%%%%%%%%%%%%%%%%%%%
%%% Appendix A. Estimator and model details
%%%%%%%%%%%%%%%%%%%%%%%%%%%%%%%%%%%%%%
\clearpage
\section{Estimator and model details}

%%%%%%%%%%%%%%%%%%%%%%%%%%%%%%%%%%%%%%
%%% A.3. Dirichlet process priors
%%%

\subsection{Dirichlet process priors}\label{subsec:A-3}

%%% $G_0$ and $\alpha$ - A weighted mixture mechanism

$G_0$ provides an initial best guess of the shape of the prior distribution $G$, which is commonly taken to be a Gaussian distribution in practice. The precision parameter $\alpha$ then controls the degree of shrinkage of $G$ to $G_0$. In other words, $\alpha$ determines the extent to which distributions in the sample space partitioned into measurable subsets $G_1, ..., G_s$ are divergent from $G_0$. To understand the role of $G_0$ and $\alpha$ more intuitively, it is helpful to refer to a form of the induced prior distribution on the site-specific parameter $\tau_j$, so-called the \textit{Polya urn scheme} \citep{dunson2007bayesian}:

\begin{equation}\label{eq:A-5_polya}
    \tau_j|G_0, \alpha, \tau_1, ..., \tau_{j-1} \sim \left(\frac{\alpha}{\alpha + j -1} \right) \cdot G_0 + \left(\frac{1}{\alpha + j -1} \right) \cdot \sum_{k = 1}^{j - 1}{\delta(\tau_k)}, 
\end{equation}

\noindent where $\delta(\tau_k)$ denotes a point mass at $\tau_k$. This conditional prior distribution for $\tau_j$ is a weighted mixture of the base distribution $G_0$ and probability masses at the previous site’s parameter values, that is, the EDF of $(\tau_1, ..., \tau_{j-1})$. In this scheme, the first site’s treatment effect $\tau_1$ is drawn from $G_0$. Then the second site’s treatment effect $\tau_2$ is drawn from $G_0$ with probability of $\alpha/(\alpha+1)$ or a new empirical distribution $\delta(\tau_1)$ with probability of $1/(\alpha+1)$. This sampling rule continues, and for the $j$th site, $\tau_j$ is drawn from $G_0$ with probability proportional to $j-1$, the number of previous sites which already have realized site-specific parameters, or is sampled from the new empirical distribution of $\sum_{k=1}^{j-1}{\delta(\tau_k)}$ with probability proportional to $\alpha$ \citep{gelman2013bayesian}.

%%% $\alpha$ as a prior sample size

$\alpha$ can be viewed as a prior sample size in some sense \citep{gelman2013bayesian}, as opposed to the sample size of empirical data $J$. Thus, a huge $\alpha$ value implies an extreme weight on the (prior) base distribution $G_0$. In that case, the joint distribution of $\tau_j$’s tends to be the product of $J$ independent draws from $G_0$ \citep{antonelli2016mitigating} and the second-stage model in equation~(\ref{eq:12_DPM}) converges to the \cite{rubin1981estimation} model with a Gaussian prior. On the other hand, a zero $\alpha$ value imposes a null weight on $G_0$, which leads to the distribution of $\tau_j$ being a point mass of $\delta(\tau_1)$. Then, the second-stage model in equation~(\ref{eq:12_DPM}) collapses to a model with all sites sharing the common value of $\tau_1$.

%%% $\alpha$ determines the number of clusters

Hence, we can infer that $\alpha$ determines the number of distinct values of $\tau_j$, often referred to as the unique number of \textit{clusters} $K$ generated by the DP. The $K$ is not necessarily an exact representation of the number of \textit{mixture components} $C$ (latent subpopulation with substantive meaning) as specified in finite mixture models, but $K$ can be considered as an upper bound of the $C$ \citep{ishwaran2000markov}. The expected number of $K$ is a function of $\alpha$ and $J$, given by the sum of the weights of $G_0$ in equation~(\ref{eq:A-5_polya}) over all $J$ sites:

\begin{equation}
\mathrm{E}\left(K \mid G_0, \alpha, J\right)=\sum_{j=1}^J \frac{\alpha}{\alpha+j-1}
\end{equation}

%%% The hyperprior for $\alpha$

The hyper prior for $\alpha$ plays an essential role in determining the expected number of clusters and therefore in controlling the posterior distribution over clusters. In practice, it is a standard approach to use a $\text{Gamma}(a, b)$ distribution with fixed hyperparameters, the shape parameter $a$ and the rate parameter $b$, to capture the uncertainty in $\alpha$ \citep{escobar1995bayesian}. A key issue is whether the choice of $a$ and $b$ may have a substantial impact on the posterior distribution of $\alpha$, and in turn on the clustering behavior. There exist a group of studies arguing that the choice of hyperparameters is a less of concern because the data tend to be quite informative, resulting in a concentrated posterior even with a high variance prior for $\alpha$ \citep{leslie2007general, gelman2013bayesian}. On the other hand, another group of studies report that estimation or inference can be sensitive to the specific choice of the hyperparameters and in general to the strategies for selecting $\alpha$ \citep{dorazio2009selecting, paddock2006flexible, murugiah2012selecting}. Our interest is to evaluate the sensitivity under two different options, diffuse and informative DP priors, particularly in the context of recovering the EDF of $\tau_j$’s.

%%% The first option to specify the $\alpha$ prior: Diffuse prior

The first option is to specify a diffuse Gamma distribution when a priori knowledge on $\alpha$ or $K$ is absent. \cite{antonelli2016mitigating} chose values of $a$ and $b$ such that $\alpha$ is centered between 1 and $J$ with a large variance to assign a priori mass to a wide range of $\alpha$ values. If $J = 50$, for example, we can assign 25 as the mean of the $\alpha$ distribution and 250 as a variance which is ten times the magnitude of the mean. Given these a priori values for the mean and variance of $\alpha$, we can obtain the corresponding values of $a = 2.5$ and $b = 0.1$ based on the moments of a Gamma distribution: $\text{E}(\alpha|a, b) = a/b$ and $\text{Var}(\alpha|a, b) = a/b^2$. 

%%% The second option to specify the $\alpha$ prior: Informative prior

This study suggests the second option, using $\chi^2$ distribution to construct an informative prior for $\alpha$. This strategy is based on the probability mass function for the prior distribution of $K$ induced by a $\text{Gamma}(a, b)$ prior for $\alpha$ and the number of sites $J$ \citep{dorazio2009selecting, antonelli2016mitigating}:

\begin{equation}\label{eq:A-7_probK}
    \text{Pr}(K|J, a, b) = \frac{b^a \cdot S_1(J, K)}{\Gamma(a)} \cdot \int_{0}^{\infty}{\frac{\alpha^{K+a-1} \cdot \text{exp}(-b\alpha) \cdot \Gamma(\alpha)}{\Gamma(\alpha + J)}} d\alpha,
\end{equation}

\noindent where $S_1(J, K)$ is the unsigned Stirling number of the first kind and $K = 1, ..., J$. Suppose $\text{Pr}(K)$ that encodes our prior information for the distribution of the expected number of clusters $K$. We can obtain a solution for $a$ and $b$ by minimizing the discrepancy between the encoded prior $\text{Pr}(K)$ and the $\alpha$-induced prior $\text{Pr}(K|J, a, b)$, defined by the following Kullback-Leibler (KL) divergence measure: 

\begin{equation}\label{eq:A-8_KLmeasure}
    D_{\text{KL}}(a, b) = \sum_{K = 1}^{J}{\text{Pr}(K) \cdot \text{log} \Bigg\{ \frac{\text{Pr}(K)}{\text{Pr}(K|J, a, b)} \Bigg\}}.
\end{equation}

\noindent \cite{dorazio2009selecting} proposed specifying $a$ and $b$ to be the values for which $\text{Pr}(K|J, a, b)$ most closely matches the discrete uniform distribution to reflect the absence of explicit prior information. This method attempts to mimic a noninformative prior for $K$. 

%%% Using $\chi^2$ distribution to encode prior knowledge on $K$

\begin{figure}[t]
    \centering
    \includegraphics[width=\textwidth]{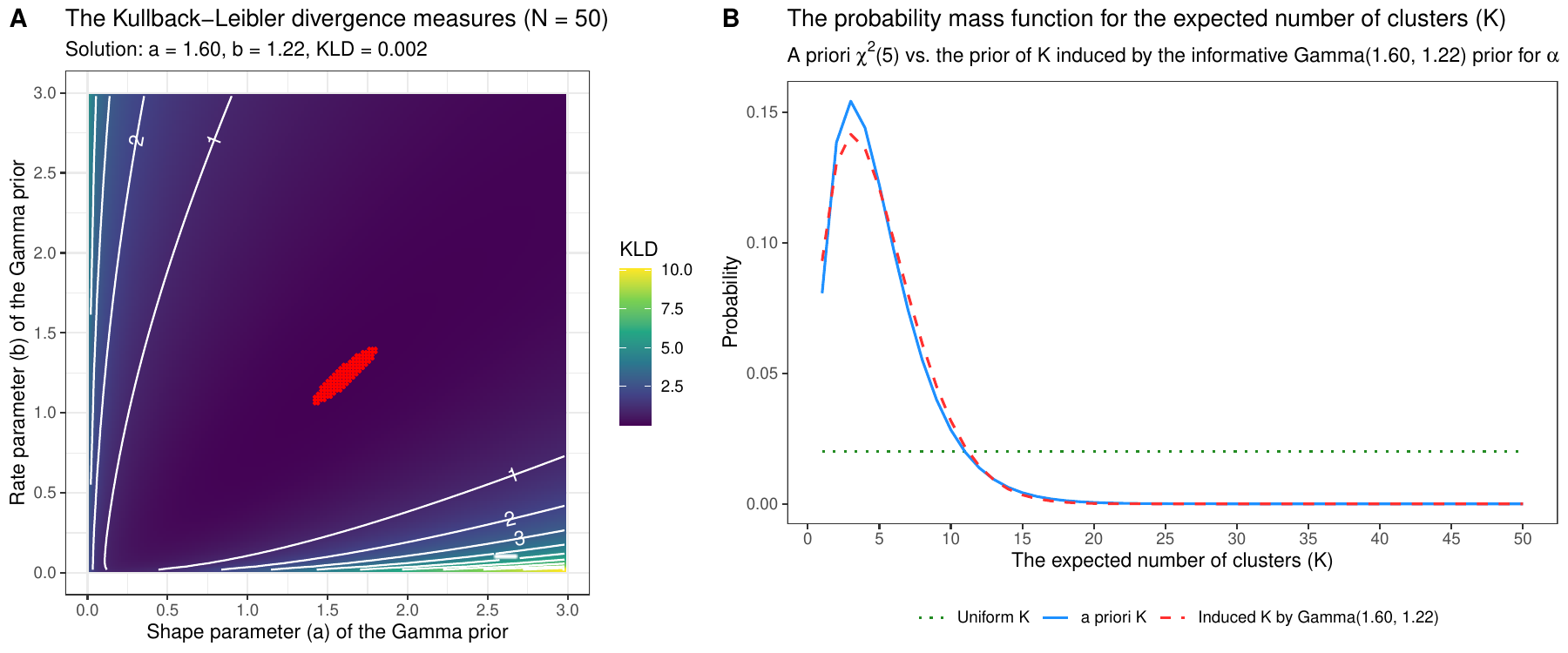}
    \caption{The derivation of the informative prior for the precision parameter $\alpha$ by approximating a $\text{Gamma}(a, b)$ to $\chi^2(5)$. \textit{Note}: The area highlighted by red dots corresponds to the region where the KL divergence measures are minimized.}
    \label{fig:figure_a01}
\end{figure}

Our proposal is to take $\text{Pr}(K) \sim \chi^2(\text{df} = u)$ to more intuitively encode our prior knowledge on the expected number of clusters $K$ and its uncertainty. The $\chi^2$ distribution has only one parameter: a positive integer $u$ that specifies the number of degrees of freedom. Our framework is mainly motivated by the feature of the $\chi^2$ distribution that its mean and variance are $u$ and $2u$, respectively. If it is expected that there are approximately five clusters ($K = 5$) and $J = 50$, then one can simply assume that $\text{Pr}(K)$ follows $\chi^2(5)$ distribution and specify a $\text{Gamma}(a, b)$ that closely matches $\chi^2(5)$ using equation~(\ref{eq:A-7_probK}) and (\ref{eq:A-8_KLmeasure}). Panel A of Figure~\ref{fig:figure_a01} shows the result of the numerical analysis based on a grid search algorithm designed to identify the global minimum of the KL divergence measure defined in equation~(\ref{eq:A-8_KLmeasure}). The Gamma distribution with $(a, b) = (1.60, 1.22)$ obtained as the solution that minimizes the KL closely matches the $\chi^2(5)$ distribution as shown in Panel B of Figure~\ref{fig:figure_a01}. This strategy is useful for constructing an informative prior for $K$, particularly when one wants to impose near-zero probabilities beyond a certain threshold ($K = 25$ in the example shown in Figure~\ref{fig:figure_a01}) and to be clear about the prior mean and variance of $K$. 

%%%%%%%%%%%%%%%%%%%%%%%%%%%%%%%%%%%%%%
%%% A.1. Constrained Bayes (CB) estimator
%%%

\subsection{Constrained Bayes (CB) estimator}\label{subsec:A-1}

The choice of rescaling for the CB estimator is derived as follows. \cite{shen1998triple} showed that the optimal EDF estimator that minimizes the ISEL in equation~(\ref{eq:8_ISEL}) is

\begin{equation}
\bar{G}_J=\mathrm{E}\left[G_J(t) \mid \hat{\boldsymbol{\tau}}\right]=\frac{1}{J} \cdot \sum \operatorname{Pr}\left(\tau_j \leq t \mid \hat{\tau}_j\right)
\end{equation}

\noindent where $\hat{\boldsymbol{\tau}}=(\hat{\tau}_1, ..., \hat{\tau}_J)$. Let the posterior mean of $\tau_j$, $\text{E}(\tau_j|\widehat{\tau}_j)$, be $\eta_j$ and the posterior variance of $\tau_j$, $\text{Var}(\tau_j|\widehat{\tau}_j)$, be $\lambda_j$. Then, the marginal mean of $\bar{G}_J$ and the finite population version of the marginal variance of $\bar{G}_J$ can be written as follows \citep{shen1998triple}:

\begin{equation}
\mathrm{E}\left[\bar{G}_J\right]=\int t d \bar{G}_J(t)=\frac{\sum \eta_j}{J}=\bar{\eta},
\end{equation}

\begin{equation}\label{eq:A-3.variance}
\widehat{\text{Var}}[\bar{G}_J]=\int t^2d\bar{G}_J(t)- \bar{\eta}^2=\frac{\sum\lambda_j}{J}+\frac{\sum(\eta_j-\bar{\eta})^2}{J-1}.
\end{equation}

The finite population variance of posterior means $\eta_j$ appears in the second term of equation~(\ref{eq:A-3.variance}). PMs tend to be under-dispersed because their variance lacks the first term in the estimated marginal variance of $\bar{G}_J$, $\sum\lambda_j/J$. The goal of the CB estimator is to adjust the posterior means to have a variance equal to the estimated marginal variance specified in equation~(\ref{eq:A-3.variance}). As a result, the CB estimate $a_j^{CB}$ is defined as follows \citep{ghosh1992constrained}:

\begin{equation}
    a_j^{CB} = \bar{\eta} + (\eta_j - \bar{\eta}) \cdot \sqrt{1 + \frac{J^{-1}\sum{\lambda_j}}{(J-1)^{-1}\sum{(\eta_j - \bar{\eta})^2}}}.
\end{equation}

%%%%%%%%%%%%%%%%%%%%%%%%%%%%%%%%%%%%%%
%%% A.2. Triple-Goal (GR) estimator
%%%

\subsection{Triple-goal (GR) estimator}\label{subsec:A-2}

The triple-goal estimator aims to obtain a single set of estimates that simultaneously targets all three inferential goals. In practice, however, the triple-goal estimator is designed to minimize the losses for two of the goals: estimating the EDF of $\tau_j$ values, $G_J$, and estimating the rank of $\tau_j$, $R_j$. The abbreviation ``GR" reflects the two direct inferential targets and is often used to denote the triple-goal estimator in the literature \citep[e.g.,][]{paddock2006flexible}.

\cite{shen1998triple} show that $\bar{G}_J(t)$ minimizes the expected ISEL, which is the loss function that targets the EDF of $\tau_j$ values. The GR estimator therefore uses the sample estimate of $\bar{G}_J(t)$ to minimize ISEL. To parallel the EDF estimates made using PMs or the CB estimator, the sample estimate of $\bar{G}_J(t)$ is then discretized to contain only $J$ mass points.

Similarly, some basic calculus shows us that the posterior expected ranks $\bar{R}_j = \sum_{k = 1}^{J} \text{Pr}(\tau_j \leq \tau_k | \widehat{\boldsymbol{\tau}})$ minimize MSELR. The GR estimator therefore ranks sites using the sample estimated ranks to minimize MSELR. Again, to parallel the EDF estimates made using PMs or the CB estimator, the sample estimated ranks are discretized to $J$ discrete rank values. 

As a result, the GR estimator intuitively minimizes both ISEL and MSELR, since it makes use of optimal estimators for both metrics. The GR estimator pays no explicit attention to reducing the MSEL of the individual site-specific parameters $\tau_j$. However, \cite{shen1998triple} argued that the GR estimator tends to produce small MSEL for the individual $\tau_j$’s because assigning the $\widehat{U}$ values by a permutation vector $\mathbf{z}$ to minimize $\sum{(\widehat{U}_{z_j}-\eta_j)^2}$ results in the same assignments as assigning them to minimize $\sum{(a_j - \eta_j)^2}$. Regardless, the focus of the GR estimator is on estimating the EDF and ranks of $\tau_j$’s.

%%%%%%%%%%%%%%%%%%%%%%%%%%%%%%%%%%%%%%
%%% Appendix C. Detailed simulation results
%%%%%%%%%%%%%%%%%%%%%%%%%%%%%%%%%%%%%%

\section{Detailed simulation results}\label{section:C}

\begin{figure}[H]
    \centering
    \includegraphics[width=\textwidth]{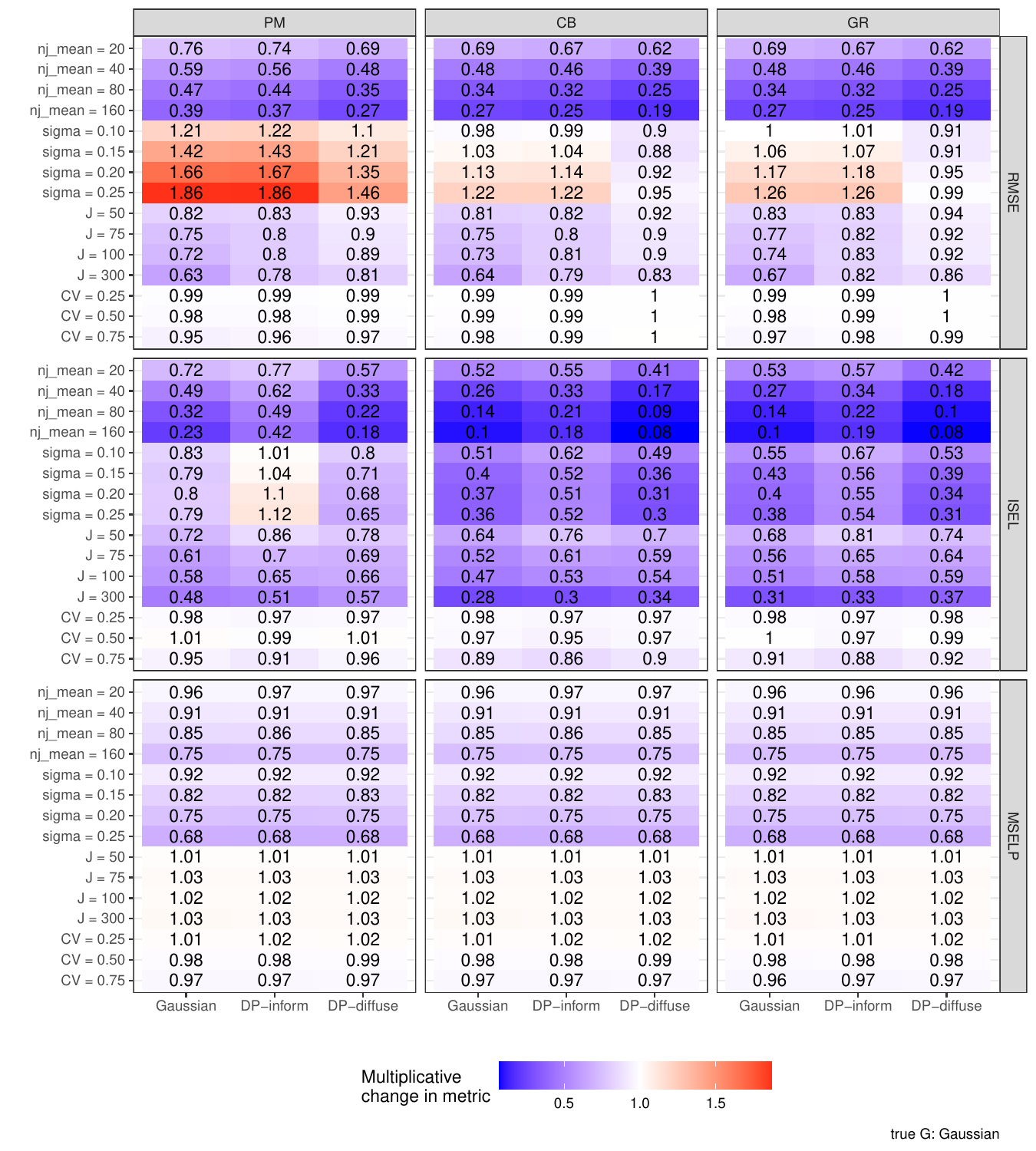}
    \caption{The meta-model predicted multiplicative change in average RMSE, ISEL, and MSELP for a single data-generating factor change from the base condition ($J=25$, $\bar{n}_j=10$, $\sigma=0.05$, and $\text{CV}=0.00$), when the true $G$ is Gaussian. In terms of performance metrics such as RMSE and MSELP, the Gaussian and Gaussian Mixture models yielded analogous patterns. The only discrepancy was observed in the ISEL metric, which was unique to the Gaussian model. The Gaussian model demonstrated superior efficacy in recovering the EDF of $\tau_j$'s, specifically when the true $G$ conformed to a Gaussian distribution.}
    \label{fig:figure_c00_1}
\end{figure}

\clearpage

\begin{figure}[H]
    \centering
    \includegraphics[width=\textwidth]{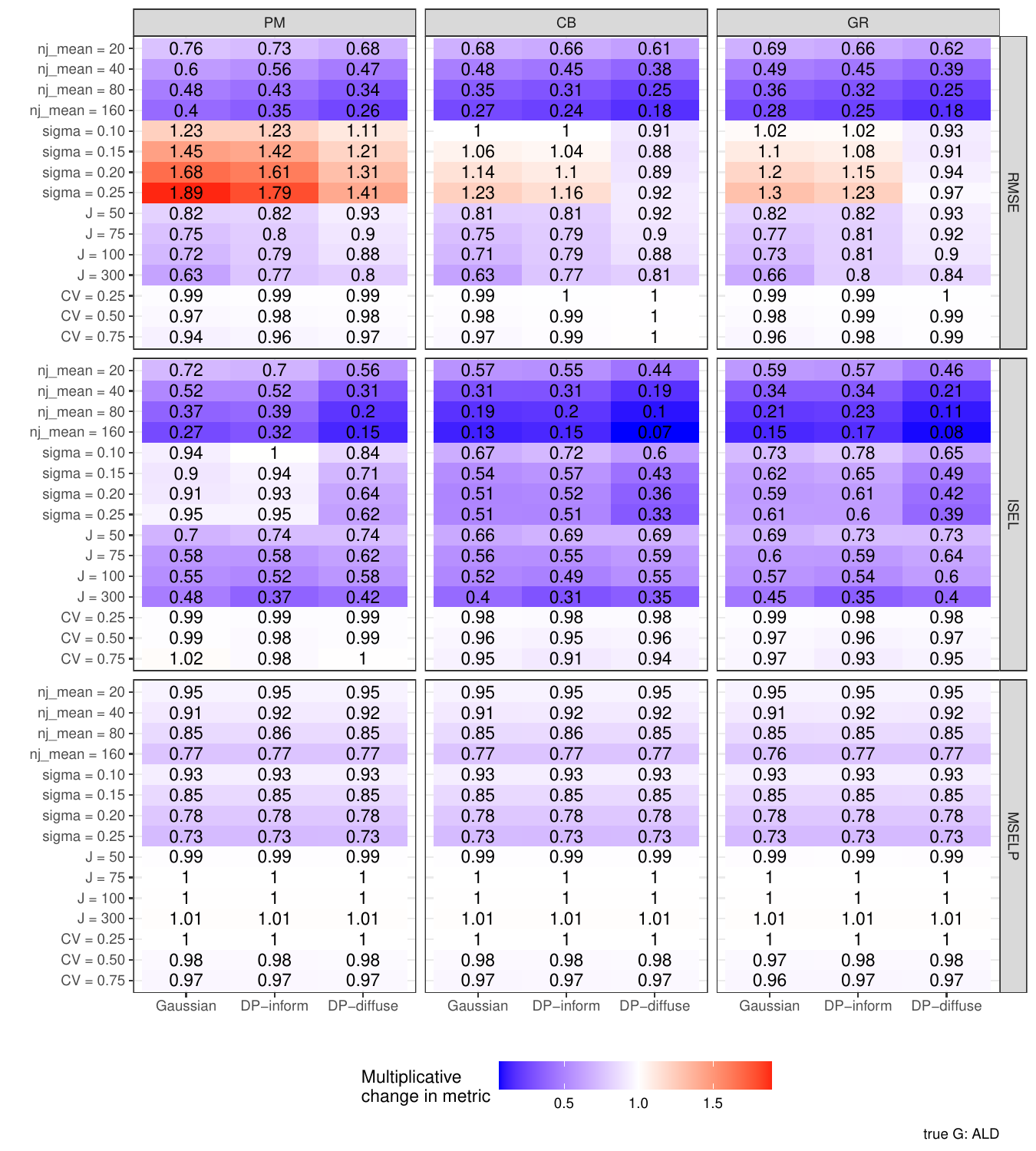}
    \caption{The meta-model predicted multiplicative change in average RMSE, ISEL, and MSELP for a single data-generating factor change from the base condition ($J=25$, $\bar{n}_j=10$, $\sigma=0.05$, and $\text{CV}=0.00$), when the true $G$ is Asymmetric Laplace (AL) distribution with $\rho= 0.1$. No distinguishable patterns emerged in the results when comparing the Gaussian Mixture and AL data-generating distributions.}
    \label{fig:figure_c00_2}
\end{figure}

\clearpage

\begin{figure}[ht]
    \centering
    \includegraphics[width=\textwidth]{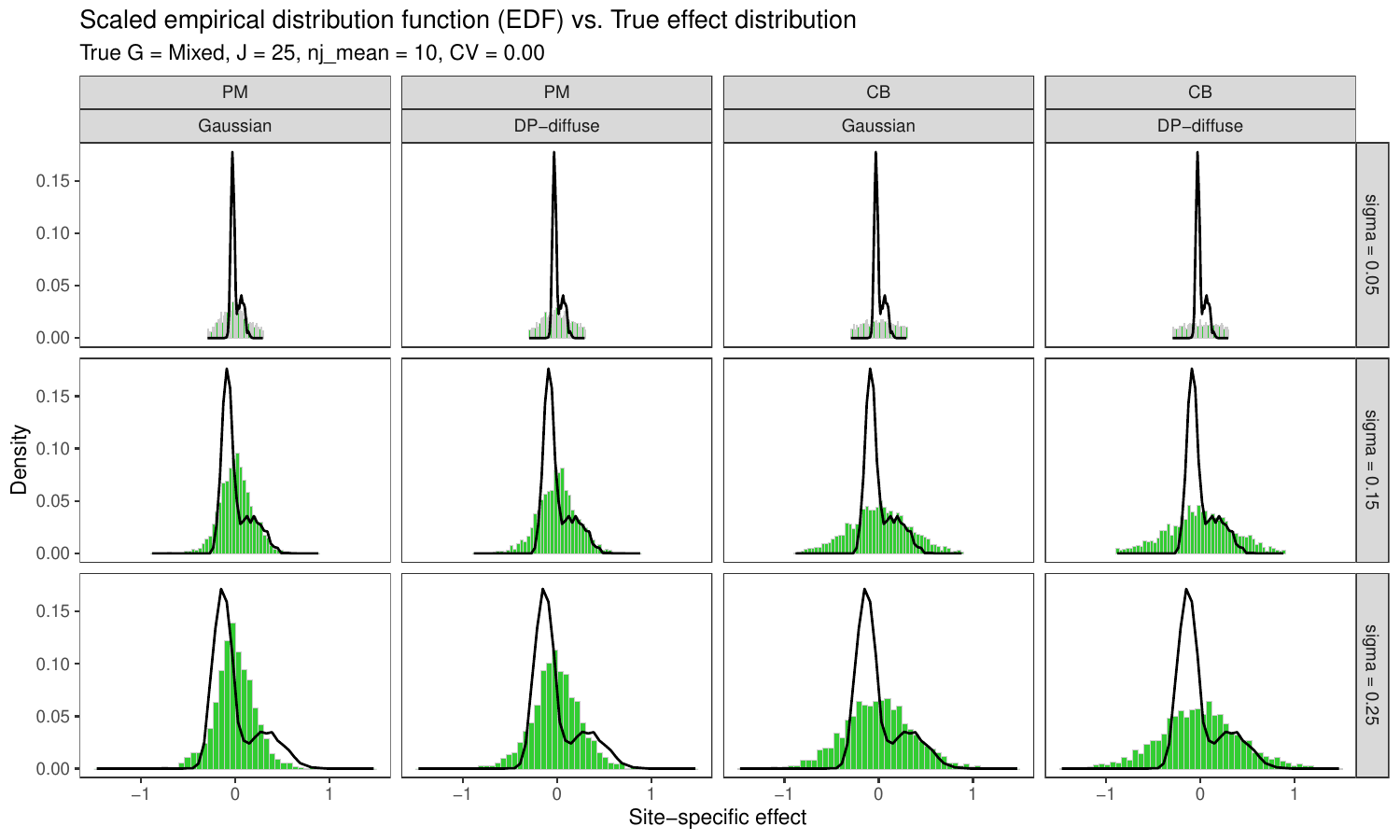}
    \caption{Changes in the empirical distribution function (EDF) of the estimated site-specific effects ($\hat{\tau}_j$’s), as the cross-site impact standard deviation $\sigma$ increases from the base condition ($J = 25$, $\bar{n}_j=10$, and $\text{CV}=0.00$), when true $G$ is Gaussian mixture. With a fixed average site size and number of sites, an increase in $\sigma$ adversely affects RMSE as it reduces the effect of shrinkage towards the prior mean.}
    \label{fig:figure_c01}
\end{figure}

\clearpage

\begin{figure}[ht]
    \centering
    \includegraphics[width=\textwidth]{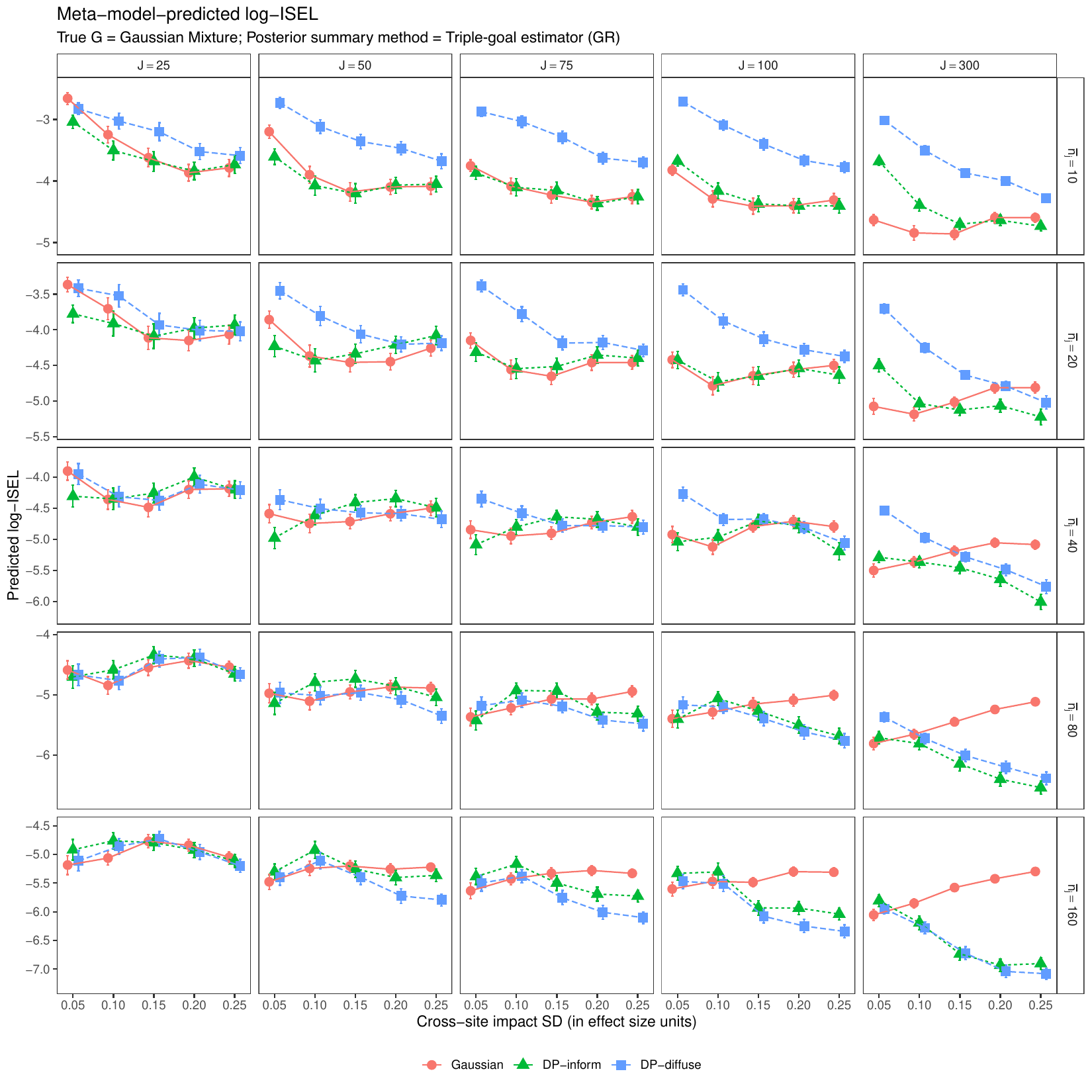}
    \caption{The meta-model predicted integrated squared error loss (ISEL) on a natural log scale, when the true $G$ is a Gaussian mixture and the triple-goal estimator (GR) is used as the posterior summary method. Plots include 95\% prediction intervals around predicted values.}
    \label{fig:figure_c02}
\end{figure}

\clearpage

\begin{figure}[ht]
    \centering
    \includegraphics[width=\textwidth]{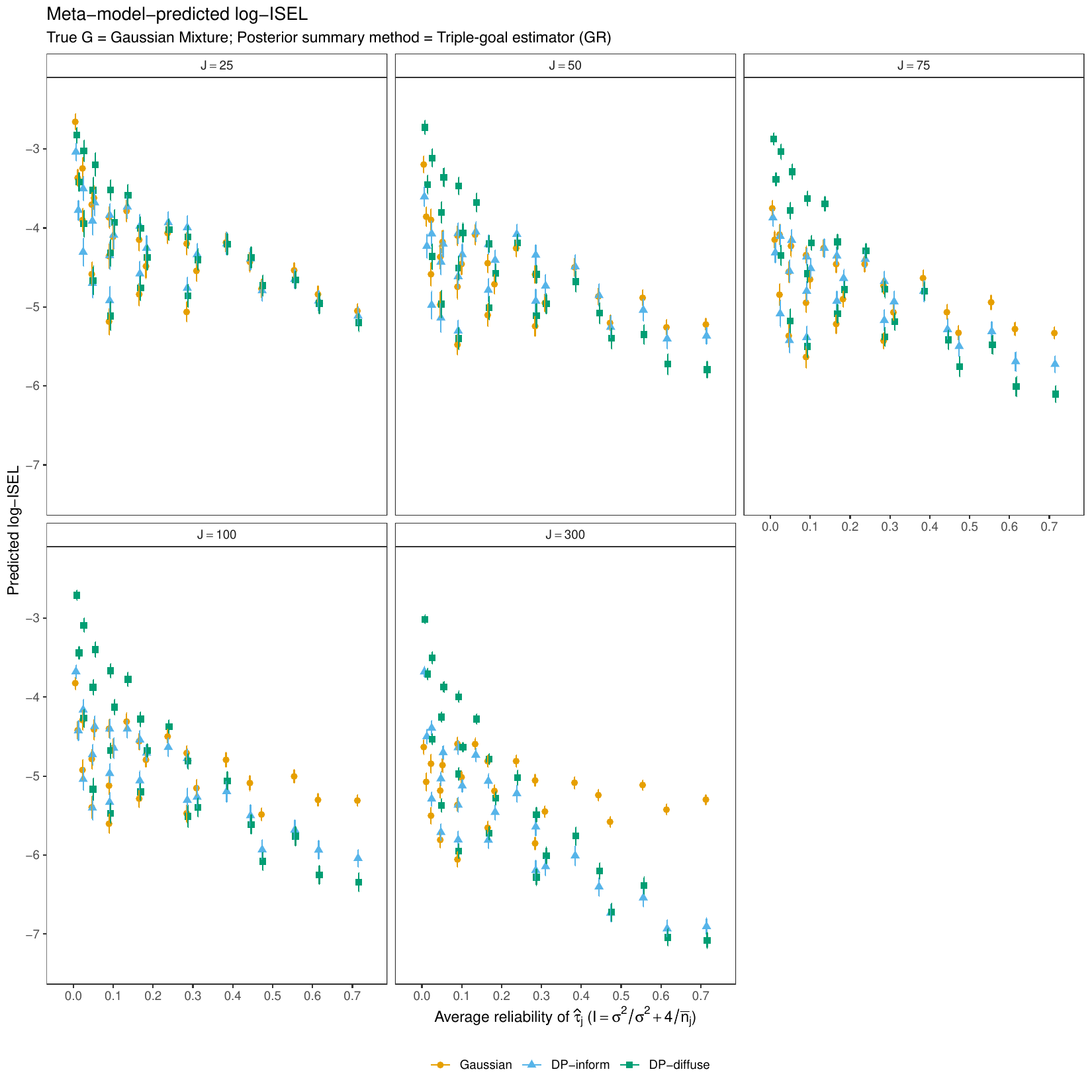}
    \caption{The meta-model predicted integrated squared error loss (ISEL) on a natural log scale, when the true $G$ is a Gaussian mixture and the triple-goal estimator (GR) is used as the posterior summary method. Plots include 95\% prediction intervals around predicted values. This figure outlines the $I$ ranges within which the simulations recommend selecting DP models for each of the 25 combinations of $\sigma$ and $\bar{n}_j$ levels, with the aim of accurately recovering the shape of site-specific effect distributions.}
    \label{fig:figure_c02b}
\end{figure}

\clearpage

\begin{figure}[ht]
    \centering
    \includegraphics[width=\textwidth]{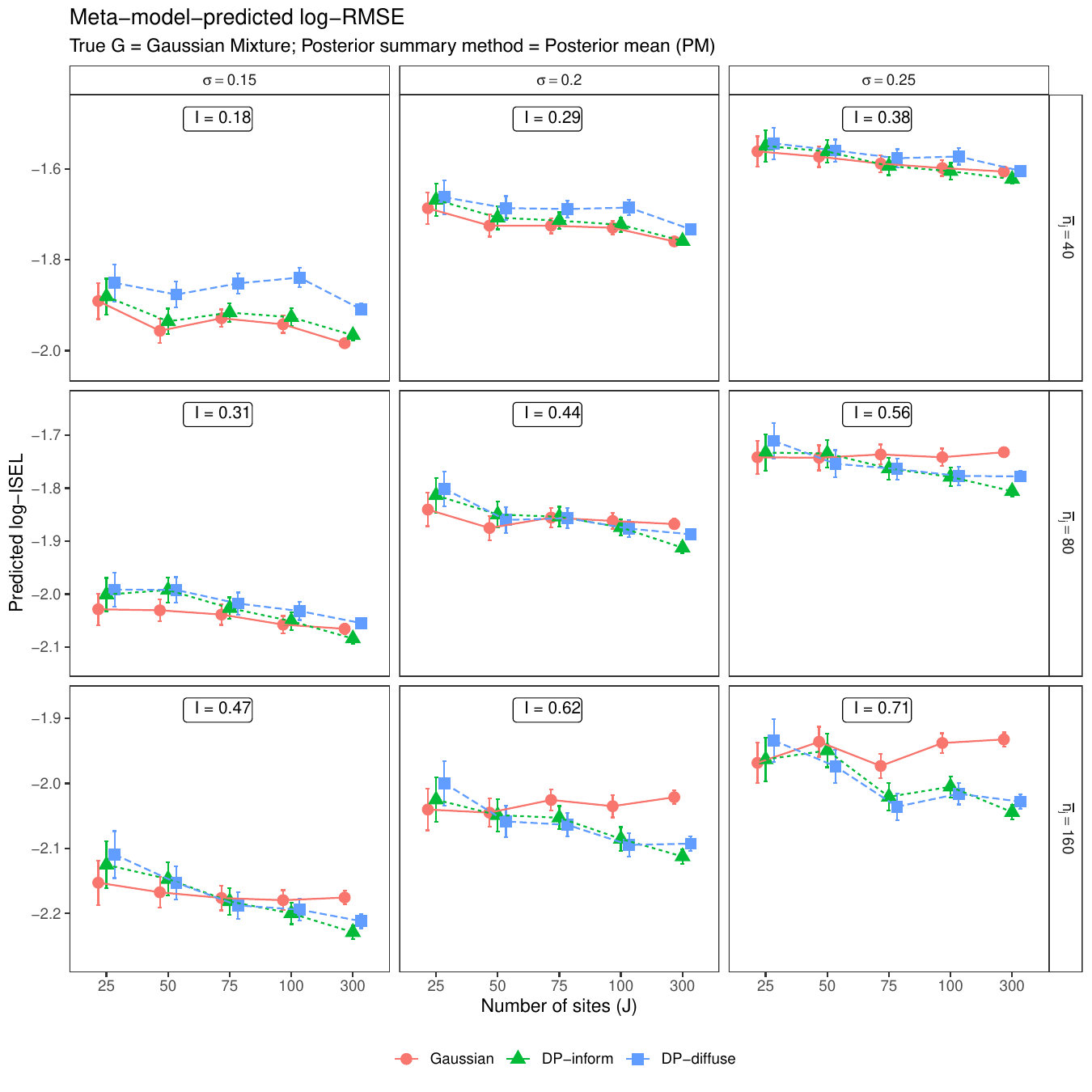}
    \caption{The meta-model predicted root mean squared error loss (RMSE) on a natural log scale, when the true $G$ is a Gaussian mixture and the posterior mean (PM) is used as the posterior summary method. Plots include 95\% prediction intervals around predicted values.}
    \label{fig:figure_c03}
\end{figure}

\clearpage

\begin{figure}[ht]
    \centering
    \includegraphics[width=\textwidth]{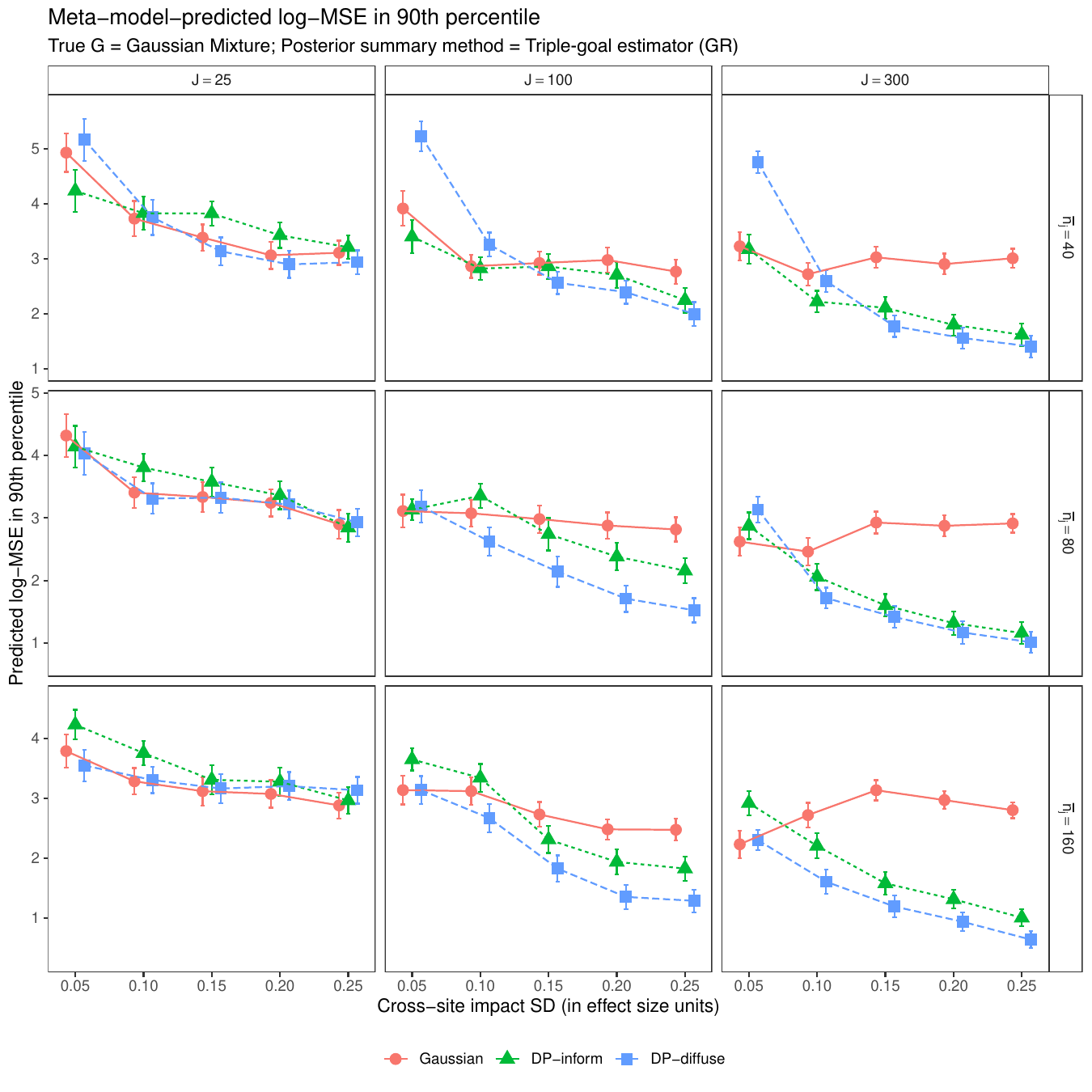}
    \caption{The meta-model predicted mean squared error (MSE) in 90th percentile on a natural log scale, when the true $G$ is a Gaussian mixture and the triple-goal estimator (GR) is used as the posterior summary method. Plots include 95\% prediction intervals around predicted values. In scenarios characterized by a considerable amount of between-site information, the DP models' adaptive clustering facilitates the recovery of site-effect distributions with significantly higher accuracy compared to the Gaussian model. The performance of DP models for tail area percentile estimation (specifically the 90th percentile) follow patterns similar to those found for ISEL. Especially under conditions of high between-site information, the DP-diffuse model were especially effective in tail percentile estimation.}
    \label{fig:figure_c06b}
\end{figure}

\clearpage

\begin{figure}[ht]
    \centering
    \includegraphics[width=\textwidth]{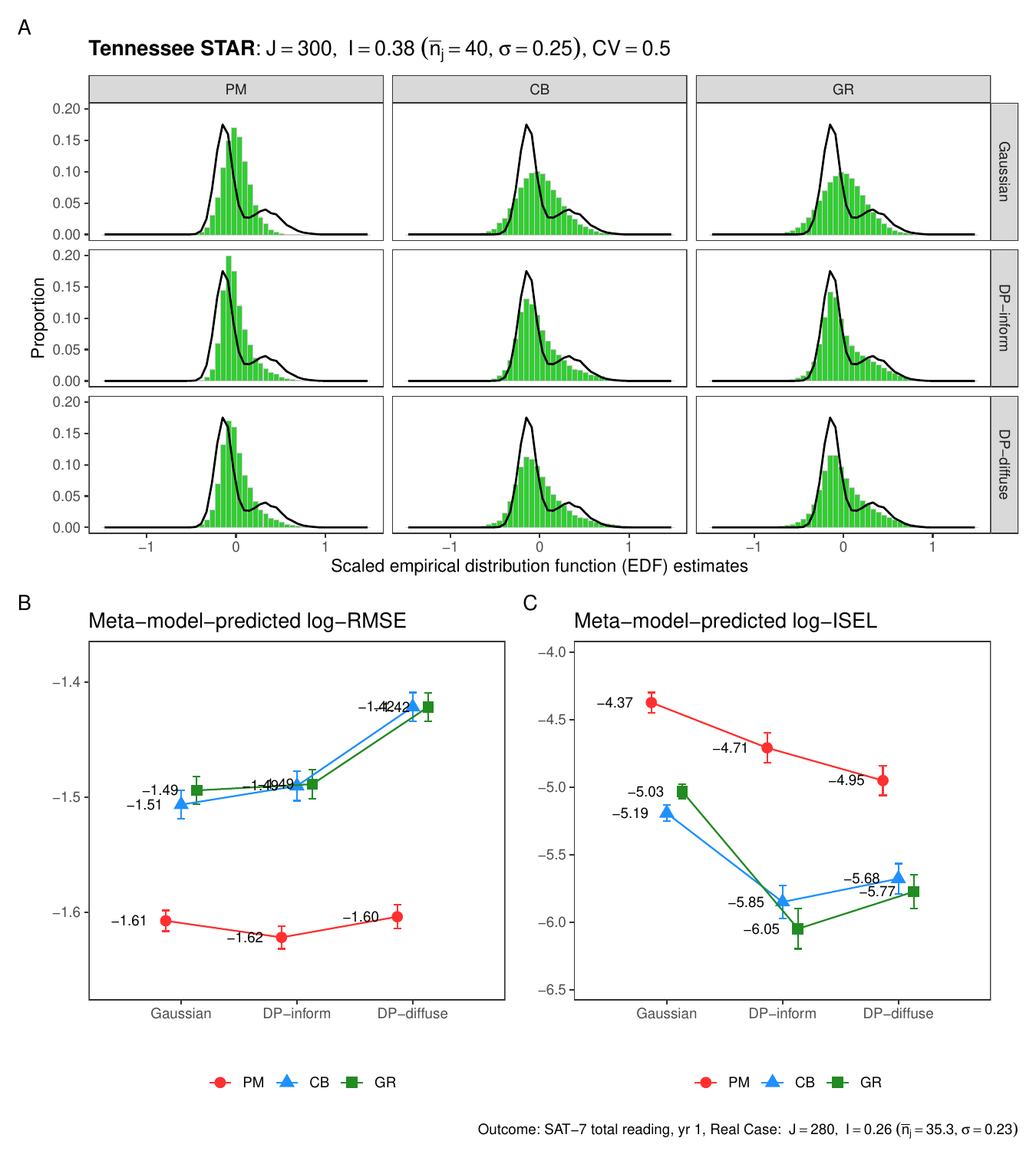}
    \caption{Scaled EDF estimates and meta-model-predicted RMSE and ISEL on log-scale for each model-method combination, when the data-generation is based on the scenario simulating the Tennessee STAR data ($J = 280$, $\bar{n}_j=35.9$, and $\sigma=0.26$ for the outcome, SAT-7 Total Math at the end of a student’s ﬁrst year in the study). True $G$ is Gaussian mixture, while $\text{CV}$ is fixed at 0.5. Plots include 95\% prediction intervals around predicted values.}
    \label{fig:figure_c04}
\end{figure}

\clearpage

\begin{figure}[ht]
    \centering
    \includegraphics[width=\textwidth]{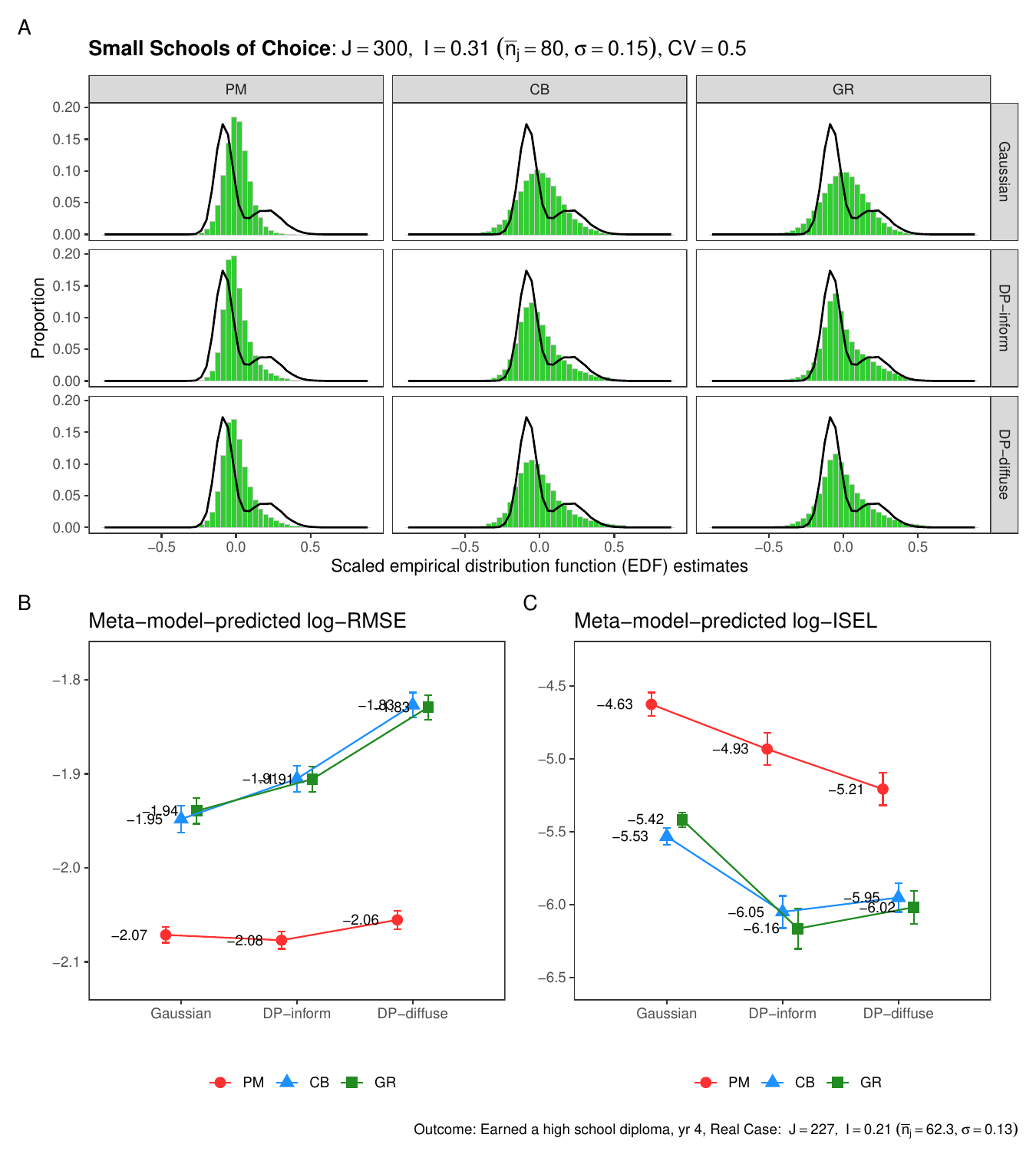}
    \caption{Scaled EDF estimates and meta-model-predicted RMSE and ISEL on log-scale for each model-method combination, when the data-generation is based on the scenario simulating the Small Schools of Choice data ($J = 356$, $\bar{n}_j=72.8$, and $\sigma=0.16$ for the outcome, Ninth-grade ``on-track" indicator). True $G$ is Gaussian mixture, while $\text{CV}$ is fixed at 0.5. Plots include 95\% prediction intervals around predicted values.}
    \label{fig:figure_c05}
\end{figure}

\clearpage

\begin{figure}[ht]
    \centering
    \includegraphics[width=\textwidth]{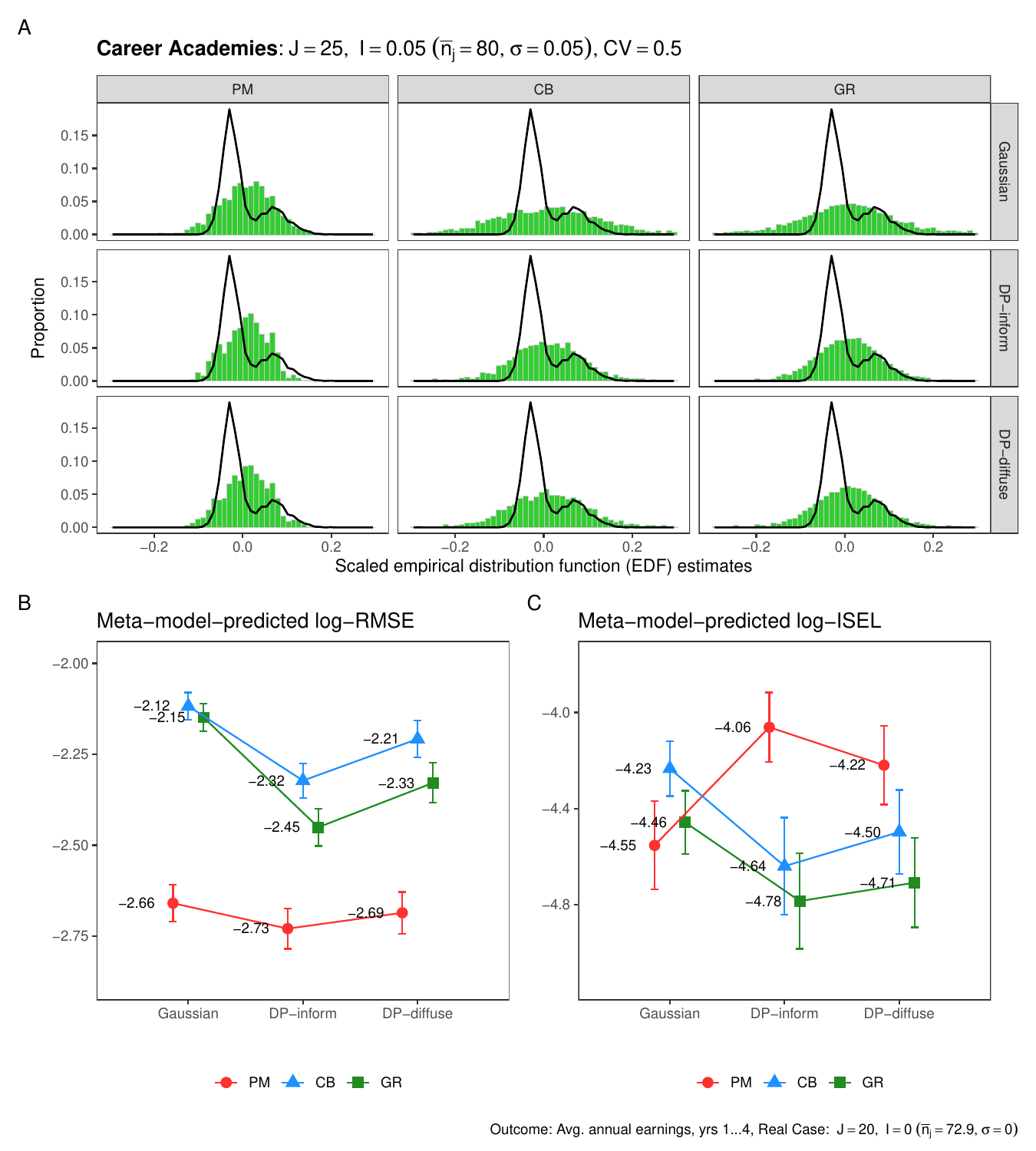}
    \caption{Scaled EDF estimates and meta-model-predicted RMSE and ISEL on log-scale for each model-method combination, when the data-generation is based on the scenario simulating the Career Academies study ($J = 20$, $\bar{n}_j=74.1$, and $\sigma=0.07$ for the outcome, enrolled in postsecondary measured 14 months after expected high school graduation). True $G$ is Gaussian mixture, while $\text{CV}$ is fixed at 0.5. Plots include 95\% prediction intervals around predicted values.}
    \label{fig:figure_c06}
\end{figure}

\clearpage

\begin{figure}[ht]
    \centering
    \includegraphics[width=\textwidth]{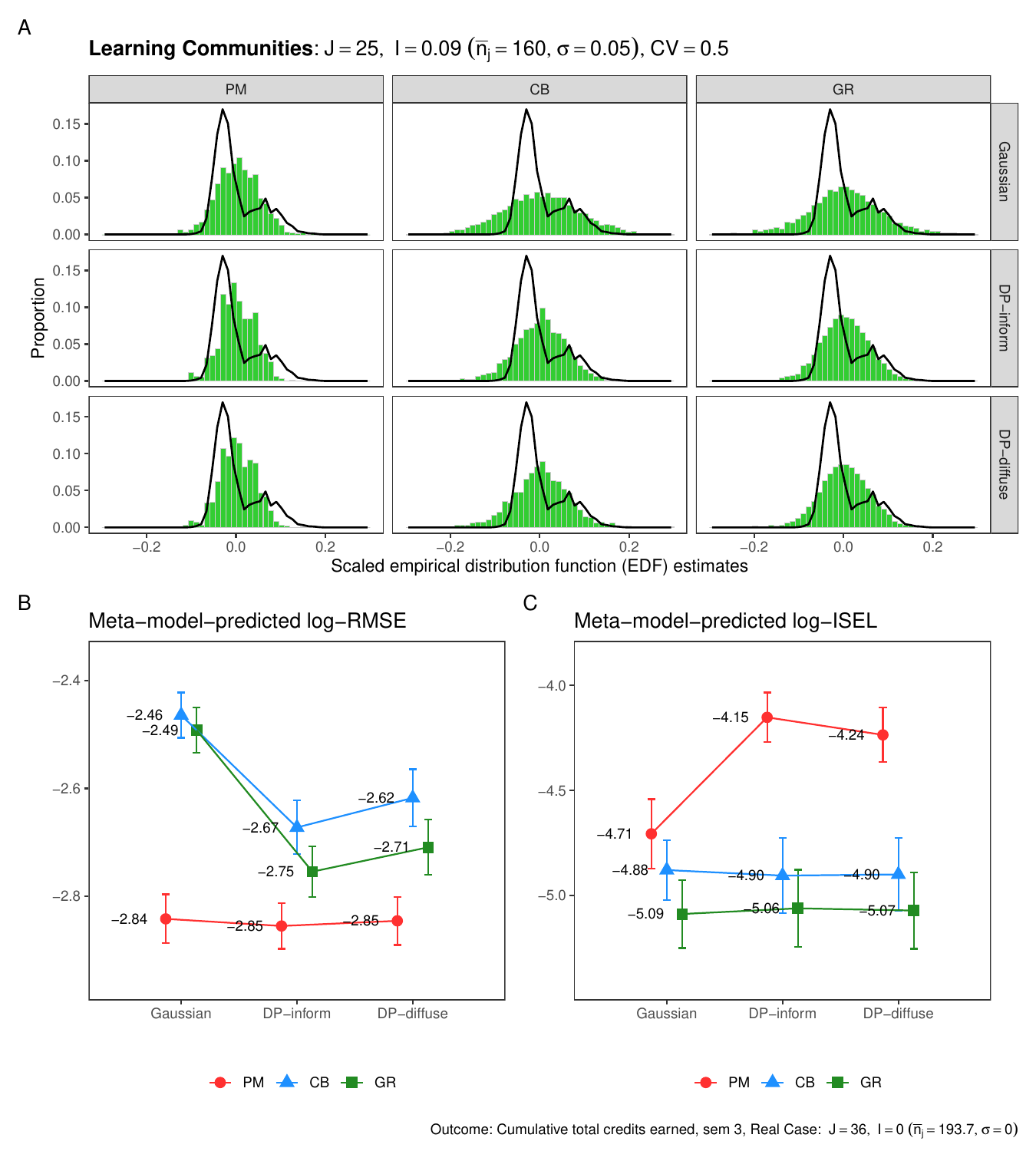}
    \caption{Scaled EDF estimates and meta-model-predicted RMSE and ISEL on log-scale for each model-method combination, when the data-generation is based on the scenario simulating the Learning Communities study ($J = 36$, $\bar{n}_j=193.7$, and $\sigma=0.04$ for the outcome, credit accumulation at the end-of-the-program semester). True $G$ is Gaussian mixture, while $\text{CV}$ is fixed at 0.5. Plots include 95\% prediction intervals around predicted values.}
    \label{fig:figure_c07}
\end{figure}

\clearpage

\begin{figure}[ht]
    \centering
    \includegraphics[width=\textwidth]{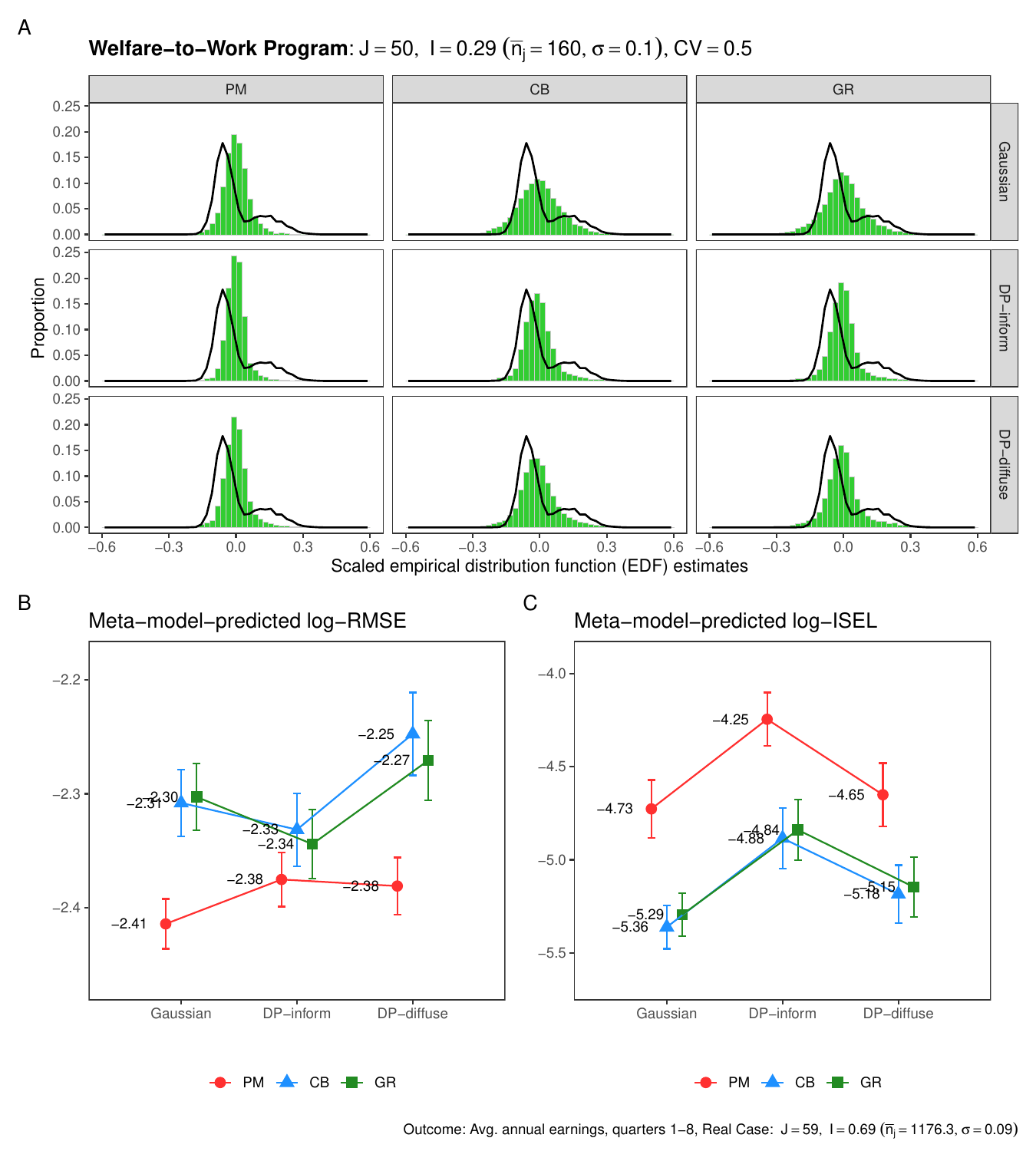}
    \caption{Scaled EDF estimates and meta-model-predicted RMSE and ISEL on log-scale for each model-method combination, when the data-generation is based on the scenario simulating the Welfare-to-Work Program study ($J = 59$, $\bar{n}_j=1176.3$, and $\sigma=0.09$ for the outcome, average annual earnings over two years). True $G$ is Gaussian mixture, while $\text{CV}$ is fixed at 0.5. Plots include 95\% prediction intervals around predicted values.}
    \label{fig:figure_c08}
\end{figure}

\clearpage

\begin{figure}[ht]
    \centering
    \includegraphics[width=\textwidth]{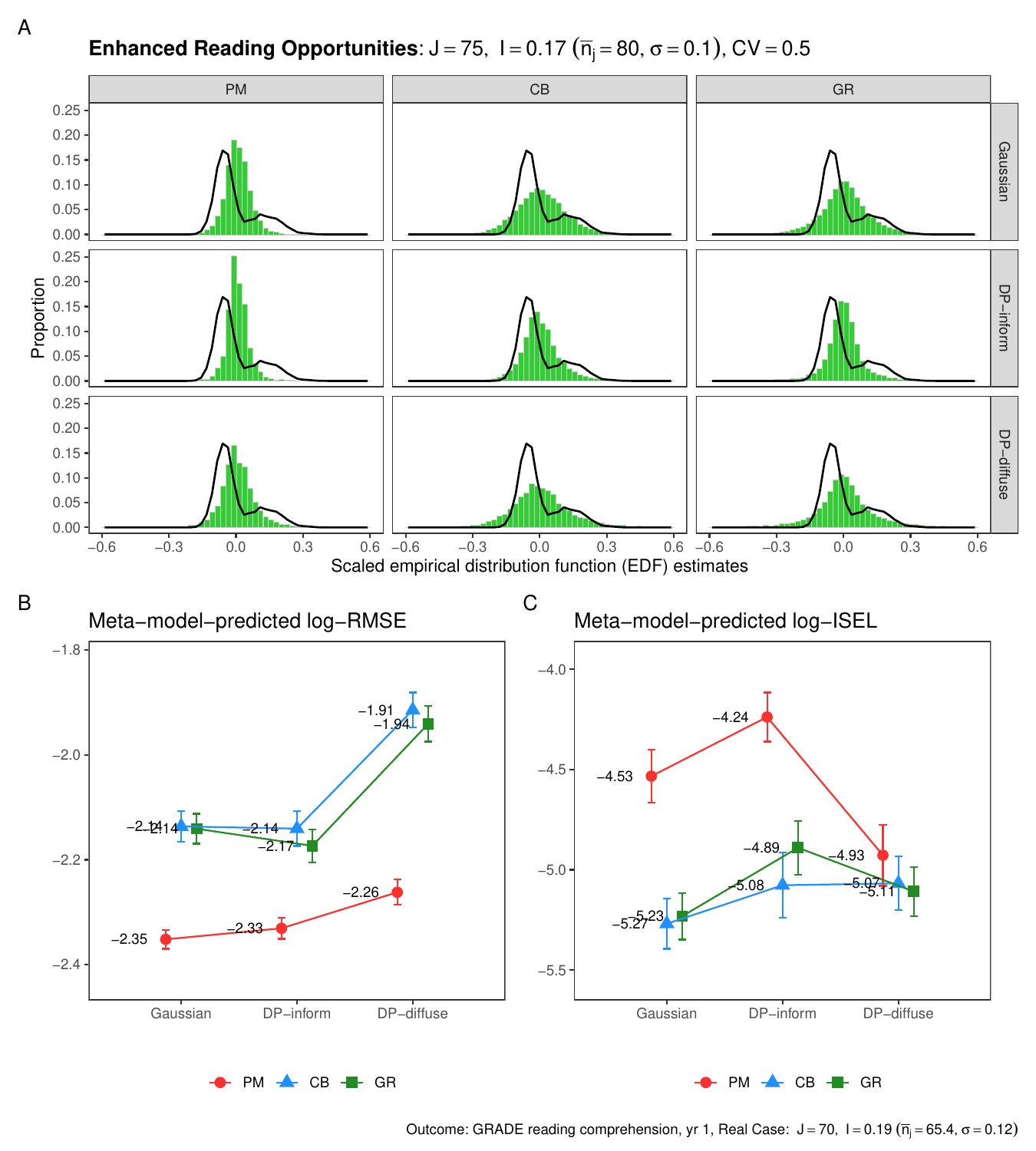}
    \caption{Scaled EDF estimates and meta-model-predicted RMSE and ISEL on log-scale for each model-method combination, when the data-generation is based on the scenario simulating the Enhanced Reading Opportunities study ($J = 70$, $\bar{n}_j=65.4$, and $\sigma=0.12$ for the outcome, Reading comprehension and reading vocabulary GRADE assessment). True $G$ is Gaussian mixture, while $\text{CV}$ is fixed at 0.5. Plots include 95\% prediction intervals around predicted values.}
    \label{fig:figure_c09}
\end{figure}

\clearpage

\begin{figure}[ht]
    \centering
    \includegraphics[width=\textwidth]{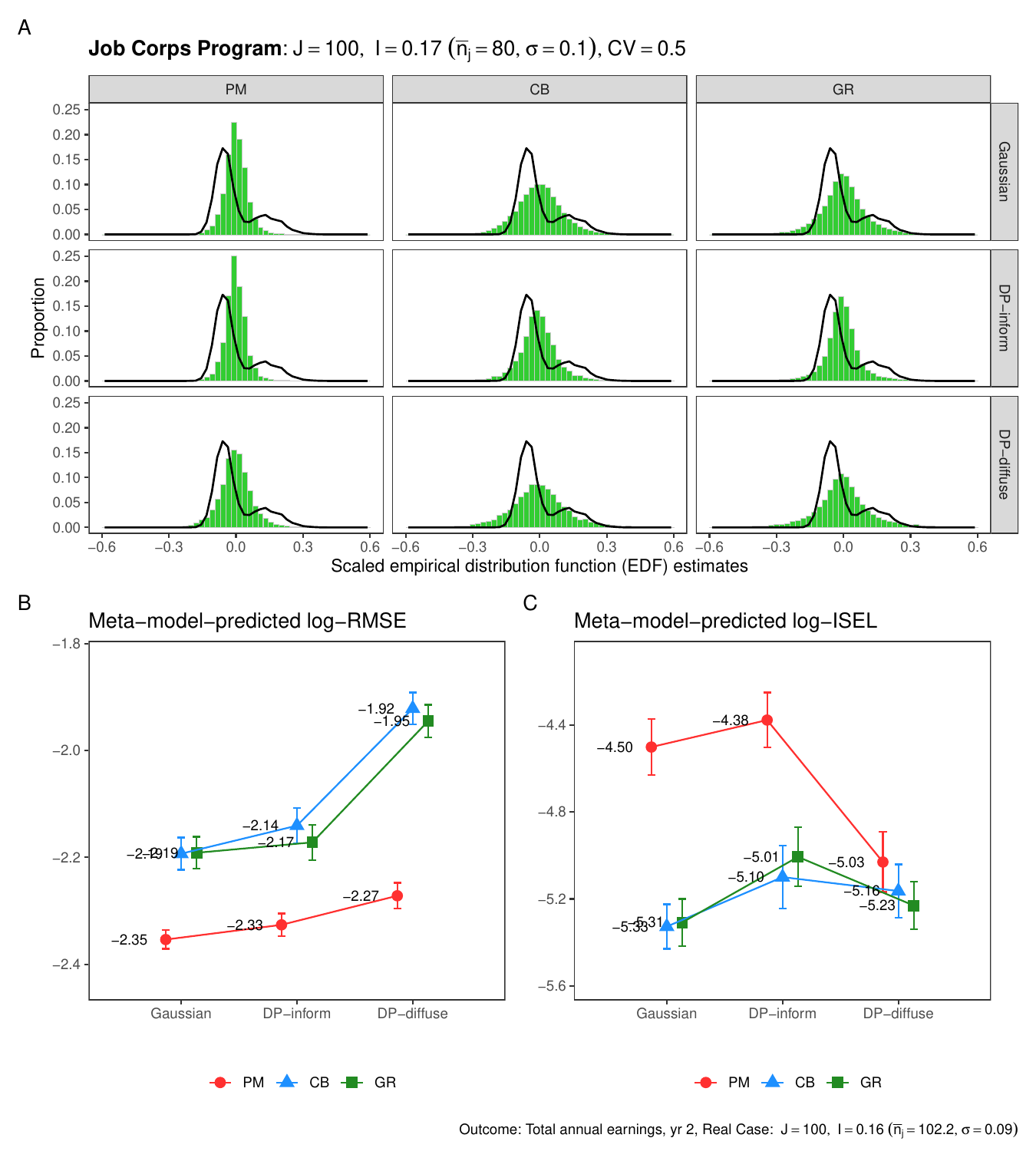}
    \caption{Scaled EDF estimates and meta-model-predicted RMSE and ISEL on log-scale for each model-method combination, when the data-generation is based on the scenario simulating the Job Corps Program study ($J = 100$, $\bar{n}_j=102.2$, and $\sigma=0.09$ for the outcome, Annual earnings outcomes at the second year after random assignment). True $G$ is Gaussian mixture, while $\text{CV}$ is fixed at 0.5. Plots include 95\% prediction intervals around predicted values.}
    \label{fig:figure_c10}
\end{figure}

\clearpage

%%%%%%%%%%%%%%%%%%%%%%%%%%%%%%%%%%%%%%
%%% Appendix D. Detailed analysis results of real-world data
%%%%%%%%%%%%%%%%%%%%%%%%%%%%%%%%%%%%%%

\clearpage

\section{Detailed analysis results of real-world data}\label{section:D}

\begin{figure}[ht]
    \centering
    \includegraphics[width=\textwidth]{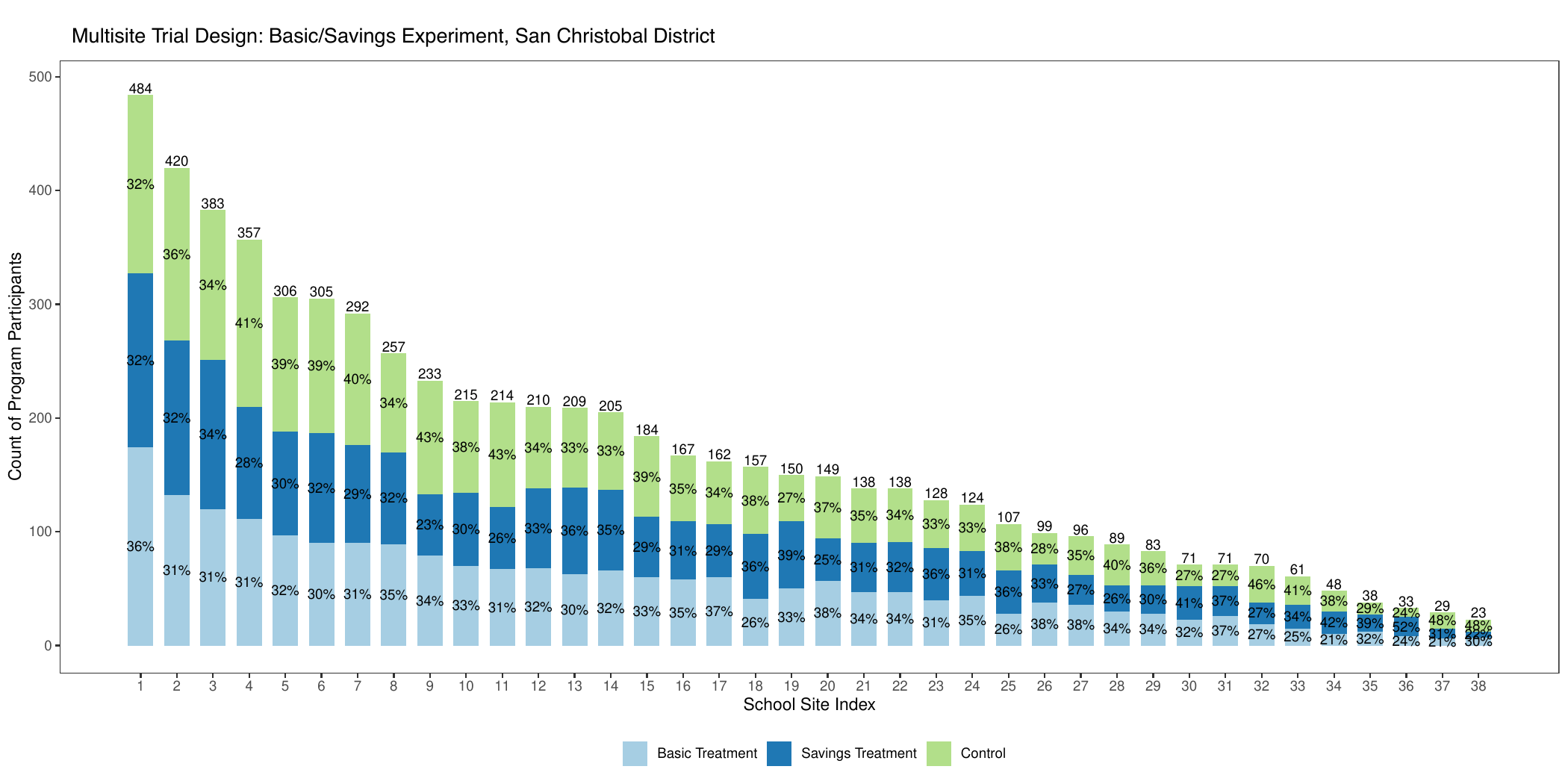}
    \caption{Multisite trial design of Basic/Savings conditional cash transfer program in San Cristobal district}
    \label{fig:figure_d01}
\end{figure}

\clearpage

\begin{figure}[ht]
    \centering
    \includegraphics[width=\textwidth]{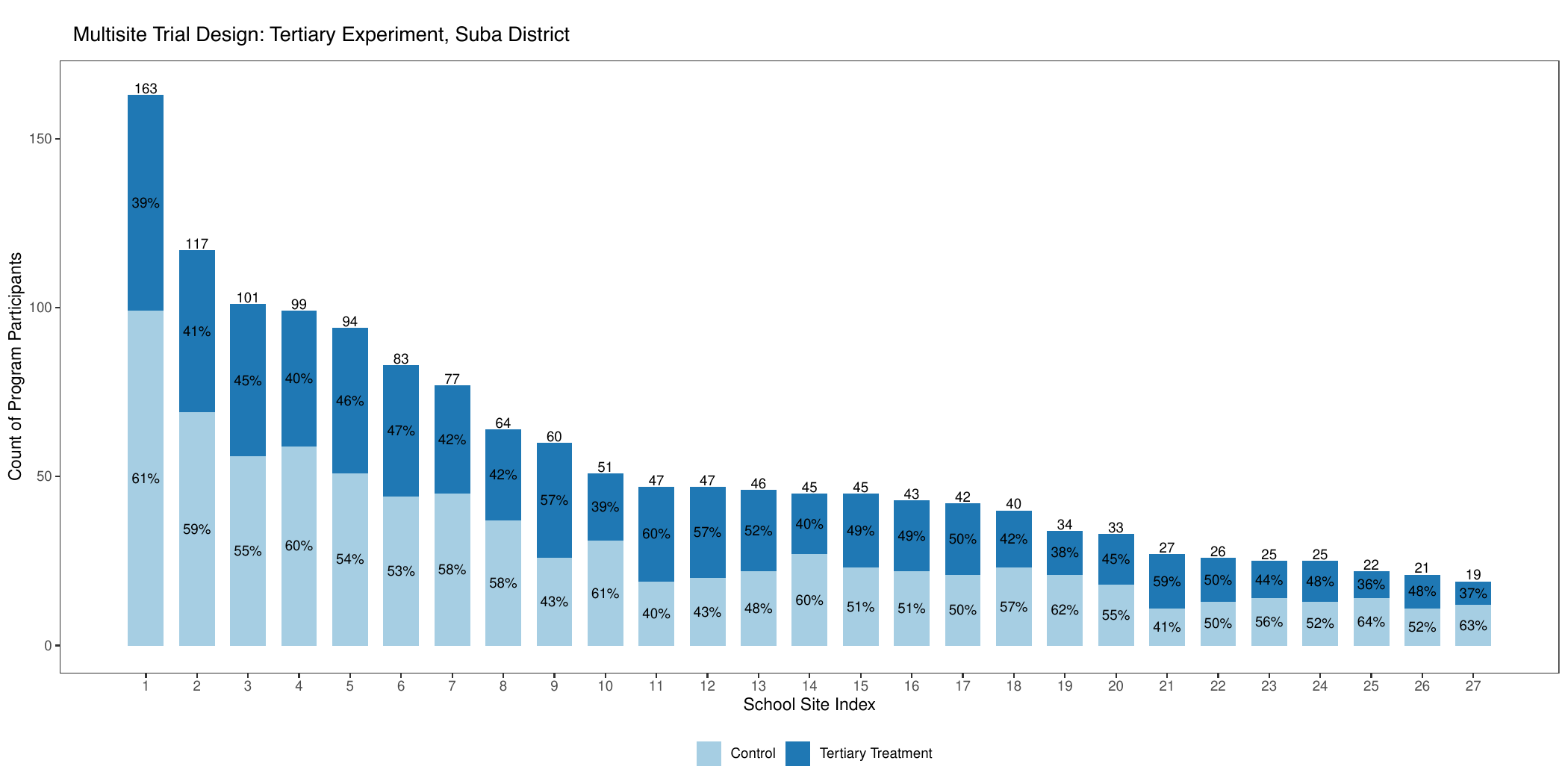}
    \caption{Multisite trial design of Tertiary conditional cash transfer program in Suba district}
    \label{fig:figure_d02}
\end{figure}

\clearpage

\begin{table}[ht]\label{table2}
\centering
\begin{threeparttable} % Wrap the tabular environment with threeparttable
\caption{Multisite trial design characteristics of the Conditional Subsidies for School Attendance study and estimates of the mean and standard deviation of the distribution of site-specific treatment effects in effect size units (Suba district).}
\begin{tabular}{ccc}
\toprule
 & \multicolumn{2}{c}{Tertiary experiment} \\
 & Tertiary enrollment,              & Tertiary enrollment,   \\
 & any time                     & University  \\ \midrule
Average treatment effect $\left(\hat{\tau}_d\right)$ & 0.08 & -0.06 \\
Cross-site effect SD $\left(\hat{\sigma}_d\right)$ & 0.08 & 0.10 \\
Geometric mean of $\widehat{s e}_j^{\prime} s$ & 0.35 & 0.45 \\
Average reliability of $\widehat{\tau}_j(I)$ & 0.05 & 0.05 \\
Total sample size $(N)$ & \multicolumn{2}{c}{1,496} \\
Number of $\operatorname{sites}(J)$ & \multicolumn{2}{c}{27} \\
Average site size $\left(\bar{n}_j\right)$ & \multicolumn{2}{c}{55.4} \\
Coefficient of variation of site sizes (CV) & \multicolumn{2}{c}{0.63} \\
Range of site sizes & \multicolumn{2}{c}{$(19,163)$} \\
Average proportion of units treated $\left(\bar{p}_j\right)$ & \multicolumn{2}{c}{0.46} \\
\bottomrule
\end{tabular}
\begin{tablenotes} % Add notes to the table
\item\textit{Note}: We excluded sites with extreme probabilities, that is, we required both $n_j p_j$ and $n_j (1 - p_j)$ to be at least 8, where $n_j$ represents the site size and $p_j$ denotes the proportion of participants treated in site $j$.
\end{tablenotes}
\end{threeparttable}
\end{table}

\clearpage

\clearpage

\begin{figure}[ht]
    \centering
    \includegraphics[width=\textwidth]{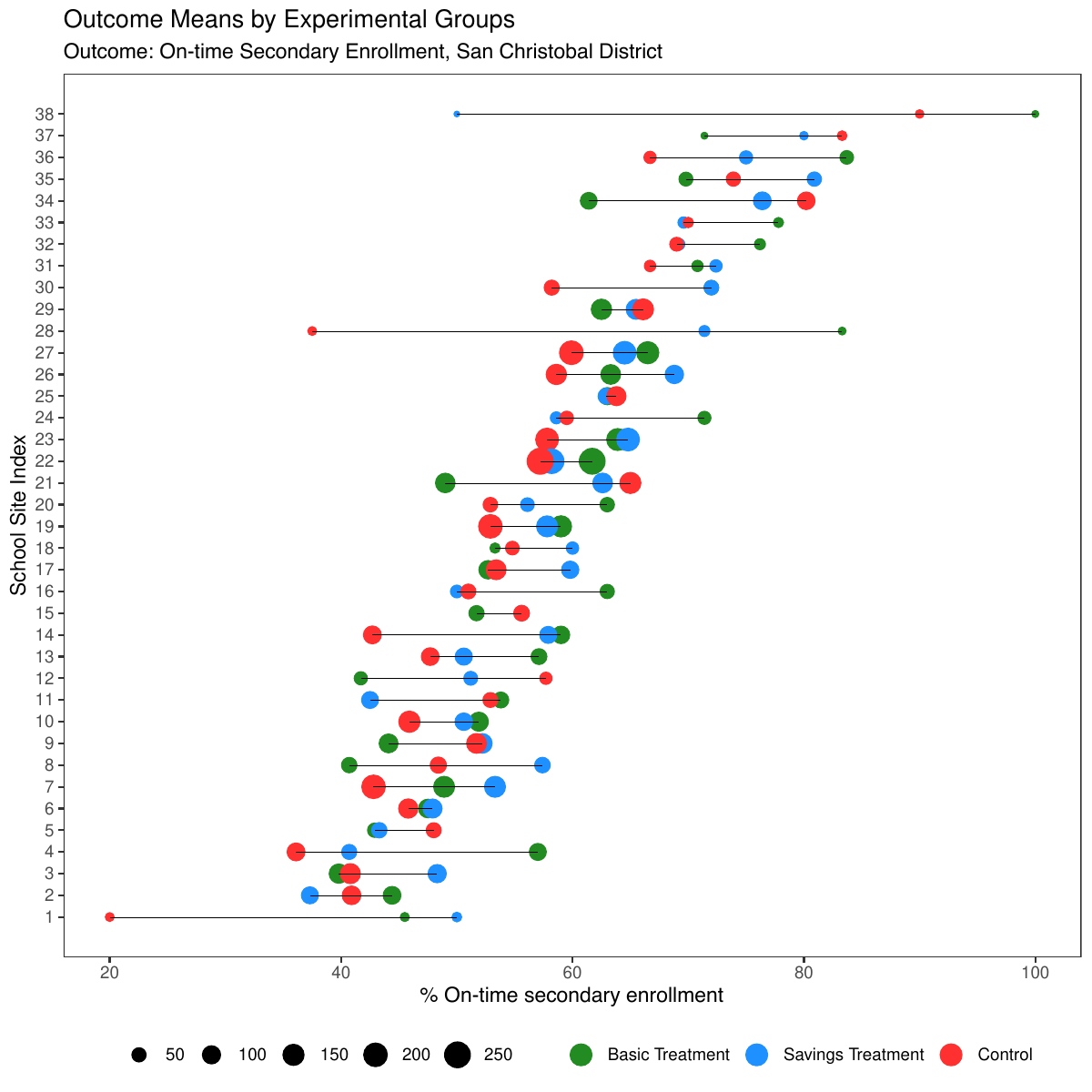}
    \caption{On-time secondary enrollment outcome mean (percentage) by experimental group and site in San Cristobal district}
    \label{fig:figure_d03}
\end{figure}

\clearpage

\begin{figure}[ht]
    \centering
    \includegraphics[width=\textwidth]{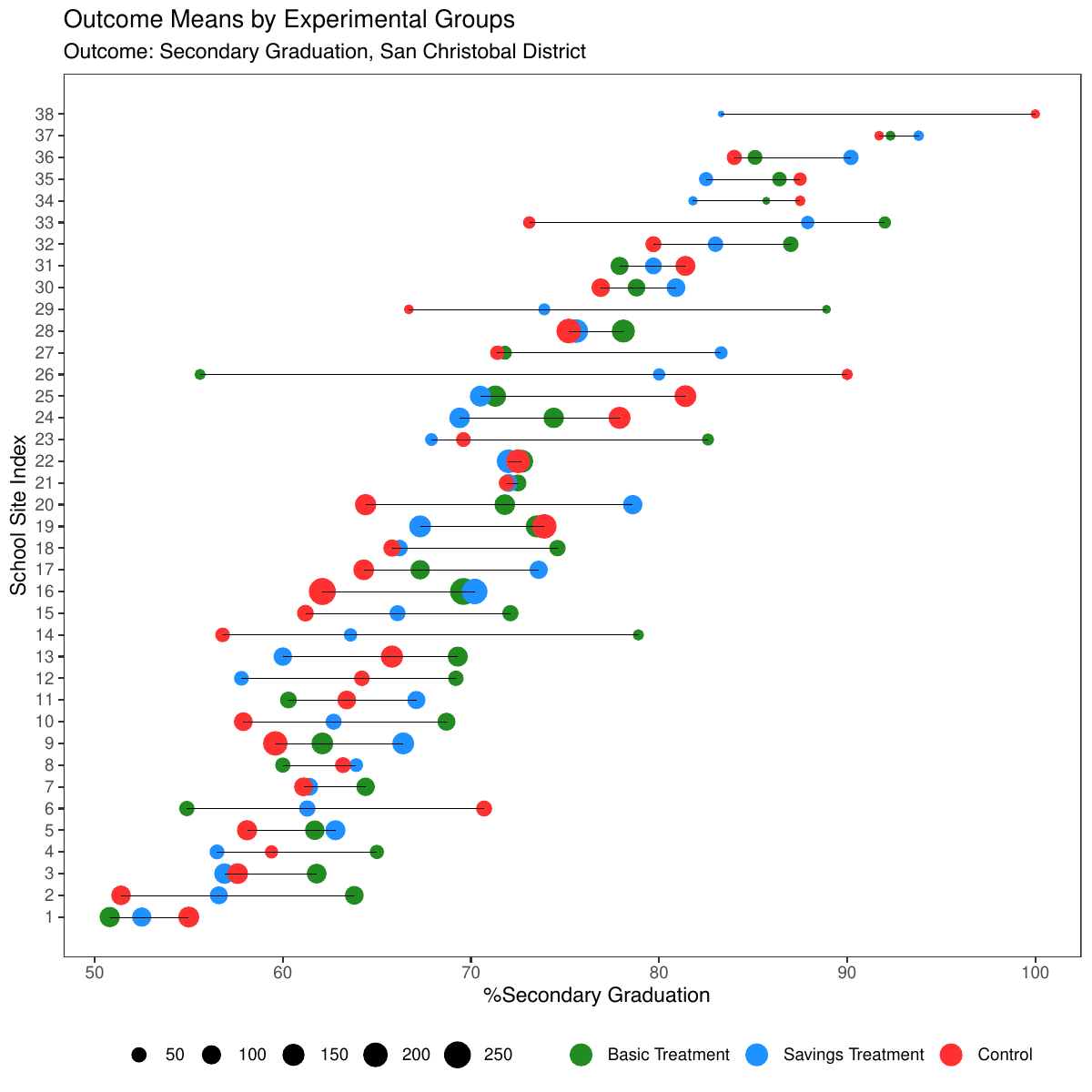}
    \caption{Secondary graduation outcome mean (percentage) by experimental group and site in San Cristobal district}
    \label{fig:figure_d04}
\end{figure}

\clearpage

\begin{figure}[ht]
    \centering
    \includegraphics[width=\textwidth]{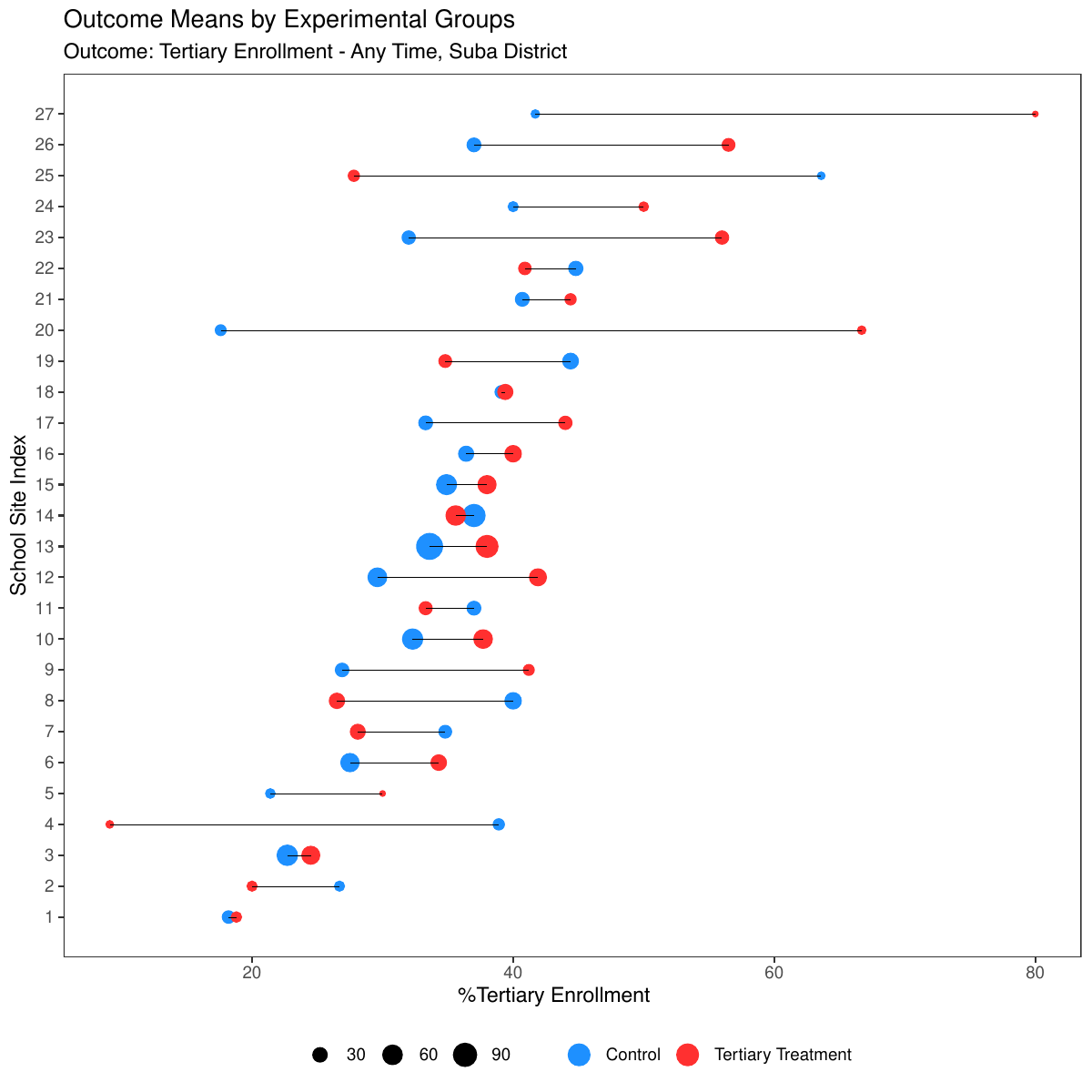}
    \caption{Tertiary enrollment outcome mean (percentage) by experimental group and site in Suba district}
    \label{fig:figure_d05}
\end{figure}

\clearpage

\begin{figure}[ht]
    \centering
    \includegraphics[width=\textwidth]{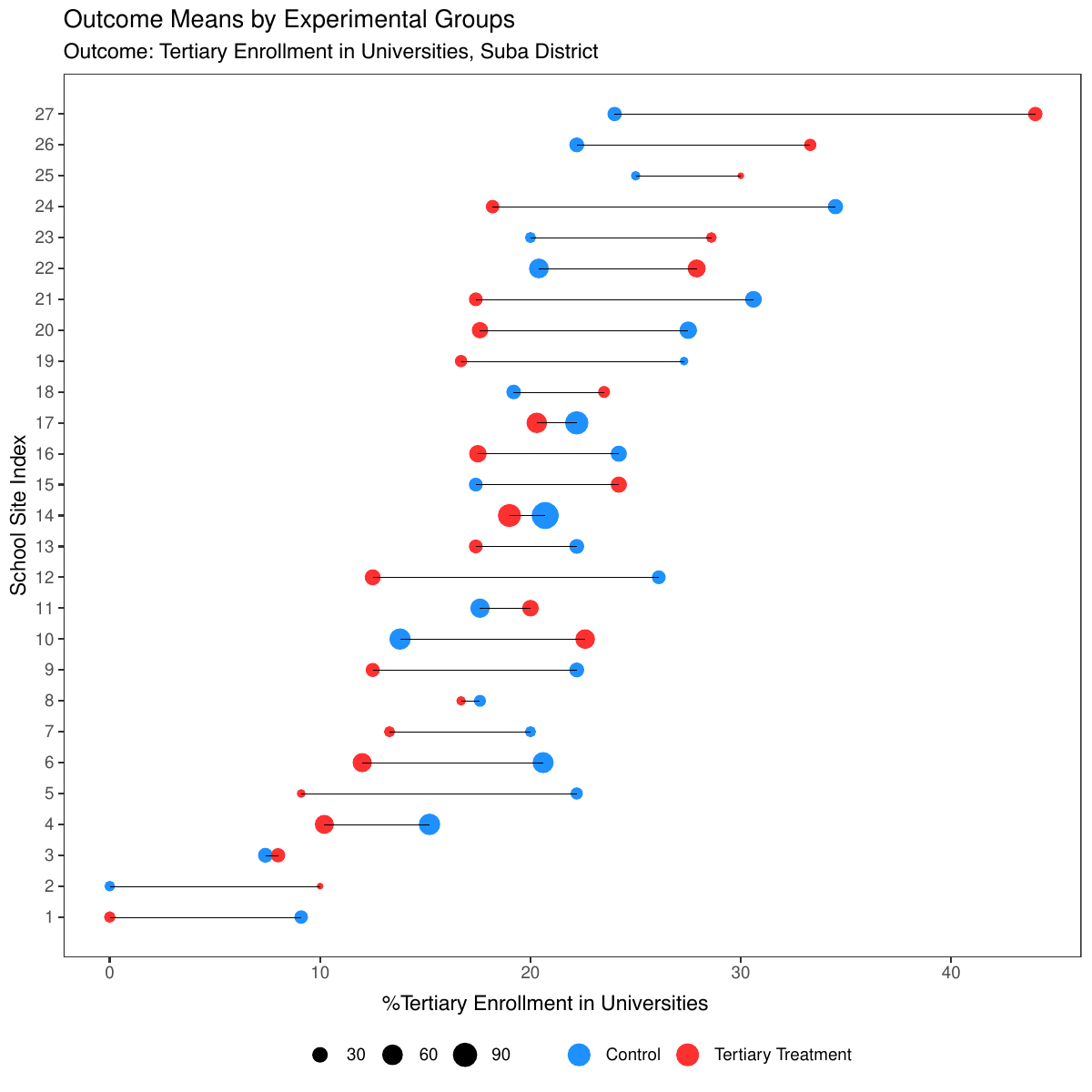}
    \caption{Tertiary enrollment in universities outcome mean (percentage) by experimental group and site in Suba district}
    \label{fig:figure_d06}
\end{figure}

\clearpage

\begin{figure}[ht]
    \centering
    \includegraphics[width=\textwidth]{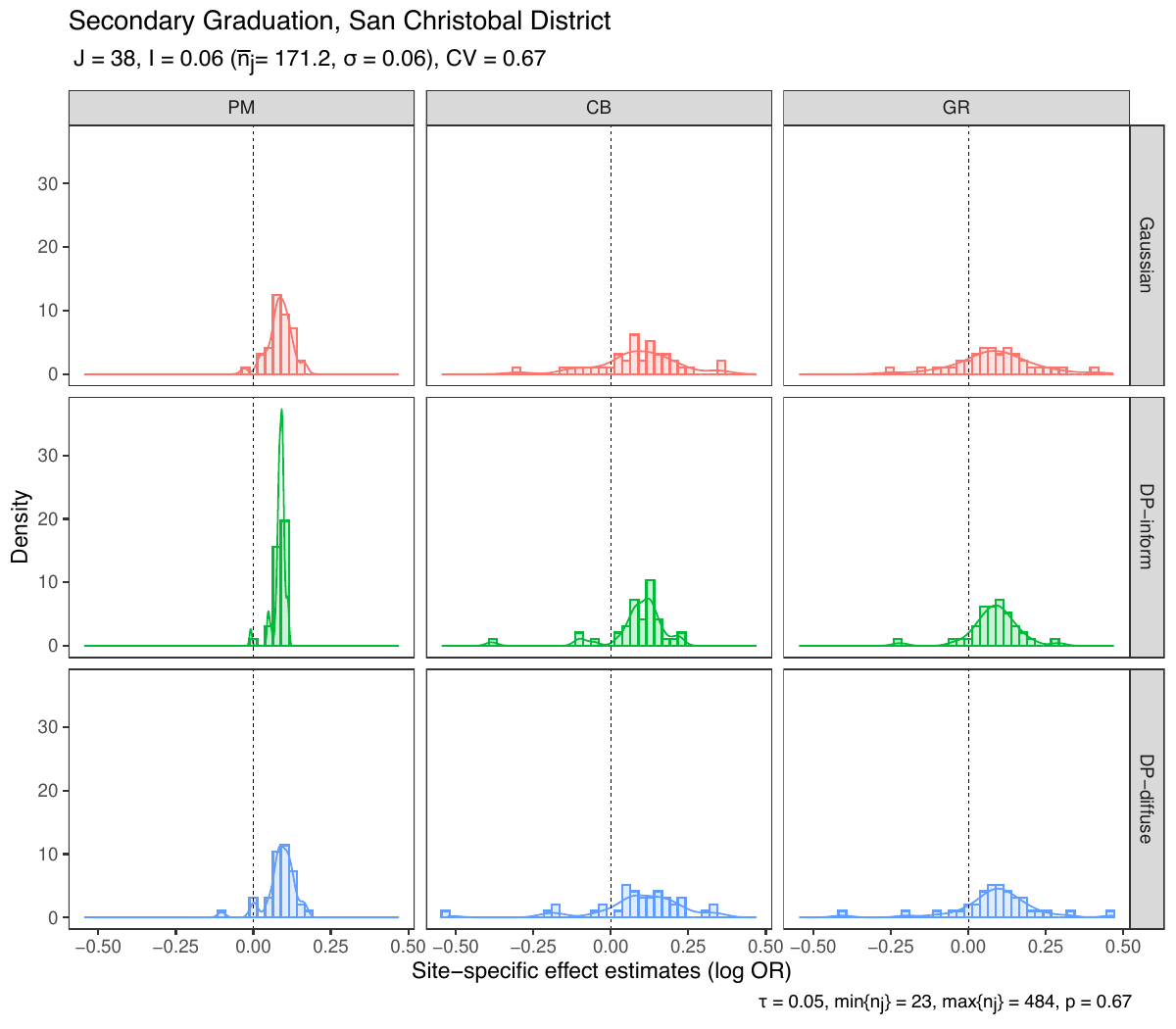}
    \caption{Distribution of site-speciﬁc treatment effect estimates on secondary graduation in San Cristobal district}
    \label{fig:figure_d07}
\end{figure}

\clearpage

\begin{figure}[ht]
    \centering
    \includegraphics[width=\textwidth]{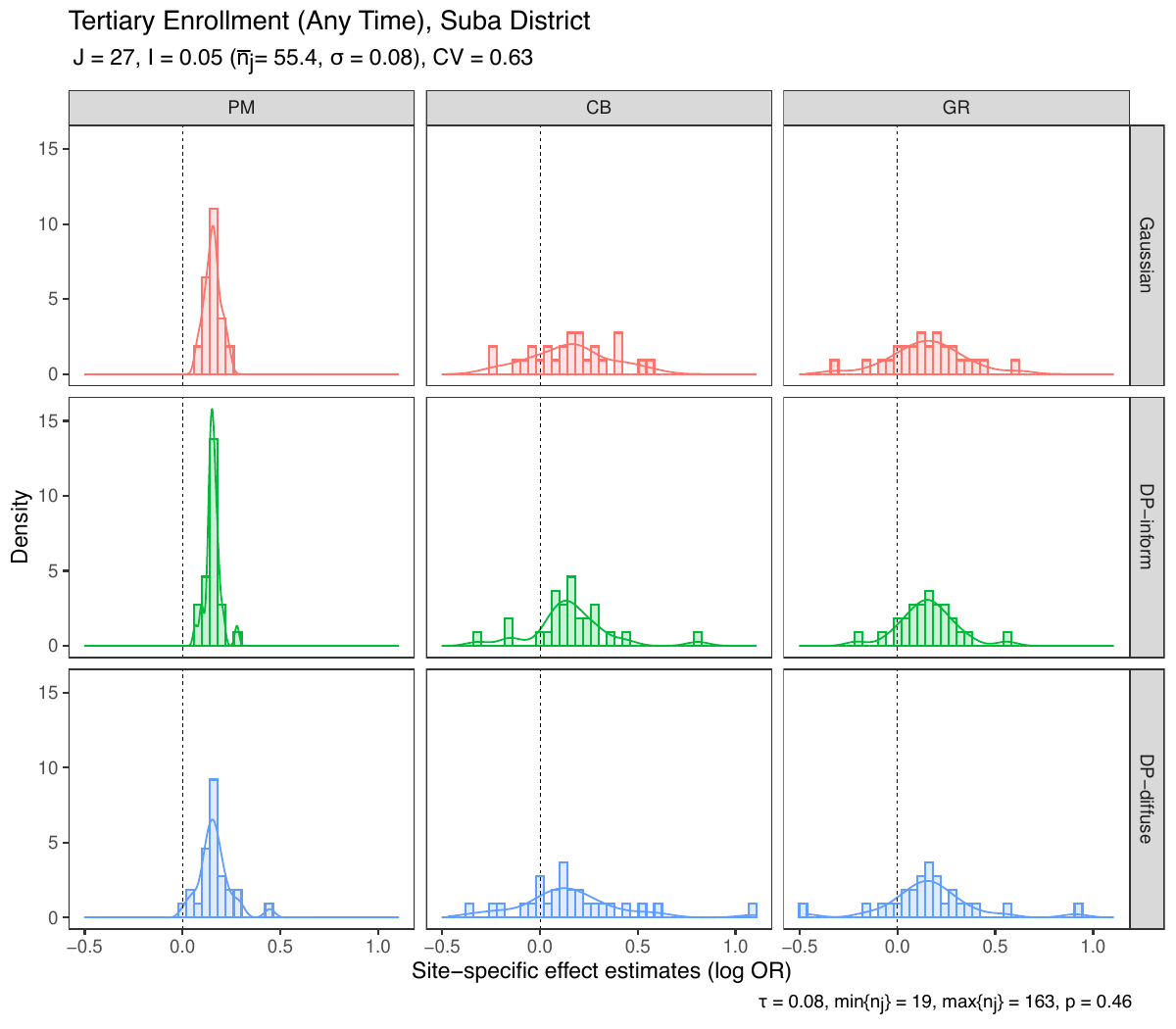}
    \caption{Distribution of site-speciﬁc treatment effect estimates on tertiary enrollment in Suba district}
    \label{fig:figure_d08}
\end{figure}

\clearpage

\begin{figure}[ht]
    \centering
    \includegraphics[width=\textwidth]{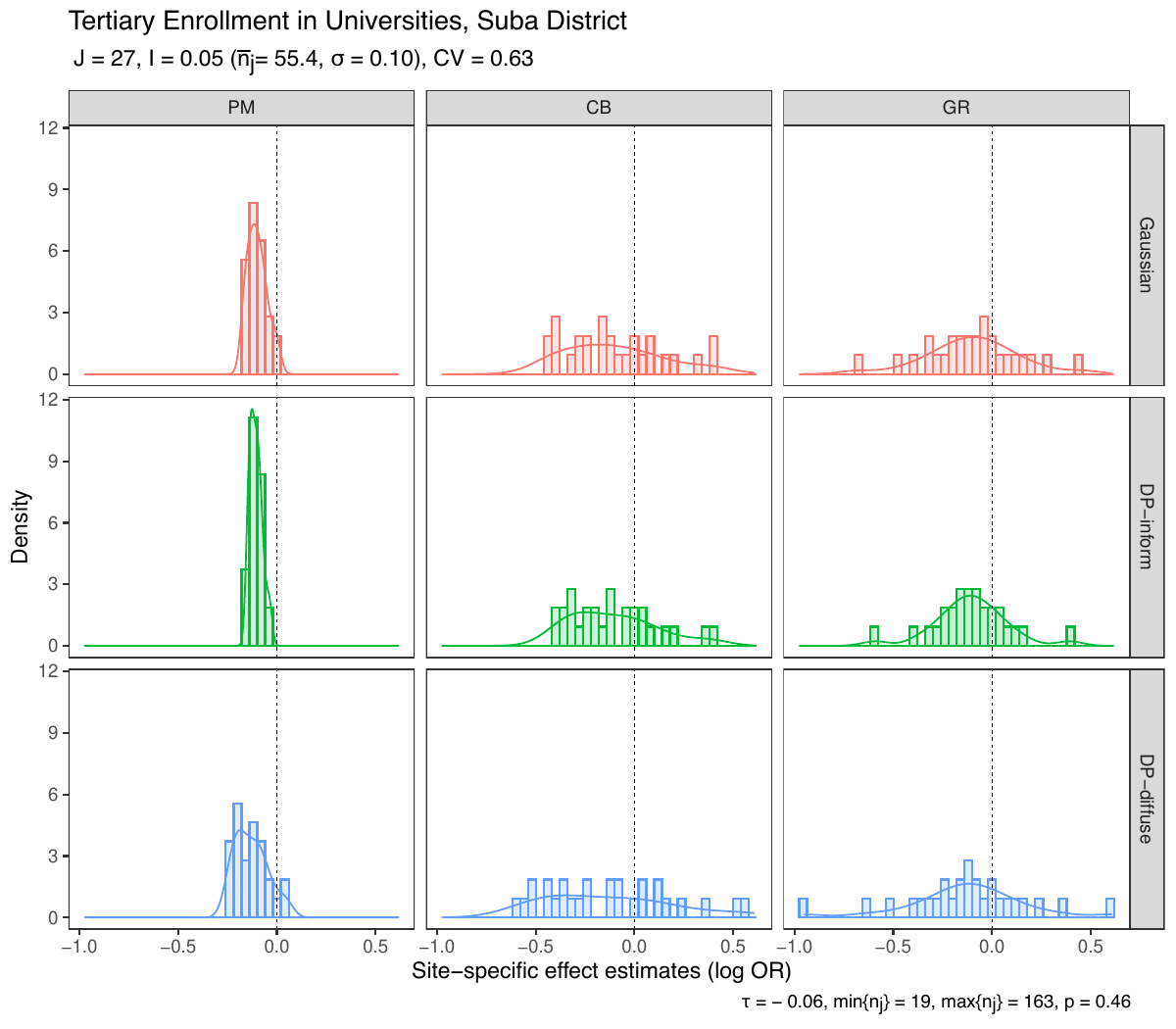}
    \caption{Distribution of site-speciﬁc treatment effect estimates on tertiary enrollment in universities in Suba district}
    \label{fig:figure_d09}
\end{figure}

\clearpage

\begin{figure}[ht]
    \centering
    \includegraphics[width=\textwidth]{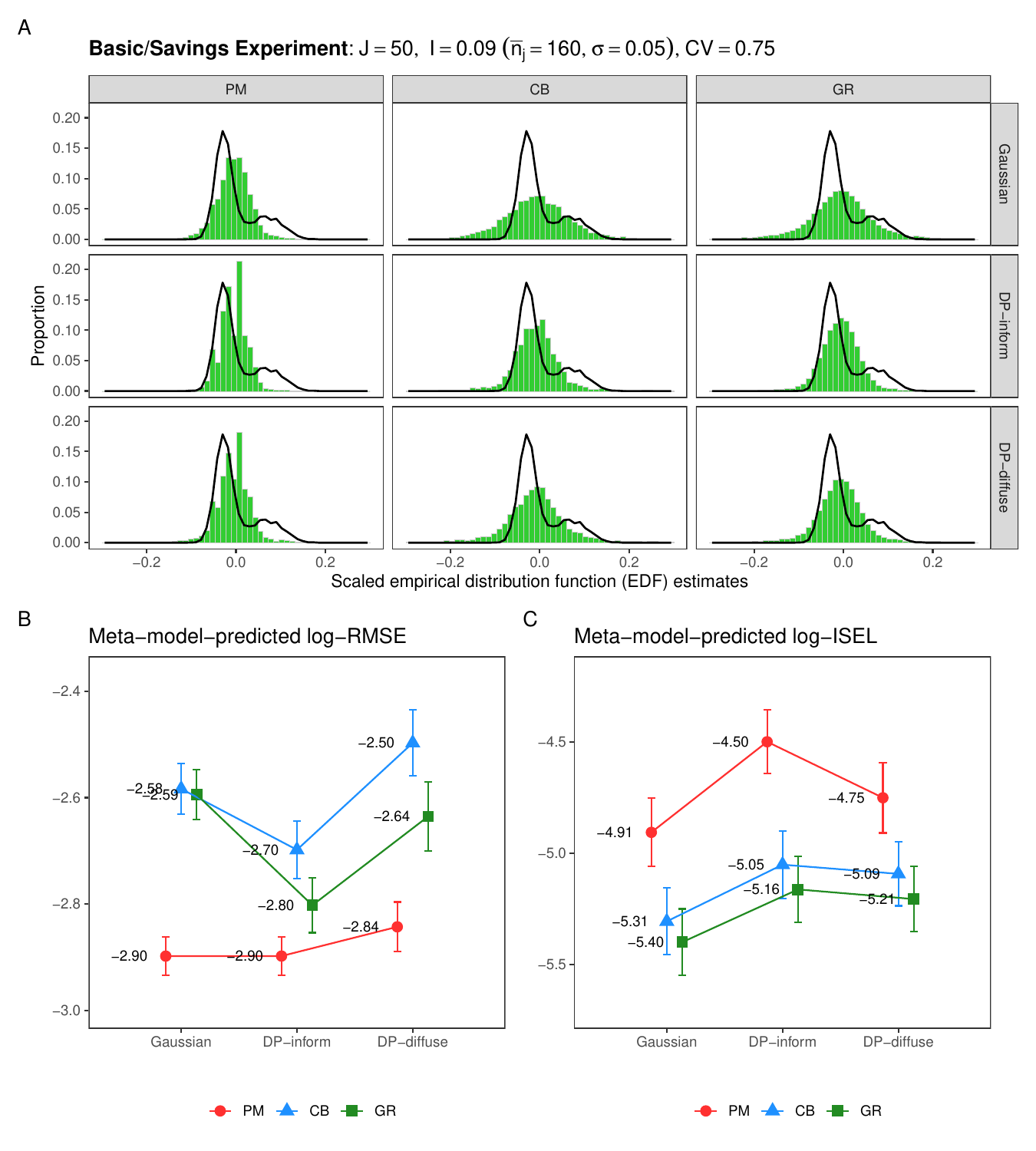}
    \caption{Scaled EDF estimates and meta-model-predicted RMSE and ISEL on log-scale for each model-method combination, when the data-generation is based on the scenario simulating the Basic/Savings Experiment in San Cristobal district ($J = 38$, $\bar{n}_j=171.2$, $\sigma=0.04$, and $\text{CV}=0.67$ for the outcome, on-time secondary enrollment). True $G$ is assumed to be Gaussian mixture. Plots include 95\% prediction intervals around predicted values.}
    \label{fig:figure_d10}
\end{figure}

\clearpage

\begin{figure}[ht]
    \centering
    \includegraphics[width=\textwidth]{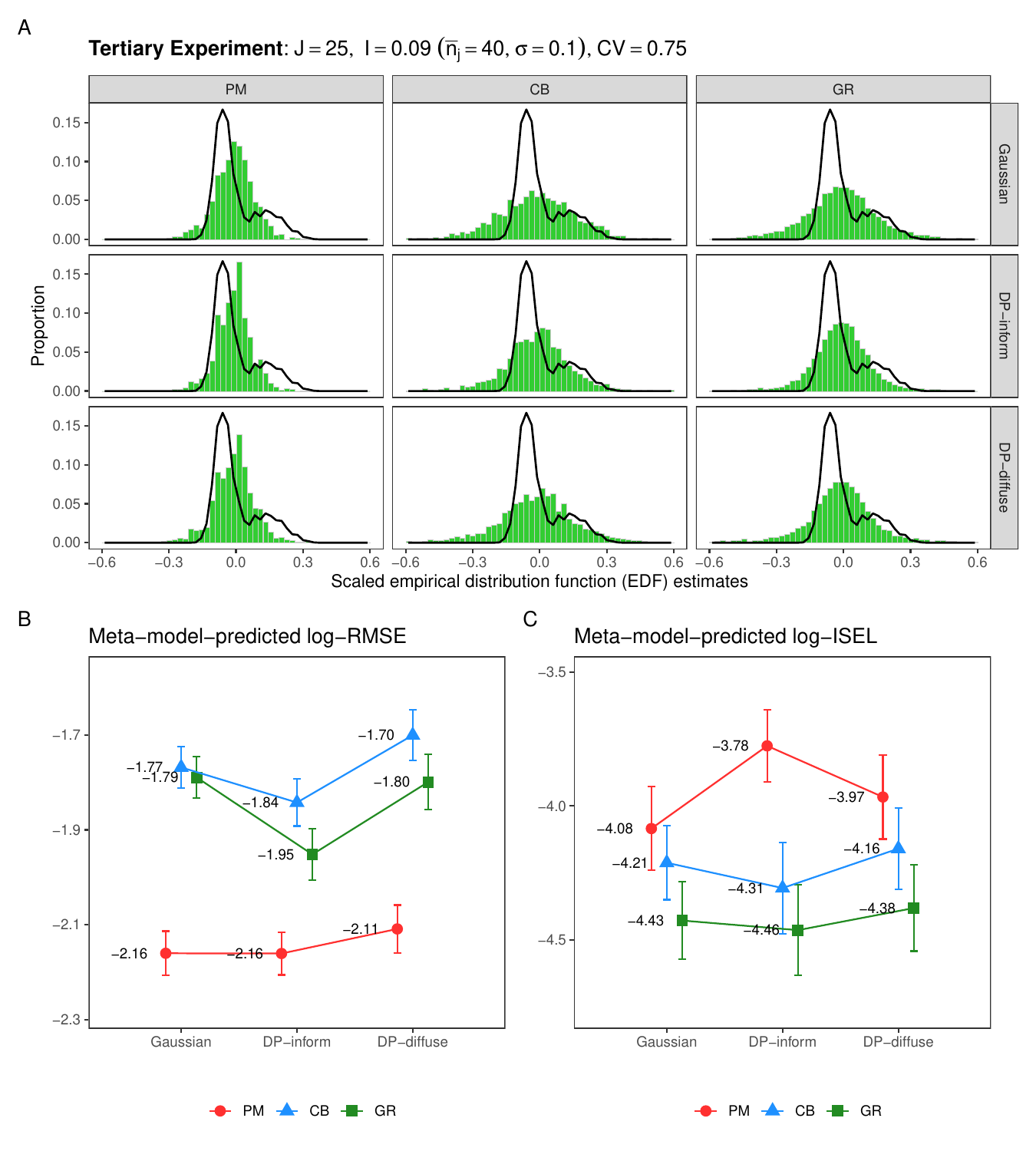}
    \caption{Scaled EDF estimates and meta-model-predicted RMSE and ISEL on log-scale for each model-method combination, when the data-generation is based on the scenario simulating the Tertiary Experiment in Suba district ($J = 27$, $\bar{n}_j=55.4$, $\sigma=0.10$, and $\text{CV}=0.63$ for the outcome, tertiary enrollment in universities). True $G$ is assumed to be Gaussian mixture. Plots include 95\% prediction intervals around predicted values.}
    \label{fig:figure_d11}
\end{figure}

%%%%%%%%%%%%%%%%%%%%%%%%%%%%%%%%%%%%%%
%%% End of document
%%%%%%%%%%%%%%%%%%%%%%%%%%%%%%%%%%%%%%
\end{doublespace}
\end{document}